\documentclass[12pt,preprint]{aastex}
\usepackage{graphicx}
\usepackage{color}           

\newcommand{\be}{\begin{equation}}
\newcommand{\ee}{\end{equation}}
\newcommand{\bea}{\begin{eqnarray}}
\newcommand{\eea}{\end{eqnarray}}
\newcommand{\ben}{\begin{enumerate}}
\newcommand{\een}{\end{enumerate}}

\newcommand{\grad}{ {\bf \nabla } }

\begin{document}
\title{Flux Accretion and Coronal Mass Ejection Dynamics}

\author{Brian T.~Welsch}
\email{welschb@uwgb.edu}
\affil{Dept. of Natural \& Applied Sciences, Univ. of Wisconsin -
  Green Bay, Green Bay, WI 54311}
\affil{Space Sciences Laboratory, University of California, Berkeley,
  CA 94720-7450}

\begin{abstract}
Coronal mass ejections (CMEs) are the primary drivers of severe space
weather disturbances in the heliosphere.
Models of CME dynamics have been proposed that do not fully include
the effects of magnetic reconnection on the forces driving the ejection.
Both observations and numerical modeling, however, suggest that
reconnection likely plays a major role in most, if not all,
fast CMEs.
Here, we theoretically investigate the accretion of magnetic flux
onto a rising ejection by reconnection involving the ejection's
background field.
This reconnection alters the magnetic structure of the ejection and
its environment, thereby modifying the forces acting upon the
ejection, generically increasing its upward acceleration. The modified
forces, in turn, can more strongly drive the reconnection.  This
feedback process acts, effectively, as an instability, which we refer
to as a reconnective instability.
Our analysis implies that CME models that neglect the effects of
reconnection cannot accurately describe observed CME dynamics.
Our ultimate aim is to understand changes in CME acceleration in terms
of observable properties of magnetic reconnection, such as the amount
of reconnected flux.  This flux can be estimated from observations
of flare ribbons and photospheric magnetic fields.
\end{abstract}
%

%


\section{Introduction}
\label{sec:intro}

In a coronal mass ejection (CME), magnetic forces in the low corona
accelerate a hot ($\sim 1$ MK) mass ($\sim 10^{15}$ g) of magnetized
plasma at high speed (from a few hundred km sec$^{-1}$ at the slow end
of the observed range to 2000 km sec$^{-1}$ or more at the fast end)
into interplanetary space.
These events are the primary drivers of severe space weather
disturbances at Earth \citep{Kahler1992}.
While CMEs are believed to be driven by the release of magnetic energy
stored in electric currents in the solar corona ({\it e.g.},
\citealt{Forbes2000}), key aspects of their initiation and subsequent
evolution are not well understood.
Characterizing the processes at work during the eruption process is 
therefore essential to understand their dynamics.

Here, we present a model of the development of fast CMEs, the
``flux accretion'' CME model, which describes how reconnection affects
the dynamics and structure of fast CMEs as they form.
Simply put, the central thesis of the model is that magnetic
reconnection underneath a rising ejection can accelerate the ejection
both: (i) directly, by momentum transfer from the reconnection outflow
jet; and (ii) indirectly, from the reconnection-modified magnetic
structure of the ejection and its surrounding magnetic field.
As explained in detail in subsequent sections, we believe the second, indirect
acceleration mechanism is the dominant influence.
The model is not directly focused on CME initiation, but rather on the
evolution of CMEs after their rise has been triggered.

The effect of reconnection on CME dynamics is ignored in models of CME
acceleration that only include forces that would be present if
  reconnection did not modify the ejection and its background
  field. For instance, \citet{Kliem2006} model CMEs as the expansion
  of a torus. Their model requires reconnection to occur for the torus
  to ``slide'' through overlying field, but they neglect dynamic
  effects due to changes in the ejection's magnetic field arising from
  this reconnection.  Most importantly, reconnection would cause
  poloidal flux that was initially external to the expanding torus to
  become entrained with it. This additional flux would produce a hoop force
  leading to greater acceleration of the torus than would occur
  without the added poloidal flux.


We argue that reconnection should fuel further reconnection, as
magnetic flux external to the ejection is pulled into the low-density
void created by the ejection's increasingly rapid rise.
Note that this ``pull'' reconnection ({\it e.g.}, \citealt{Yamada1997,
  Kusano2012}) is driven by the evolution of the global system; no
local driver of reconnection inflow is needed.
We remark that hints of the CME's rapid acceleration driving further
reconnection to occur were seen in simulations that exhibited
numerically problematic cavitation in the wakes of CMEs after
reconnection began \citep{MacNeice2004}: the numerical difficulties
arose when very low densities occurred near strong Lorentz forces in
the outflow region.
Note also that the sites of reconnection will tend to rise with time,
as the low-density region trails the rising ejection.

The reinforcing interplay between the upward motion of the ejection
and the reconnection --- feedback --- is effectively a macroscopic
instability, which we which we refer to as a {\em reconnective
  instability}.
We adopt this term because it is distinct from resistive
instabilities, such as the tearing mode \citep{Furth1963}, since this
feedback need not arise from the presence of resistivity {\em per se}:
in some magnetic field configurations, the presence of resistivity
(which enables reconnection) might not produce any feedback between
reconnection and large-scale dynamics.
For example, a current sheet (or ``current ribbon'' in 3D) should
develop when the footpoints of an initially potential pair of bipoles
are displaced a small distance, causing the coronal field, subject to
fixed field-line connectivity, to evolve slightly away from the
potential state ({\it e.g.}, \citealt{Longcope1996d}).  Onset of
reconnection in such a system should decrease the current in the
sheet.\footnote{It should be noted that near-potentiality is not required
  for reconnection to be quenched by large-scale dynamics. For
  instance, consider two bipolar, parallel, twisted flux systems in an
  ideal MHD equilibrium. With no flux shared between the two systems,
  a separatrix surface exists between them.  Now imagine that their
  footpoints were displaced slightly.  This would cause a current
  sheet to develop between them (or enhance any already present).
  Onset of reconnection between the two systems should lead to
  reduction of current in the sheet.}
In contrast, \citet{Longcope2014} found that ``kinematic and
quasi-static'' flux transfers via reconnection across a current sheet
in some configurations that they investigated ``resulted in an
increase in the size and strength of the current sheet.''  They noted
that if ``large-scale dynamics'' intensify some initial, local
reconnection process, then a ``resistive eruption mechanism'' could
operate.
\citet{Sturrock1989} qualitatively discussed a similar possibility.
This suggests that, depending on how a large-scale system responds to
a differential change in magnetic connectivity within the system, the
configuration might be ``reconnectively stable'' or ``reconnectively
unstable.''
Because this feedback effect would be inherently dynamic and
nonlinear, it is not easily tractable with analytic methods.
A large-scale reconnective instability would be distinct from both
ideal MHD instabilities, which assume the field's connectivity is
fixed, and resistive instabilities, which do not necessarily produce
feedback to the reconnection process.
In a sense, the breakout configuration proposed by \citet{Antiochos1999a}
was intentionally engineered to become reconnectively unstable as the
large-scale field is evolved by shearing on its boundary.
While reconnective stability could be investigated in equilibrium
configurations, a reconnective instability in CMEs might arise as a
secondary instability when the system is already evolving away from
equilibrium due to some other (perhaps ideal), primary instability
({\it e.g.}, a kink instability).

It is possible that global-scale dynamics might depend sensitively
upon the details of the reconnection process. In the pull-reconnection
framework, the ejection will rise regardless of how quickly flux
reconnects. The {\em speed} of this rise, however, likely {\em will}
be affected by the rate at which the operating reconnection mechanism
processes flux, because both the reconnection outflow and ejection's
structure will differ from cases in which reconnection proceeded at 
different rates.
This motivates theoretical or numerical investigations of
eruption dynamics with varied reconnection parameters.

Our model fits within the overall framework of the ``standard model''
of eruptive flares, often called the CSHKP model
(after Carmichael-Sturrock-Hirayama-Kopp-Pneumann, by \citealt{Svestka1992};
for a more recent review, see \citealt{Webb2012}),
%
%
which was essentially created to explain the formation and evolution
of the two-ribbon emission pattern frequently associated with CMEs in
terms of reconnection beneath a rising ejection.
In the model, this reconnection leads, either directly or indirectly,
to the acceleration of particles, which propagate downward along the
newly-reconnected magnetic fields toward the transition region and
upper chromosphere.  Here, they interact with the denser plasma to
create hard X-rays via bremsstrahlung and heat the surrounding plasma.
(\citealt{Hudson2011} and \citealt{Benz2008} provide excellent
reviews.)  The resulting intensity enhancements typically form two
parallel, elongated emission structures, known as flare ribbons.
Ribbons are often observed in H$\alpha$ and UV images ({\it e.g.}, in
the 1600 \AA~ channel of the TRACE satellite; \citealt{Handy1999}),
although some CMEs occur without ribbons or other obvious emission
signatures ({\it e.g.}, \citealt{Robbrecht2009}).
Flare ribbons always straddle the polarity inversion line (PIL; where
the photospheric radial magnetic field changes sign) beneath the
ejection. It is presumed that the ribbon emission occurs on
conjugate footpoints of newly reconnected magnetic flux.
Observations typically show ribbons moving apart from each other
(perpendicular to the PIL); in the CSHKP model, this arises naturally
as reconnection successively acts on flux tubes with footpoints
that are farther and farther apart.

Flare ribbons overlie strongly magnetized areas of the photosphere
within and near active regions (ARs), and it was long ago hypothesized
that the rate of (unsigned) magnetic flux being swept out by the
ribbons as they move apart should be related to the rate of coronal
magnetic reconnection (\citealt{Forbes1984, Poletto1986}; for a recent
observational study, see \citealt{Hinterreiter2018}).
%
%
In a study of 13 halo CMEs with flares, \citet{Qiu2005} reported a very
strong correlation between the total unsigned magnetic flux swept out
by ribbons over the course of each event --- which we hereafter refer
to as the ribbon flux --- and the CME speed: they found a linear
correlation coefficient of 0.89. In a related study, \citet{Qiu2004}
reported an association between the {\em rate} of flux being swept out
by ribbons and CME acceleration.  These findings are consistent: the
time integral of acceleration yields velocity, and the time integral
of the rate of magnetic flux being swept by ribbons ($\propto$
acceleration) yields the total magnetic flux swept by ribbons
($\propto$ velocity).
\citet{Bein2012} suggested a ``feedback relationship''
between CME acceleration and the rate of magnetic reconnection that,
in principle, is directly related to the rate of flux being swept out
by ribbons.
In numerical simulations of a CME's formation and acceleration,
\citet{Karpen2012} also found that their model ejection's acceleration
was closely tied to the onset of fast reconnection beneath the rising
ejection.
More recently, \citet{Lynch2016} found a similar association between
the onset of reconnection underneath a ``stealth'' ejection and
acceleration of that ejection.

The idea that reconnection can affect the acceleration of CMEs dates
at least to work by \citet{Anzer1982}, who noted that reconnection
under the eruption ``produces an increasing outward magnetic pressure
gradient'' across the CME.
In addition, as \citet{Moore2001} later noted, reconnection under an
eruption can also cut ``magnetic tethers'' (inward magnetic tension)
restraining an eruption.
%
%
\citet{Vrsnak2004} observed that a ``two-ribbon flare appears as a
consequence of ... fast magnetic field reconnection,'' and then
explicitly referred to a ``feed-back relationship between the CME
motion and the flare energy release.''
More recently, \citet{Inoue2018} invoked ``a nonlinear positive
feedback process between the flux tube evolution and reconnection,''
in which their numerically simulated ejection's rise drove tether
cutting reconnection that, in turn, enabled the ejection to rise
farther.
\citet{Zhang2006} also explored the idea of positive feedback between
reconnection and CME acceleration.  A concise passage in that paper's
concluding section discusses several ideas that we explore further
below.  We therefore quote this succinct paragraph in full:
\begin{quote}
  The temporal coincidence [between CME acceleration and flare
    radiative flux] also suggests that the CME run-away process and
  the magnetic reconnection process may mutually feed each other. The
  two processes not only start at the same time, but also end at
  almost the same time based on observations. The magnetic
  tether-cutting process is known to be very effective in accelerating
  the CME flux rope \citep{Vrsnak2004}. First, it reduces the tension
  of the overlying restraining field by cutting the tie with the
  photosphere. Second, it increases the magnetic pressure below the
  flux rope by adding the poloidal flux through reconnection. Third,
  it enhances the outward hoop force due to the curvature of the flux
  rope thanks to the poloidal flux added. The escape of the flux rope
  reduces the magnetic pressure below, and thus induces an inflow
  toward the central current sheet; the current sheet is caused by the
  magnetic stretch as a result of the rising flux rope. Therefore, a
  faster rise of the flux rope causes a faster inflow, which results
  in a faster tether-cutting reconnection. At the same time, a faster
  reconnection causes a stronger outward force and thus a faster CME
  acceleration. The whole process, through flux rope rising, magnetic
  field stretching underneath, inflow tether-cutting reconnection, and
  further accelerating, forms a closed loop of positive feeding, which
  leads to the simultaneous CME acceleration and flare energy release.
\end{quote}
\citet{Vrsnak2006} described the ejection process in very similar
terms.  Both pictures are mostly consistent with our view.
 In the pull-reconnection scenario that we envision, there is actually
 cavitation behind the rising ejection, between it and the
 reconnection site.  Nearer to the trailing edge of the rising
 ejection, concave-up post-reconnection flux can exert additional
 magnetic pressure on the ejection, thereby modifying the hoop force.
 While Zhang and Dere separate the pressure and hoop effects in their
 second and third items above, in our view these two points
 essentially refer to the same effect.
(In the context of CMEs, the hoop force was probably first introduced by
either \citet{Anzer1978} or \citet{Mouschovias1978}.)
In addition to these ideas, we also believe that upward-moving
post-reconnection flux below the core of the ejection exerts a
force on the ejection arising directly, from the momentum flux in the
reconnection outflow.
We discuss these issues in greater detail below.

Our chief aims here are (1) to highlight additional observations of
CME properties consistent with the key role of reconnection in
producing fast CMEs, and (2) to extend foregoing qualitative picture
by quantitatively characterizing how reconnection affects CME
dynamics.
%
%
What is new in this work?  To our knowledge, estimates
of reconnection-related changes to CME dynamics have not been analyzed in
the terms we consider: as functions of the properties of reconnecting
fields.
The remainder of this paper is organized as follows.  
In Section \ref{sec:observations}, we outline four observations that
motivate our view of magnetic reconnection as a key driver of many
eruptions, rather than a by-product of some other mechanism ({\it e.g.}, an
ideal instability) that powers ejections.
In Section \ref{sec:alternatives}, we discuss our model in the
framework of the CSHKP model, and distinguish aspects of the flux accretion 
model from other models of CME initiation and dynamics.
In Sections \ref{sec:outflow} and Sections \ref{sec:lorentz}, we use
simplistic models of reconnection and changes in CME structure,
respectively, to characterize the effects that reconnection can have
on CME dynamics.
In Section \ref{sec:conc}, we conclude with a brief discussion of the
significance of flux accretion for CMEs.

\section{Review of Relevant Observations}
\label{sec:observations}

The key observations implicating reconnection in the development of CMEs are:

\ben

\item {\bf Reconnection is typical when CMEs occur:} Patterns of
  emission in the solar atmosphere interpreted as signatures of
  reconnection (flare ribbons and post-flare arcades) are generic
  features of many, if not most, CMEs.  Observers have reported the
  existence of ``stealth CMEs'' ({\it e.g.}, \citealt{Robbrecht2009,
    Lynch2010}), which lack low coronal signatures (LCSs) such as
  dimmings, coronal waves, and flares.  Stealth events are believed to
  be a minority of CMEs. In a study of 34 events with STEREO,
  \citet{Ma2010} found that about 1/3 had no LCSs, and that the speeds
  of their stealth events were typically slow, below 300 km s$^{-1}$.   

\item {\bf Reconnection fluxes and CME speeds are correlated:} As
  noted above, CME velocity is correlated with ribbon flux.  In
  addition to the study by \citet{Qiu2005}, \citet{Gopalswamy2017b}
  estimated reconnection fluxes using post-eruption arcades and report
  correlation coefficients between reconnection fluxes and CME
  speeds near 0.6 in a sample of about four dozen CMEs.  In a sample
  of 16 events, most of which were also in the sample analyzed by
  \citet{Qiu2005}, \citet{Deng2017} computed a linear fit of CME speed
  as a function of ribbon flux, and found the no-ribbon-flux intercept
  to be about 550 km s$^{-1}$, roughly the speed of the slow solar
  wind.  This suggests that reconnection might be a key factor in
  accelerating CMEs to speeds much faster than the slow solar
  wind. Further, the close temporal association between CME
  accelerations and flare energy release reported by \citet{Zhang2006}
  --- major acceleration starts and ends as the flare energy release
  starts and ends --- implicates reconnection in the acceleration
  process.
  
  
\item {\bf CME masses and speeds are correlated:} The velocities of
  CMEs are correlated with their masses.  This mass-speed relationship
  was noted long ago, as in Figure 4 of \citet{Vourlidas2002}, which
  shows that the average mass of CMEs tends to be higher when their
  accelerations are larger.  (Since the fitted accelerations were
  assumed constant [\citealt{Vourlidas2000}], higher acceleration
  directly implies higher velocities.)  \citet{Vrsnak2008} also noted
  that ``the driving force is greater in more massive CMEs.''  The
  mass-speed relationship can also be verified directly from the CDAW
  CME
  catalog.\footnote{http://cdaw.gsfc.nasa.gov/CME\_list/UNIVERSAL/text\_ver/univ\_all.txt}
%
%
\begin{figure}[ht] 
  \centerline{\includegraphics[width=\textwidth]{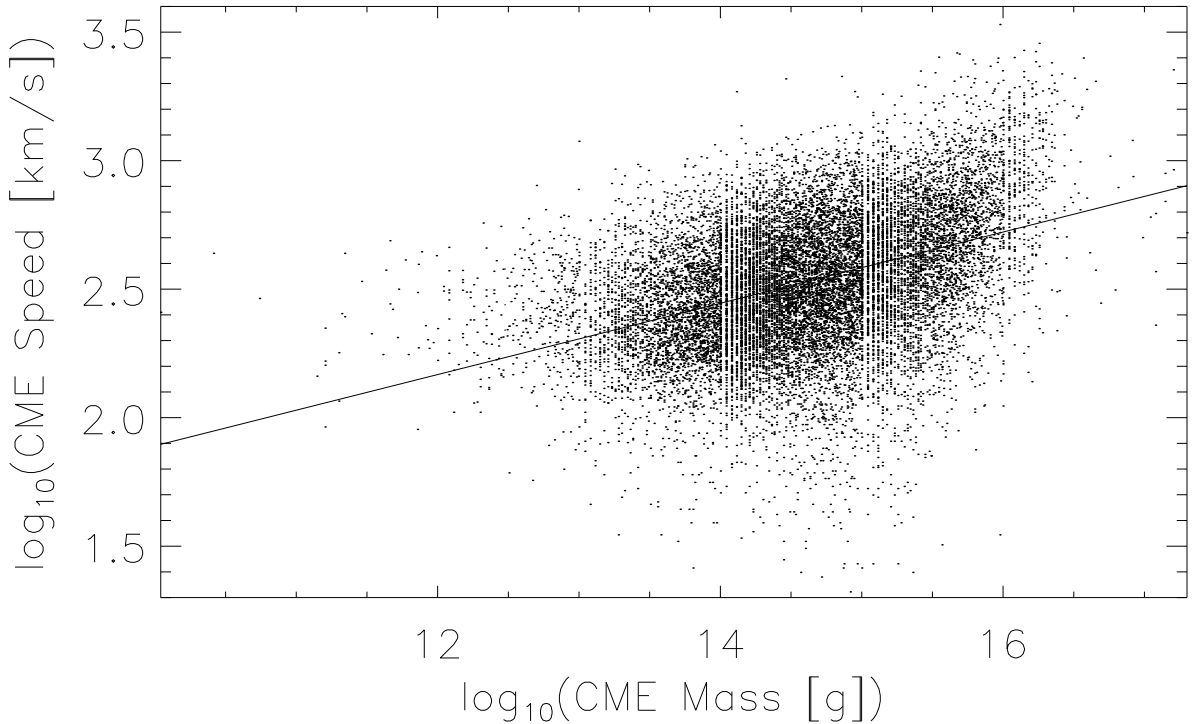}}
  \caption{A scatterplot of CMEs' linear speeds {\it versus} their masses,
    from the CDAW CME catalog. A least-absolute-deviation fit to the
    logarithms of each is overplotted.  Although a large amount of
    scatter is present, a trend is clear. (Quantization visible in
    this plot arises from two-digit precision used for masses in the
    catalog.  The effect is most visible where changes in the tenths
    digit are fractionally largest -- {\it i.e.}, just to the right of each
    power of 10.)  }
\label{fig:speed_vs_mass}
\end{figure} 
  Figure \ref{fig:speed_vs_mass} shows a scatter plot of speeds {\it versus}
  masses, and a relationship between the two can be discerned.  A
  least-absolute deviation fit to the logarithms of each was
  performed, and is overplotted.  For the $N$ = 17792 CMEs with masses
  and linear speeds listed in the catalog's text version from 1996 Jan
  - 2016 May, the linear and rank-order correlation coefficients
  between CME linear speeds and logarithms of mass were $0.42 \pm
  0.01$ and $0.40 \pm 0.1$, respectively, where the 1-$\sigma$
  confidence intervals were computed using Fisher's $z$-transform.
  While this correlation is statistically significant, it is not
  strong, indicating that factors other than mass strongly influence
  CME speed.
  Uncertainties in estimating CME masses and true speeds, made by
  assuming plane-of-sky position and velocity, respectively, introduce
  some scatter into this correlation.  We note that for halo
  events, the linear and rank-order correlations are stronger: 0.48
  $\pm 0.03$ and 0.55 $\pm 0.03$, respectively.  (We also note that
  these correlations do not arise from CMEs' velocities scaling with
  their sizes: the linear and rank-order correlations between non-halo
  CMEs' linear speeds and angular widths are weaker, at 0.29 and 0.18,
  respectively.)  Because magnetic flux is frozen to the plasma
  outside the small diffusion region where the reconnection occurs,
  magnetic flux that is accreted onto an ejection should increase its
  mass.  This process was modeled by \citet{Lin2004b}, who found that
  reconnection approximately doubles the ejection mass compared to the
  mass of the pre-eruptive structure in their model.  They also found
  more mass added when a stronger background magnetic field was
  present, due to the higher inflow speed.  An example of mass loading
  from flare reconnection can be seen in the lower panels of Fig. 4 of
  the breakout simulations described by \citet{MacNeice2004}.
  It should be noted, though, that causality might run the other
  way: the mass-speed correlation might arise because faster CMEs
  become more massive by collecting ambient material at their leading
  edges, via a ``snow-plow'' effect.  Entrainment of solar wind plasma
  onto the CME is, however, uncertain ({\it e.g.}, \citealt{Feng2015}.)  One
  difference between the processes would be the density distribution:
  snow plowing should increase density at the CME's front, while 
    mass loading from the reconnection outflow would increase the
  density in the trailing half of the CME. It is probable that both
  processes are at work. (Reconnection also adds mass in what was
    previously external flux at the eruption's leading edge to the
    ejection, but would not, by itself, strongly affect the mass
    density at that leading edge.)
  But this does suggest an observational question: how does mass
  distribution within CMEs compare between fast and slow CMEs?

  \item {\bf Reconnection fluxes and ICME poloidal fluxes are
    correlated:} As reported by \citet{Qiu2007}, the ribbon fluxes in
    CMEs are correlated with the poloidal magnetic flux inferred from
    {\em in situ} flux-rope fitting.  Gopalswamy {\em et al.}  (2017a,
    2017b) \nocite{Gopalswamy2017a, Gopalswamy2017b} found similar
    results. These associations are consistent with the picture that
    the key properties of the structures of interplanetary CMEs
    (ICMEs) that are measured {\em in-situ} are produced during the
    eruption process.  The physical picture is that flux overlying an
    eruption becomes entrained with the eruption via magnetic
    reconnection beneath the rising ejection, and this entrained flux
    forms the poloidal flux in interplanetary flux ropes. If true,
    then the flux in flare ribbons, which reflects the amount of flux
    reconnected in the corona, should closely match the poloidal flux
    in interplanetary flux ropes.  We also remark that {\em in situ}
    observations suggest that interplanetary flux ropes possess
    several turns ({\it e.g.}, eight in a case studied by
    \citealt{Larson1997}), far beyond the number expected to be stable
    in a pre-eruptive structure.
      
\een

\section{Flux Accretion and Existing CME Models}
\label{sec:alternatives}

Many models of CME initiation have been proposed, and we discuss some
below.  While some of these focus on the initiation of CMEs, and not
necessarily their further development and acceleration, it is still
worthwhile to consider their consistency with the observations above.
All share a common property: the pre-eruption state is marked by a
balance between outward magnetic pressure from the confined fields
that will erupt and inward magnetic tension from overlying
fields.
In each case, evolution of some component of the system disrupts this
balance.
Because magnetic reconnection will generally modify the structure of
an erupting flux system, we refer to the core of the fields that will
erupt as a proto-ejection --- a not-yet-fully formed eruptive
structure.  After an eruption has begun, the reconnection process
causes the magnetic structure of the material that is erupting to
evolve.  

\ben

\item The tether-cutting model \citep{Moore1992, Moore2001} is
  focused on the initiation of CMEs.  In it, initial reconnection
  within sheared fields of a highly stressed magnetic configuration
  removes downward tension that was inhibiting the upward expansion of
  the stressed fields.  Having thus cut some of its tethers, the
  proto-ejection that was poised for rapid rise is free to do so.  
  Reconnection then plays an ongoing role in the eruption.
  \citet{Moore1992} state that continuing reconnection:
  \begin{quote}
    ``further untethers the core field, providing a positive feedback
    that sustains the magnetic explosion ... The whole process of
    coordinated eruption and reconnection is driven by the magnetic
    pressure of the unleashed core field.''
  \end{quote}

  The role of reconnection is described similarly in the later paper by
  \citet{Moore2001}:
  \begin{quote}
    ``During the explosion, the reconnection within the core of the
    sigmoid progressively cuts more and more of the tethers, allowing
    the unleashed part of the core field to expand upward, the new
    short loops to implode downward, and the crossed arms of the
    sigmoid and surrounding inner envelope field to flow into the
    reconnection site.''
  \end{quote} 

  
  These descriptions do suggest a basis for the observed correlation
  between ribbon fluxes and CME speed: more reconnection cuts more
  tethers, enabling faster escape of the proto-ejection.
  But they ignore any effect from the increased hoop force on the
  ejection due to flux added by the reconnection.  As with the model
  of \citet{Kliem2006} mentioned above (and discussed further below),
  this view holds that reconnection removes downward forces
  restraining the ejection but does not result in any additional
  upward forces driving the ejection. The acceleration of the ejection
  predicted by the tether-cutting view would differ from that if the
  hoop force were increasing during the eruption.  The
  tether-cutting model also does not account the observed correlation
  between CMEs' masses and speeds.  If the sole effect of reconnection
  is to cut tethers, why should more massive CMEs be faster?

\item In the framework of the breakout model \citep{Antiochos1999a},
  reconnection occurs in two regions: initially, it occurs only above
  a proto-ejection, which removes {\em strapping fields} that were
  inhibiting its rise.  Strapping fields essentially act as tethers,
  but instead of being cut by reconnecting below the proto-ejection,
  as in the tether cutting model, they are cut from above by
  reconnection with fields external to the erupting system.
  Subsequently, after the proto-ejection has begun to rise, {\em flare
    reconnection} begins beneath the rising ejection.  This flare
  reconnection is a generic feature of many CME simulations, including
  those outside the breakout framework, such as the flux cancellation
  model ({\it e.g.}, \citealt{Amari2010}).  Accordingly, flare reconnection
  is not a unique characteristic of the breakout model.  Unlike the
  tether cutting model, the breakout model provides no natural
  explanation of why CME speeds should be correlated with ribbon
  fluxes.  Like the tether cutting model, the breakout model does not
  predict the correlation between CME speed and ejection mass.

\item As noted above, \citet{Kliem2006} describe an eruption evolving
  via the torus instability that involves reconnection.  But their
  description does not incorporate any changes to the force driving
  the escaping torus due to the effects of reconnection on the
  magnetic structure of the escaping field. Thus, it provides no
  explanation for the correlations between either CME speed and ribbon
  flux, or CME speed and mass.

\item A kink instability might lead to onset of an eruption ({\it e.g.},
  \citealt{Williams2005}).  Since this initiation mechanism does not,
  by itself, address the subsequent evolution of an eruption, this
  mechanism also cannot directly account for the correlations between
  CME speed and either ribbon flux or ejection mass.
  
\een

Again, these models are focused on CME {\em initiation}, not
subsequent evolution of the eruption.
Our model also operates within this conceptual framework, although our
primary focus is on the development rather than the initiation
of the eruption.
Our primary criticism of the models of CME processes above is that
they are incomplete --- {\it i.e.}, that they should be modified, not rejected.
It should also be remarked that all the models above operate within the
``storage-and-release'' paradigm, in which the coronal field possesses
enough magnetic energy prior to the onset of the eruption
to power it --- {\it i.e.}, any external forcing might trigger an eruption,
but would not supply any significant energy to it.

The consequences of reconnection for CME dynamics have been modeled
before.  \citet{Lin2000} analytically considered 2.5D reconnection in
Cartesian geometry in a current sheet that formed in the wake of a 
flux rope that had suddenly jumped to a higher position due to loss of
equilibrium at a lower, initial position.  They note that some amount
of reconnection is necessary for the flux rope to escape, and modeled
the reconnection kinematically.  With a reconnection rate near 0.1
$M_A$, where $M_A$ is the upstream Alfv\'en Mach number of the inflow,
their modeled evolution was similar to that observed in long-duration
eruptive events.  \citet{Lin2005} analyzed sequences of EUV imager,
spectrograph, and coronagraph observations to constrain the properties
of reconnection in a current sheet in the wake of a CME and reported
reconnection rates in the range 0.01 - 0.23 $M_A$, broadly consistent
with those modeled by \citet{Lin2000}.

An alternative CME paradigm was proposed by \citet{Chen1996}, in which
a pre-existing, stable coronal flux rope would be driven to erupt
ideally by a sudden increase of the poloidal flux wrapping around the
rope. In this geometry, the field along the axis of the flux rope is
toroidal, a nomenclature we will use below.  In Chen's model, this
flux increase is driven ideally, by photospheric flows at the time of
the eruption, in effect supplying power to drive the CME as it
happens.  A careful analysis of observed photospheric flows near the
time of a CME by \citet{Schuck2010} found the predictions of the model
to be strongly inconsistent with the data, with model parameters and
observed quantities differing by orders of magnitude.
As \citet{Vrsnak2004} and \citet{Forbes2006} noted, however, coronal
reconnection during a CME can add poloidal flux to an erupting flux
system much as Chen's hypothesized flows would.  If so, then aside
from its unrealistic driving mechanism, the model's predictions of CME
dynamics could be accurate in some cases ({\it e.g.}, \citealt{Chen2006}).


\section{Momentum Transfer from the Reconnection Outflow}
\label{sec:outflow}

Although many numerical models of CMEs exhibit flare-type
reconnection, relatively little is known about the significance of
this reconnection in accelerating ejections, {\it versus} acceleration that
would have occurred by ideal mechanisms alone --- {\it e.g.}, after
strapping fields were reconnected away in the breakout model, or after
sufficient photospheric flux were canceled in the flux cancellation
model.  While some amount of reconnection is essentially inevitable in
numerical MHD models, we suggest that its nearly ubiquitous presence
in simulations of CMEs is related to its key role in the eruption
process on the actual Sun.
We note that the role of reconnection {\it versus} ideal evolution could be
investigated in studies with a ``reconnection-controlled'' model, like
the FLUX finite-element code employed by \citet{Rachmeler2010}.

In this section and the next, we investigate how magnetic reconnection
in the wake of a rising ejection --- flare reconnection --- can
accelerate that ejection.

We focus in this section on how flare reconnection can directly
transfer momentum from the outflow jet to the ejection, thereby
accelerating it.
In the standard CSHKP model of an eruptive flare, magnetic
reconnection occurs below a rising ejection, with outflow jets
directed upward and downward.  The downward-moving flux eventually
forms post-flare loops, and the upward-moving flux eventually merges
with the rising ejection.
Reconnection likely occurs in a kinetic-scale diffusion region, within
which magnetic flux is not frozen into the plasma.  

\begin{figure}[ht] 
  \centerline{\includegraphics[width=\textwidth]{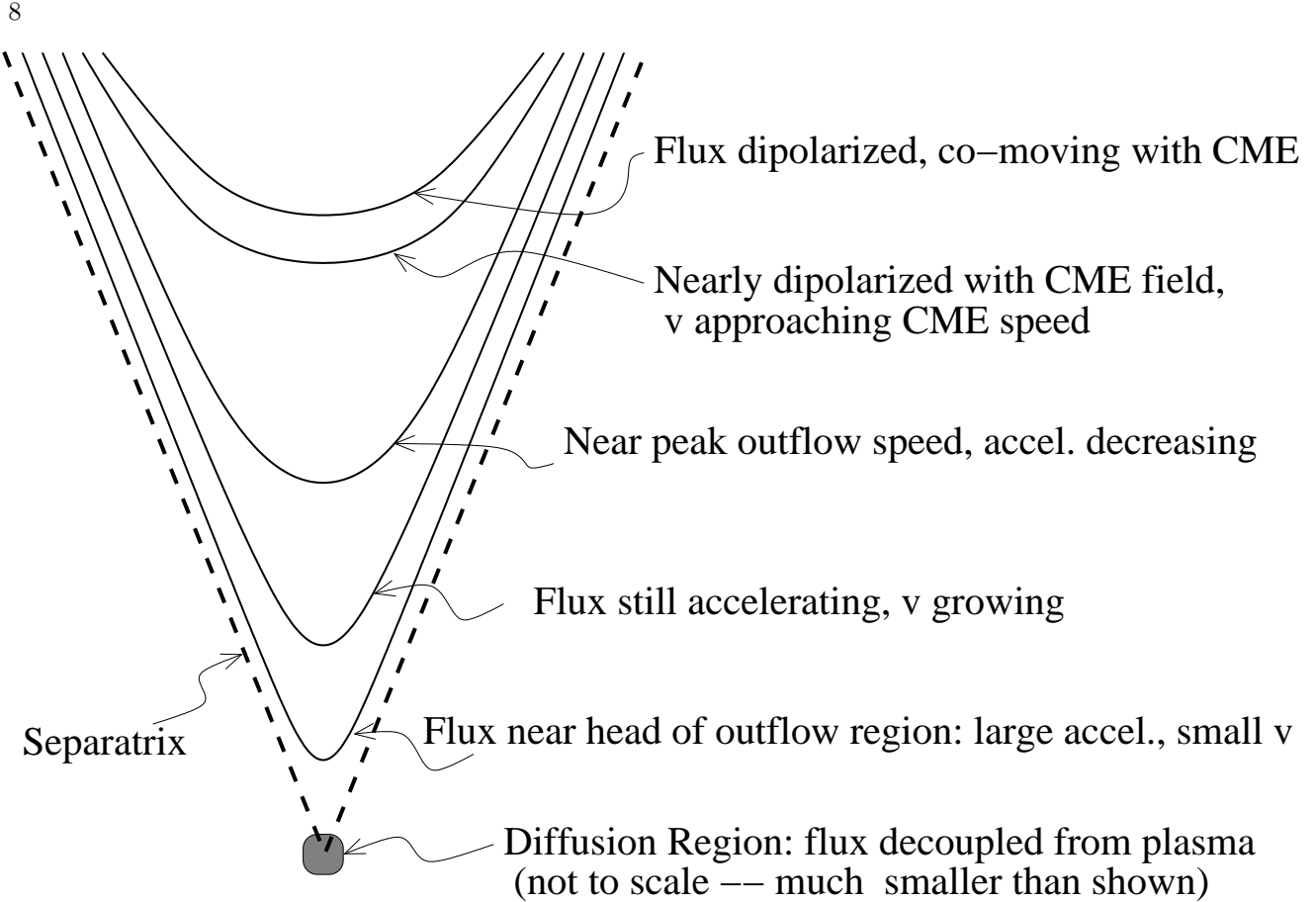}}
  \caption{A schematic illustration of post-reconnection flux moving
    through the outflow region at equally spaced time intervals. The
    diffusion region shown is unrealistically large compared to the
    scale of magnetic variations shown here: in the actual solar
    corona, kinetic scales (on the order of meters) are far below the
    scale of observable structure (on the order of a few hundred
    kilometers).  The opening angle of the outflow region was chosen
    arbitrarily.  The key point is that the dipolarization of
    post-reconnected fields occurs over a much larger area than the
    diffusion region.}
\label{fig:rx_geometry}
\end{figure}

As post-reconnection flux first enters the outflow region, field lines
are $\Lambda$- or V-shaped (for downward and upward outflows,
respectively), with sharp kinks near the diffusion region.
While the plasma flow directly exiting the diffusion region should be
essentially Alfv\'enic, magnetically connected plasma over a larger
volume will be moving much more slowly.
But the strong magnetic tension from the large-scale kink in the field
will rapidly accelerate magnetically linked plasma to the
perpendicular Alfv\'en speed, $v_{A \perp} = \sqrt{B_\perp^2/(4 \pi
  \rho)}$.
Figure \ref{fig:rx_geometry} illustrates the geometry.
(As usual, we define $B_\perp$ to be the component of the magnetic
field in the 2D plane containing the X-point associated with the
reconnection.)  This process of {\em dipolarization} ({\it e.g.},
\citealt{Priest2000}) implies
a nonzero momentum flux ${\cal F}_{\vec p}$ is present in
each outflow region, with magnitude
\be {\cal F}_{\vec p} = (\rho v_{A \perp}) v_{A \perp} = \frac{B_\perp^2}{4 \pi}
~. \label{eqn:p_flux} \ee 
%
%
A key point is that the dipolarization occurs over a much larger scale
than the diffusion region, with the acceleration coming from the
large-scale structure of the magnetic field.
Eventually the upward-moving flux catches up to the slower-moving
ejection, and slows to the proto-ejection's speed --- in the process,
transferring its excess momentum to the proto-ejection as it merges with
it.

Integrated over a solar eruption, how much momentum might be
transferred by this mechanism?  \citet{Leroy1983} and
  \citet{Casini2003} report typical axial magnetic fields of $\sim 10$
  G or more along the axes of non-active-region coronal prominences
  above the limb, with fields tilted by about 20$^\circ$ with respect
  to prominence axes.  If a prominence with a 10 G field erupts as a
  CME, a perpendicular field strength of $\sim 3$ G is plausible.
%
%
Assuming $B_\perp = 3 G$ implies a momentum transfer from the outflow
jet of roughly $1$ (g cm s$^{-1}$) per cm$^{2}$ per second.
%
%
%
While the diffusion region is likely on ion kinetic scales (perhaps on
the order of a few meters for typical fields in the corona), the
outflow only becomes Alfv\'enic after it expands and accelerates.
Hence, momentum transport from the outflow occurs over macroscopic
area, and might be a few Mm across and a few tens of Mm long, meaning
this could occur over an area of $\sim$ (3 $\times 10^8$ cm)(3 $\times
10^{9}$ cm) $ \simeq 10^{18}$ cm$^2$.
\citet{Zhang2006} studied acceleration in a few dozen CMEs, and report
the median duration of the ``main'' acceleration phase of 50 minutes,
implying the strongest burst of momentum flux might perhaps last 3000
s.  These roughly estimated parameters imply a net momentum transfer
near $3 \times 10^{21}$ g cm s$^{-1}$.

A CME of mass $10^{15}$ g and speed of 300 km s$^{-1}$ ({\it e.g.},
\citealt{Vourlidas2000}) has a momentum of $3 \times 10^{22}$ g cm
s$^{-1}$.  This suggests momentum transfer from the reconnection
outflow might supply $\sim 10$\% of the momentum of a CME with typical
mass and modest speed.  Different choices for the reconnection
  field strength, the area of the outflow region or duration of the
reconnection could substantially alter this very uncertain
estimate.

Many CMEs are much faster than this, but such CMEs typically originate
from within active regions, where coronal field strengths might be a
factor of 10 or more larger.
Given the $B_\perp^2$ scaling of ${\cal F}_{\vec p}$ in Equation
\ref{eqn:p_flux}, it is possible that momentum transfer from the
reconnection outflow supplies a substantial part of an active region
CME's final momentum.
The size of the reconnection outflow region can be larger in active
regions, too. To cite a large case, \citet{Aschwanden2001} report the
length of the post-flare arcade in the Bastille Day event of July 2000
to be about 200 Mm. The width of the upward outflow region in the
simulations of \citet{Karpen2012} is nearly the width of the sheared
PIL at the model's base.  If similar photosphere-to-coronal
scaling is present on the Sun, the width of the accelerated outflow in
the Bastille Day event might still be about 3 Mm. The axial field
might be of order 100 G.  Because the reconnection typically proceeds
from more sheared to less sheared field lines ({\it e.g.},
\citealt{Aschwanden2001, Su2007a}), the average reconnecting component
might be 30 G.
The associated GOES curves show a rise time of a bit less than 20
minutes, implying momentum transfer over $\sim$ 1000 s.  With these
parameter choices, the momentum transfer would be about $5 \times
10^{23}$ g cm sec$^{-1}$. This is about 20\% of the CME's momentum of
$2.3 \times 10^{24}$ g cm sec$^{-1}$, from the CDAW CME catalog's
linear speed and mass of 1674 km sec$^{-1}$ and 1.4 $\times 10^{16}$
g, respectively.
  
The observations discussed in Section \ref{sec:observations} imply
that both a CME's mass and speed are correlated with reconnected flux,
so its momentum should be, too.
So momentum transfer from flare-reconnection outflow is partly
consistent with the observations.  Our parameter estimates, however,
are very uncertain, so we cannot tightly constrain the significance of
the reconnection outflow for CME dynamics.
We expect that more detailed assessments of the role of the
reconnection outflow in CME acceleration can be investigated by
analyzing the results of existing numerical models of CMEs. For
instance, in the simulations performed by \citet{Karpen2012}, the
morphology of outflows where the reconnection jet meets the body of
the ejection in their Figure 14 suggests that the jet's flows distort
--- and therefore exert forces on --- the ejection.

\section{Reconnection-Induced Changes in External Forces}
\label{sec:lorentz}

In the CSHKP model, reconnection in the wake of a rising ejection
alters field line connectivity between the background magnetic field
and the ejection.  In general, this modifies the forces on the
ejection, a scenario that we investigate in this section.
We emphasize that models that do not account for changes in forces on
CMEs arising from reconnection-driven field evolution must be
incomplete: their predictions for CME height {\it versus} time, for
instance, will not be based upon valid physics.
A key point is that we explicitly assume that the eruption is already
underway: the proto-ejection is already moving upward, due to an
unbalanced force, so our focus is not on initiation. 

When flux surrounding an ejection reconnects underneath it, some of
the reconnected flux then becomes entrained with the ejection.  We
refer to this entrainment as flux accretion.  This accretion can
simultaneously increase the outwardly directed forces driving the
ejection and decrease the inwardly directed forces restraining it, for
some of the reasons outlined in the text passage from
\citet{Zhang2006} cited above.
As noted previously, because magnetic flux is frozen to the plasma
outside the small diffusion region where the reconnection occurs, flux
accretion will increase the mass and size of an ejection.
Recently, \citet{Compagnino2017} used statistical matching criteria to
associate flares and CMEs based upon their timing, and report a
correlation between flare radiative flux in the GOES 1 -- 8 \AA \, band
and CME mass. This is consistent with the idea that reconnection both
leads to flare emission and adds mass to associated CMEs.

\begin{figure}[ht] 
  \centerline{\includegraphics[width=\textwidth,clip=true]{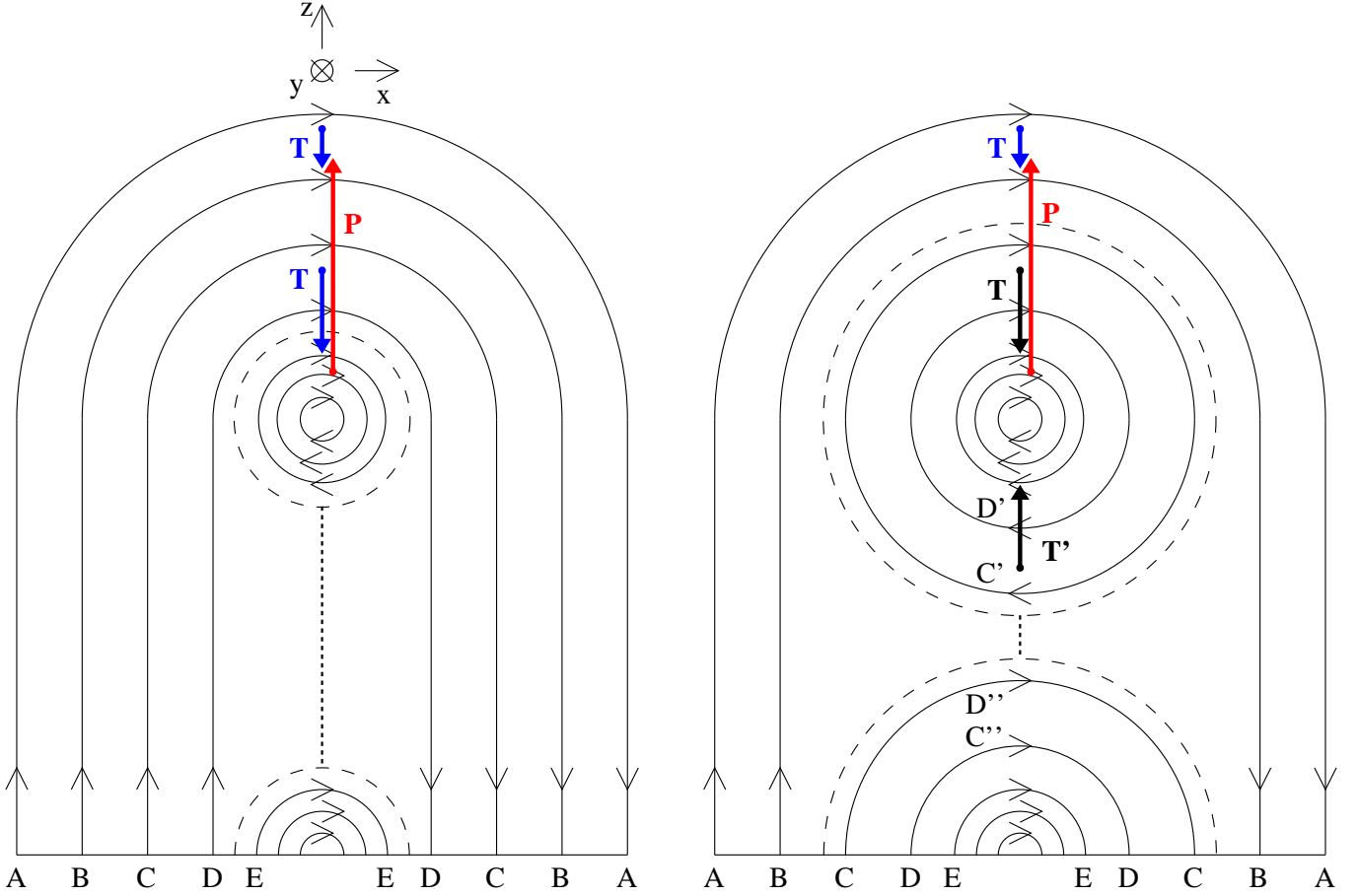}}
\caption{{\it Left:} A schematic illustration of some magnetic force
  densities present in a 2D, non-force-balanced magnetic
  configuration, at a given time $t_i$.  By hypothesis, the
  configuration is not in equilibrium, and a net upward Lorentz force
  acts upon the CME (in the $z$ direction here). The {\it solid black
    lines} show a sample of magnetic field lines, meant to
  approximately convey field strengths. The {\it dashed black lines} show
  two separatrix surfaces, one of which encloses the ejection (closed,
  dashed circle). The {\it red vector} shows the upward component of then
  magnetic pressure gradient within the ejection due to upwardly
  decreasing field strength, and {\it blue vectors} show the inward magnetic
  tension forces at two points that act to restrain the ejection. The
  {\it vertical dotted line} denotes a current sheet in the wake of the
  rising ejection.  Here, field lines' shapes are unphysical; our
  focus is on their connectivity.
    {\it Right:} An illustration of some magnetic force densities
      present in a 2D, non-force-balanced magnetic configuration, but
      at a slightly later time, $t_i + \Delta t$.  In this case,
      coronal magnetic flux that was anchored at the photosphere
      between points C and D in the {\it left} panel has reconnected in the
      wake of the rising ejection.  Hence, the flux between C and D
      that previously overlay the ejection has been accreted onto the
      ejection, and now lies within the dashed circle enclosing the
      ejection.  The newly reconnected, concave-up flux near the base
      of the ejection, between the points C' and D', exerts an upward
      magnetic tension, shown acting at the point labeled {\bf T'},
      that cancels the previously unbalanced downward magnetic tension
      from flux above the ejection, shown acting at the point labeled
      {\bf T}.  This reconnection implies that the downward magnetic
      tension {\bf T} that formerly restrained the ejection as an
      external force acting upon it is now an internal force that
      cancels with the newly created upward magnetic tension force
      {\bf T'}.  Since some restraining force was removed, the net
      upward force on the ejection is therefore larger.
    %
}
\label{fig:pre_post_2d}
\end{figure}

Figure \ref{fig:pre_post_2d} qualitatively illustrates an idealization
of this process.  In the figure, we assume that a finite amount of
reconnection has occurred, but we now ignore the dynamical relaxation
that would occur with realistic reconnection, since that effect was
discussed above.  Hence, in this sketch, the post-reconnection field
lines have already dipolarized.
This assumption of very simplistic field geometries for the pre- and
post-reconnection fields greatly simplifies our comparisons between
the two.
It should be noted that each vector shown in Figure
\ref{fig:pre_post_2d} represents a force density ({\it i.e.}, the force per
unit volume at a given point).

To quantitatively investigate the process, we now analyze the momentum
equation for the ongoing ejection.  We assume that the combined
configuration of the CME and its background field is not in
equilibrium, and a net upward Lorentz force acts upon the CME. Some
process having triggered the eruption, we then wish to consider the
effects of reconnection upon its further development.

A key challenge arising from this assumption is that simplifying
assumptions typically adopted to study pre-eruptive configurations ---
for example, that the magnetic field is ``force-free,'' {\it i.e.}, that
$(\vec{J} \, \times \, \vec{B}) = 0$ (see, {\it e.g.},
\citealt{Schrijver2008}) do not apply.  Unfortunately, it is
difficult, without resorting to numerical simulations, to model the
spatial variation of $\vec B$, as would be required to analyze how
reconnection modifies the $(\vec{J}\, \times \, \vec{B})$ force.

Consequently, we
%
%
idealize the ejection as an object distinct from its coronal
surroundings, and estimate the net forces on it. Newton's second law
for the CME at a given instant in time $t_i$ is given approximately by
\be \frac{d\vec p}{dt} =
-(P_{\rm out} S_{\rm out} - P_{\rm in} S_{\rm in})
+ \vec F_{\rm Lorentz, ext} - m \vec g
- \vec F_{\rm drag}
~, \label{eqn:n2_full} \ee
where: $\vec p$ is the CME's momentum, equal to $m_{\rm eff} \vec v$;
$\vec v$ is the CME's center-of-mass velocity,
$m_{\rm eff}$ is the CME's effective mass, equal to the CME's mass $m$
plus its virtual mass ({\it e.g.}, \citealt{Forbes2006}, equal to the mass
of the background plasma that it displaces);
%
%
%
$P_{\rm in}$ and $P_{\rm out}$ are the plasma (gas) pressures on the
CME's inner (or back, or bottom) and outer (or front, or top) 
  surfaces, respectively, and $S_{\rm in}$ and $S_{\rm out}$ are the
corresponding surface areas;
$\vec g$ is the gravitational acceleration due to the Sun at the
location of the CME's center of mass;
%
$\vec F_{\rm Lorentz,~ext}$ is
the Lorentz force {\em on} the ejection {\em by} external fields;
and $F_{\rm drag}$ is the drag on the CME.  
%
%
%
We further assume that: the coronal plasma is magnetically dominated
({\it e.g.}, \citealt{Forbes2000}), so that the gravity and pressure
gradient terms can be neglected; and that drag on the rising
proto-ejection is negligible in the high-$\beta$ plasma near the Sun
while reconnection is ongoing \citep{Cargill1996}.  We then have
\be \frac{d\vec p}{dt} = \vec F_{\rm Lorentz,~ext}
\label{eqn:n2} ~. \ee
Although we have assumed a net upward Lorentz force is exerted
on the proto-ejection, external fields exert both downward and
upward forces on it.
The dominance of Lorentz forces is only expected to hold low in the
corona.

We next consider the {\em change} $\Delta \vec F_{\rm Lorentz}$ in
net external Lorentz force on the CME due to the reconnection of a small
amount of flux $\Delta \Phi$, assumed to occur over a small time
interval (relative to the duration of the eruption) $\Delta t$.
Observations of supra-arcade downflows (SADs) thought to represent
cross-sections of magnetic structures in reconnection outflow suggest
the reconnection is patchy \citep{McKenzie2009}, and creates
structures on length scales on the order of a few Mm
\citep{Savage2011}.
Considering a single reconnection event in isolation is consistent
with assuming that the reconnection is patchy \citep{Linton2006,
  Linton2009}.
%

\subsection{Reduction in Downward Magnetic Tension}
\label{subsec:tension}

First, as emphasized in Figure \ref{fig:pre_post_2d}, some component
of the downward magnetic tension acting on the CME is canceled after
reconnection occurs.  The force density from downward magnetic tension
prior to the reconnection was
\be \vec f_{\rm tension} = \frac{1}{4\pi} (\vec B \cdot \grad) \vec B
~, \label{eqn:tension_density} \ee
integrated over the volume containing the to-be-reconnected flux
overlying the ejection.  (Here and elsewhere we will use $\vec f$ for
force densities, and $\vec F$ for forces, which are volume-integrated
force densities.) To simplify this analysis, we first assume no guide
field (a component of the field into the plane of the figure) is
present where the reconnection occurs, an assumption we will relax
later. We refer to this field as purely poloidal, and denote the
  reconnecting field's strength as $B_p$.
This means the component of the field into the page is not
contributing to the change in tension due to the reconnection that we
compute here.   Idealizing the overlying flux tube as forming a
half-toroid with major radius $R$ and constant cross-sectional area $\Delta A$
along its length, 
%
%
the pre-reconnection force $\Delta F$ in the downward direction
integrated over the sub-volume that overlies the ejection
and whose flux reconnects is then
\bea \Delta F_{\rm tension}
&=& \frac{1}{4\pi} \int dV \, | \hat z \cdot (\vec B \cdot \grad) \vec B |
\label{eqn:df_tension0} \\
&=& \frac{1}{4\pi} \int dV \, \frac{B_P^2}{R} \, \left | \hat z \cdot
\frac{\partial \hat b}{\partial \theta} \right | \\
&=& \frac{1}{4\pi} \int \frac{B_P^2}{R} dA \int_{-\pi/2}^{\pi/2} R\,
d\theta \cos \theta  \\
&\simeq& \frac{B_P \, \Delta \Phi}{4\pi} \left ( \frac{1}{R} (2R) \right )  
= \frac{\Delta \Phi B_P}{2\pi} ~, \label{eqn:df_tension} \eea
where $\hat b = \hat \theta = \cos (\theta) \hat x - \sin (\theta)
\hat z$, $\theta$ is the angular position around the ejection's axis
(with $\theta = 0$ along $\hat z$, and increasing clockwise), $dV = dA
\, R \,d\theta$, $\Delta \Phi = B_P \, \Delta A$ is the flux that will be
reconnected, and we approximate the tension term's magnitude as $B_P^2$
over the radius of curvature, $R$ \citep{Spruit1981}.

Physically, this force is not eliminated by reconnection in our
idealized model.
Rather, this small component of the overall downward force
is {\em canceled} by the
formation of concave up flux, resulting in a net {\em change} in
Lorentz force directed {\em upward} equal to $\Delta \Phi B/2 \pi.$
Strictly, the pre-reconnection downward tension force density was
present in a volume {\em external} to the ejection, so it did not act
directly on the ejection itself (as, for instance, gravity and
external pressure can).  For the ejection to rise, however, the forces
driving it would have had to also accelerate plasma in this external
volume upward, too. By both incorporating this external flux into the
erupting system and canceling its downward force, reconnection thereby
reduces this impediment to the ejection's rise.

We note that our result is a simplistic approximation, for a few
reasons.
First, it is clear that any realistic reconnection model would not
yield the geometries depicted in Figure \ref{fig:pre_post_2d}, from
which the result in Equation \ref{eqn:df_tension} was derived.
As discussed in the previous subsection, post-reconnection field lines
start with sharp vertices, then dipolarize while traversing the
outflow region and become more rounded.  The field geometry shown in
Figure \ref{fig:pre_post_2d} is meant represent the
already-dipolarized state, to enable characterizing the effect of
reconnection on forces that act on the ejection independent of the
outflow-driven momentum flux that we already analyzed.
Nonetheless, the already-dipolarized, post-reconnection fields
underneath the ejection will not precisely mirror fields above the
ejection, so the upward and downward forces will not exactly cancel as
we have described.
In fact, from the typical decrease in total pressure (gas plus
magnetic) with height in the corona, the field strength below the
ejection's core will be greater than that above the core --- meaning
the upward-directed tension force would be stronger.
This point will be revisited in Section \ref{subsec:hoop} below.

A second complication is that the ejection will rise during the time
interval $\Delta t$ that the reconnection and subsequent
dipolarization occur.
The combination of this rising motion with the decreasing total
pressure in the solar atmosphere with height, which we have ignored,
will cause the ejection's structure to evolve over $\Delta t$.
Here, we simply ignore changes in the CME's structure due to its
upward displacement during this brief interval.

\begin{figure}[htp] 
  \centerline{\includegraphics[width=\textwidth,clip=true]{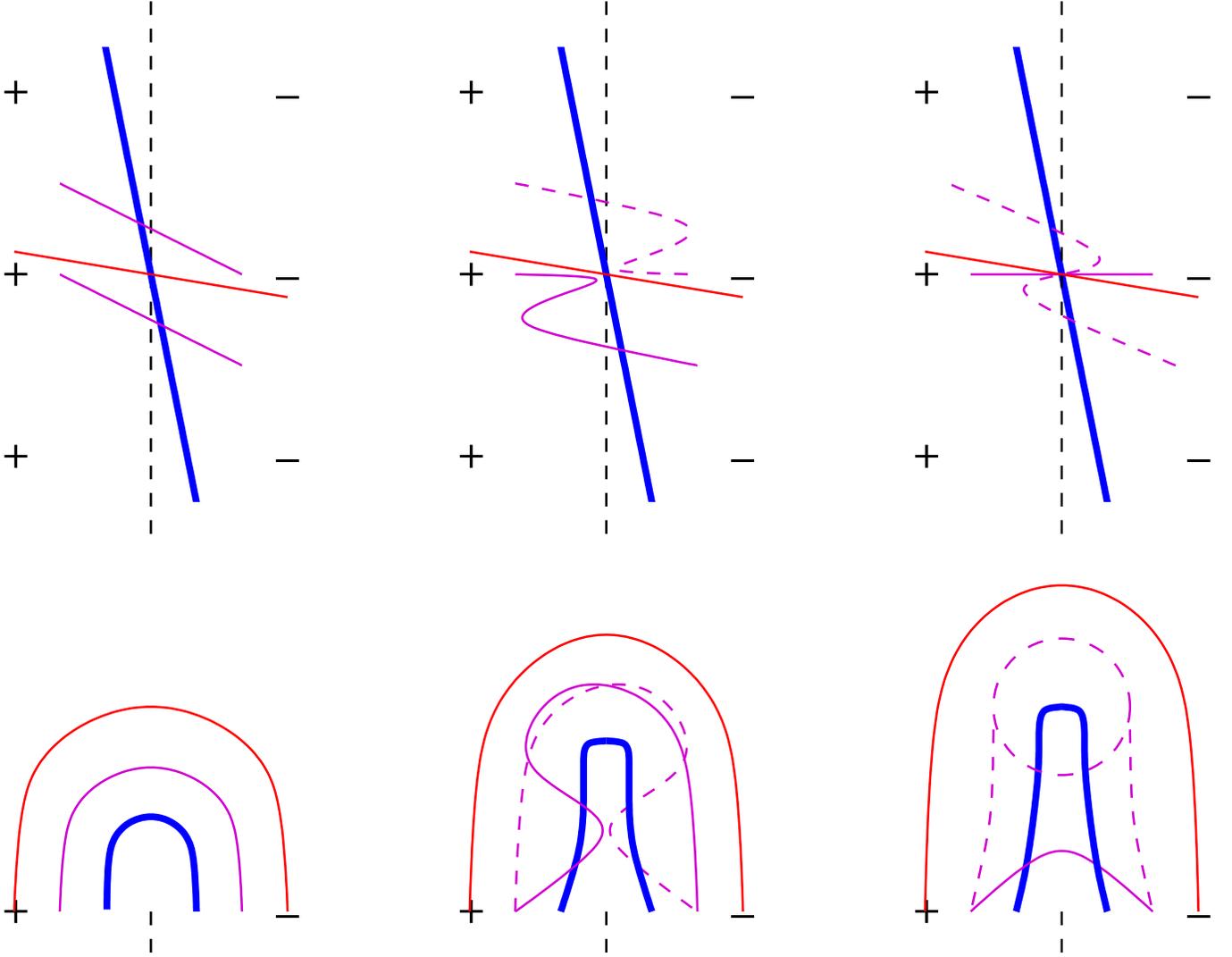}}
  \caption{An illustration of evolving
      connectivity due to magnetic reconnection in the presence of a
      guide field during an eruption.  {\it Top left:} A schematic, overhead
      view of field lines above a PIL prior to an eruption.  Flux
      represented by the {\it blue line} forms the proto-ejection, and is
      drawn thickest to aid visualization of the connectivity.
      configuration, as viewed from the bottom side of the {\it
        top-left} image. Note that magnetic shear --- how closely the
      field runs parallel to the PIL --- decreases with increasing
      height: the {\it blue} (lowest-lying) field line is the most
      sheared, the {\it red} (highest) field line is relatively
      unsheared.
      {\it Top middle:} An overhead view of the configuration after the
      proto-ejection has begun to rise.  Cavitation in its wake draws
      in flux, represented by the {\it magenta} field lines.  The {\it upper
      magenta} field line is shown as a {\it dashed line} to more clearly
      display connectivity.
      {\it Bottom middle:} an end-on view of the configuration in the
      {\it top-middle} image, as viewed from the bottom of that image.  
      {\it Top right:} overhead view of the configuration after flux
      represented by the {\it magenta} lines has reconnected, forming two
      post-reconnection flux domains: one containing flux that winds
      around the ejection {\it (dashed magenta line)} and one containing
      shorter flux {\it (solid magenta line)} that runs beneath it.
      {\it Bottom right:} an end-on view of the configuration in the
      top-right image, as viewed from the bottom of that image.
      Flux represented by the {\em dashed magenta} field line has accreted
      onto the ejection.  Note that this entrained flux is longer than
      the pre-reconnection {\it magenta} lines, and contains two segments
      with downward magnetic tension (above the {\it blue line}) and one
      with upward magnetic tension (below the {\it blue line}).  Flux
      represented by the {\it solid magenta} field line might be observed as
      a post-flare loop.     
  }
\label{fig:above_and_below}
\end{figure}

A third complication is that we have ignored the 3D structure of the
eruption.  Perhaps most importantly, our simplistic picture neglects the
proto-eruption's axial field --- the component of the magnetic field
in the direction normal to the plane containing Figure
\ref{fig:pre_post_2d}.
Observations of prominences and soft-X-ray over PILs generally
indicate the existence of an important component of the magnetic
field along the PIL.
A component of the field along the PIL corresponds to a nonzero
guide field in the ensuing reconnection above the PIL.
Accordingly, we now incorporate the effect of this guide field into
our estimate of the change in tension force due to the reconnection.
We assume the total field strength remains fixed, and focus on how
reconnection modifies the tension force as the reconnecting field's
direction varies with respect to the proto-ejection's axis.
Figure \ref{fig:above_and_below} illustrates an idealized
configuration before, during and after a reconnection event, with
  a guide field present.
The field's configuration is consistent with observations showing that
lower-lying loops are more closely parallel to the PIL
than higher loops \citep{Martin1996, Schmieder1996}.
In the figure, the presence of a guide field alters the
post-reconnection field's morphology from the purely 2D case: it leads
to accretion of helical instead of circular flux.
We define the $B_\parallel$ to point in the invariant direction (the
$y$ direction in Figure \ref{fig:pre_post_2d}, into/out of the figure
plane), with $B_P$ being perpendicular to the invariant direction
(within the plane of Figure \ref{fig:pre_post_2d}). For a field line
pitch angle $\omega$ with respect to the guide field $B_\parallel$,
and $\tan \omega = B_P/B_\parallel$, the infinitesimal of length,
$ds$, along the helical field line segment is
\be ds = \sqrt{ds_P^2 + ds_\parallel^2} = \sqrt{(R \, d\theta)^2 + (R
  \, d\theta)^2(B_\parallel/B_P)^2} = (R \, d\theta) \sqrt{1 + \cot^2
  \omega} ~. \ee
So the segment arching over the proto-ejection is longer, by an amount
$\sqrt{1 + \cot^2 \omega}$, meaning that the volume over which the
tension force acts (the volume of the reconnected flux tube) is
larger compared to that integrated in deriving Equation
\ref{eqn:df_tension}.
But, with fixed total field strength, $B_{\rm tot} = \sqrt{B_P^2 +
  B_\parallel^2}$, nonzero $B_\parallel$ means $B_P$ is
smaller, with $B_P = B_{\rm tot}/\sqrt{1 + \cot^2 \omega}$.
This, in turn, means that the force density is also smaller,
$B_P^2/R = B_{\rm tot}^2/[R (1 + \cot^2 \omega )]$.
This lower force density can be understood in terms of the larger
radius of curvature of the reconnected flux tube, which traverses a
longer distance when crossing underneath the proto-ejection.
The net effect of including the guide field is that the reduced force
density, integrated over the larger volume, is smaller than the result
in Equation \ref{eqn:df_tension}, yielding
\be \Delta F_{\rm tension} = \frac{\Delta \Phi B_{\rm tot}}{2\pi}
\frac{1}{\sqrt{1 + \cot^2 \omega}} ~.  \label{eqn:df_guide} \ee
But this result can also be understood in terms of the field's
components: $B_\parallel$ does not contribute to the tension, so the
$B_P$ in Equation \ref{eqn:df_tension} is replaced with
$B_{\rm tot}/\sqrt{1 + \cot^2 \omega}$ here.
With $\omega = 20^\circ$, the factor $1/\sqrt{1 + \cot^2 \omega}$ is
equal to 0.34.  So the reduction in tension due to tether-cutting
reconnection with a relatively small pitch angle is significantly
smaller than our estimate without a guide field (a pitch angle of
90$^\circ$).  As noted previously, the angle between the reconnecting
field and the PIL will vary from more to less parallel over the course
of an eruption.  So the decrease in restraining force on a
proto-ejection produced by tether-cutting reconnection should vary in
a typical event, as (i) the shear angle of reconnecting fields
decreases as the reconnection proceeds and (ii) the strength of
reconnecting fields decreases as flux farther from the active region
core reconnects.
  

We digress briefly to consider another significant result of this
analysis: even though a magnetic flux $\Delta \Phi$ reconnected in the
corona, there is a flux of $(4 \, \Delta \Phi)$ in the associated
photospheric footpoints.
If ribbon emission were produced at the footpoints of reconnecting
coronal fields, then {\em four} footpoints would produce ribbon
emission.  This counts flux at both ends of each reconnected flux
tube, because emission is believed to appear at both.  Thus, four
distinct ribbons could appear, a situation that \citet{Goff2007} refer
to as a quadrupolar flare.  But it is also possible that emission from
each same-polarity pair of footpoints will occur within the same,
contiguous ribbon.
%
%

Ribbon emission often precedes emission from post-flare loops,
rooted in the ribbons, which have a morphology similar to the solid magenta
line in the right column of Figure \ref{fig:above_and_below}.
But is there something different about ribbon emission at the
footpoints of helical loops -- corresponding to the dashed magenta
line in the figure's right column -- that do {\em not} subsequently
produce bright, post-flare loops?
The enhanced density from chromospheric evaporation that brightens
short post-flare loops might be too dilute to brighten longer loops,
since emission measure scales as density squared.  Or is there a
quantitative difference in ribbon emission at footpoints of these
longer loops?  If the reconnection process accelerates ribbon-causing
non-thermal particles near the reconnection site, then all four
footpoints should produce similar ribbon emission.
\citet{Chandra2009} and \citet{Zhao2016} relate observed ribbon
emission to helical field lines, based upon their J- or
reversed-J-shaped morphology ({\it e.g.}, \citealt{Williams2005,
  Green2007}), and report that emission from such footpoints is more
faint. This suggests that the acceleration of ribbon-causing
non-thermal particles does not solely involve processes near the
reconnection site.  A major difference between the long and short
post-reconnection loops is that the latter dipolarize downward into
layers of the solar atmosphere that are much more dense and contain
much higher field strengths.  This contraction itself could lead to
particle acceleration directly, by betatron acceleration ({\it e.g.},
\citealt{Somov2003}), or indirectly, via wave excitation ({\it e.g.},
\citealt{Fletcher2008}).


We have also ignored dynamic effects arising from curvature along the
axis of the erupting flux system, which we address in our discussion
of the hoop force (Section \ref{subsec:hoop}) shortly.

\subsection{Global-Scale Magnetic Pressure Variations}
\label{subsec:pressure}

In addition to the reduction in downward-directed magnetic tension
force restraining the CME, magnetic reconnection would, considered by
itself, increase the magnetic pressure underneath the CME.  This is
because the reconnection introduces additional magnetic flux into the
volume underlying the CME, without changing the amount of
magnetic flux above the CME.  If the CME were not rising, this would
enhance the difference in magnetic pressure across the CME, which
would increase the upward force upon it.
In Figure \ref{fig:pre_post_2d}, for instance, reconnection of the
flux between points C and D at the photosphere adds the fluxes between
points C' and D', and between points C'' and D'' into the space below
the CME's bottom boundary.  This change increases the average flux
density --- equivalently, magnetic field strength --- beneath the
ejection.  The higher average field strength underneath the ejection
implies an increased upward magnetic pressure acting the ejection from
below.

This increased magnetic flux density underneath the CME might be
thought to supply pressure that acts like the gas behind a bullet in a
gun barrel: the pressure difference across the bullet accelerates it.
As noted in the introduction, however, in our pull-reconnection
scenario, an ejection can rise sufficiently fast that there is a
decrease in the {\em average} magnetic pressure in its wake, rather
than an excess. So, compared to a stationary flux rope, the
  upward magnetic pressure behind a rapidly rising rope could be smaller.
Note, however, our emphasis on ``average'' here: to the extent
that the upward-dipolarizing, post-reconnection flux catches up with
the CME, $B^2$ near the CME's trailing edge would be higher than
farther into its wake.  So the force from the magnetic pressure
gradient is not necessarily monotonically upward.  We need to consider
the increase in magnetic pressure at the trailing edge of the CME due
to the reconnection, our focus in the next subsection.


\subsection{CME-Scale Magnetic Pressure Variations: Changed Hoop Force}
\label{subsec:hoop}

We now consider how magnetic reconnection affects the hoop force on an
erupting flux system. We first briefly review the hoop force (Section
  \ref{subsubsec:hoop_review}).  Then, treating a proto-ejection as a
  torus with a purely poloidal external field, we derive expressions
  for the hoop force (Section \ref{subsubsec:hoop_hoop}) and the {\em
    change} in this force due to the reconnection (Section
  \ref{subsubsec:dhoop}).  Finally, we consider the effect of a
  toroidal component in field external to the torus (Section
  \ref{subsubsec:bp_and_bt}).

\subsubsection{The Hoop Force in CMEs}
\label{subsubsec:hoop_review}

Erupting flux systems must be, in a rough sense, $\Omega$-shaped: the
photospheric footpoints of fields participating in an eruption do not
move significantly during the event, but the flux in the corona
balloons outward into the heliosphere.
Curvature of the magnetic field along the axis of the $\Omega$-shaped
erupting flux system --- the axial field being due to the flux
system's toroidal component --- implies that a large-scale, downward
magnetic tension force is present.  
(This large-scale, downward tension force from axial curvature is not
expected to change directly due to reconnection, so we do not consider
it further.)

\begin{figure}[htp] 
  \centerline{\includegraphics[width=0.6\textwidth,clip=true]{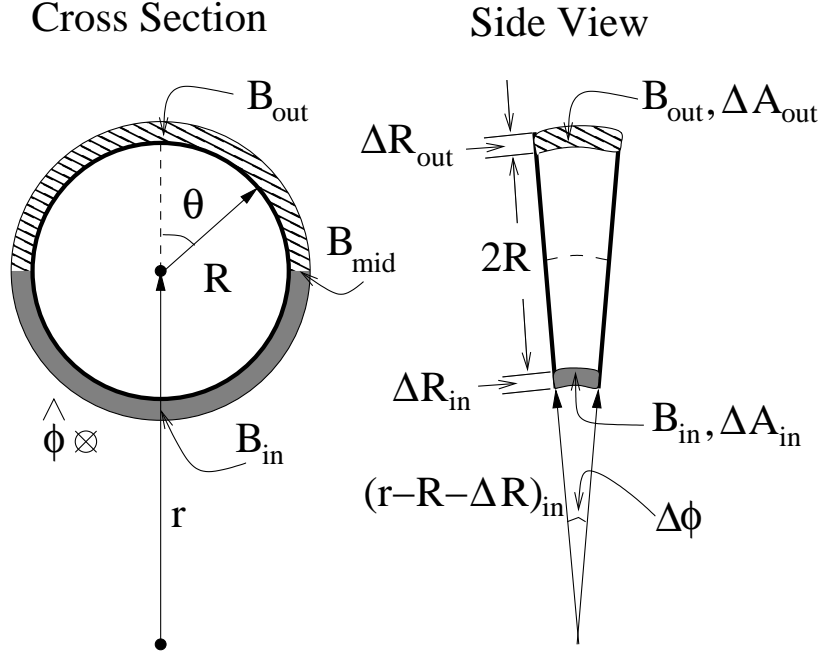}}
  \caption{An illustration of quantities used in describing the torus
    geometry used in hoop force calculations. {\it Left:} This end-on
    view of a cross section of the torus's volume shows major and
    minor axes, $r$ and $R$, respectively.  The poloidal angle,
    $\theta$, about the torus's axis increases clockwise, and the
    azimuthal angle, $\phi$, increases into the page.  Newly
    dipolarized flux is depicted as a {\it shaded semicircle} at the torus's
    inner edge, and its conjugate flux fills the {\it hashed area} at the
    outer edge.  Magnetic flux densities at the inside, middle, and
    outside surfaces are labeled as $B_{\rm in}, B_{\rm mid}$, and
    $B_{\rm out}$, respectively. {\it Right:} This side (edge-on) view shows
    a segment of the torus. Newly dipolarized flux is depicted as a
    {\it shaded patch} at the torus's inner edge, and its conjugate flux
    fills the {\it hashed area} at the outer edge.}
  \label{fig:torus}
  \end{figure}

Set against this inward force from curvature of the eruption's
toroidal field, the component of the field winding around the
eruption's axis --- the poloidal field, $\vec B_P$ --- tends to exert
an outward, hoop force ({\it e.g.}, \citealt{Anzer1978}) on the ejection.
We analyze the hoop force arising from poloidal flux that winds around
a toroidal volume. We assume the cross section of the torus is
circular, with the geometry illustrated in figure \ref{fig:torus}.
%
%
The hoop force can be understood qualitatively in the following way
\citep{Freidberg1987}: (i) poloidal flux passing through each
differential area in the torus's hole $(\Delta A_{\rm in} = \Delta r_{\rm in}
r_{\rm in} \, \Delta \phi)$ must pass through a conjugate differential area
outside the torus $(\Delta A_{\rm out} = \Delta r_{\rm out} r_{\rm out} \,
\Delta \phi)$;
(ii) the area of the outside half of the torus's surface,
$S_{\rm out}$, is larger than the inside half of its surface, $S_{\rm
  in}$, since the differential surface area, $dS$, increases linearly
with distance from the center of the hole ($dS \propto r \, d\phi$);
(iii) if $\Delta r_{\rm out} = \Delta r_{\rm in}$, then $B_P$ decreases as $1/r$
(and $B_P$ falls off more rapidly if $\Delta r_{\rm out} > \Delta r_{\rm in}$),
meaning that $B_P$ must be weaker on the outside half of the torus's
surface than on its inside half;
(iv) but the inward and outward magnetic pressure forces on the torus
scale as $B_P^2$ times $S_{\rm out}$ and $S_{\rm in}$, respectively,
so the inverse-square decay of magnetic pressure with distance implies
that the surface-integrated inward pressure force on $S_{\rm out}$ is
less than the outward force on $S_{\rm in}$. Hence, the poloidal field
exerts a net outward force.
If this outward force were sufficient to overcome inward tension from
curvature of the toroidal field, and no other forces were present,
then the major radius of the torus would increase.  This expansion
might be inhibited by magnetic fields external to the torus.  But if
fields external to the torus decay sufficiently rapidly with distance,
then the torus will accelerate outward \citep{Kliem2006}.

As a starting point for our analysis, we assumed that our flux system
was not in force balance: it was already erupting.  So a hoop force,
or something akin to it, must already be present. But how does
reconnection affect this force?

The key assumption that we make about the evolution of the
post-reconnection field is that the concave-up flux from the flare
reconnection under the proto-ejection does, indeed, catch up to it.
Stated another way: the reconnection adds flux to the ejection faster
than the ejection's flux moves outward.
That flux catches up to the ejection is evident in the lower panels of
Fig. 4 of the breakout simulations described by \citet{MacNeice2004},
which show that the erupting flux system maintains a circular cross
section as flux is added to it.  (In Fig. 3d of the paper by
\citet{Karpen2012}, the simulated eruption's cross section is deformed
by the reconnection jet to become somewhat concave, suggesting that
post-reconnection flux easily overtakes the outward-moving eruption.)
The catch-up of post-reconnection flux is consistent with our earlier
treatment of momentum transfer to the ejection from the reconnection
outflow.
Essentially, this flux dipolarizes as it catches up to the CME's
trailing edge, and its dipolarization is arrested (in the
co-moving frame) by backwards-directed magnetic pressure at that edge.
Given this assumption, magnetic pressure from the post reconnection
flux will exert an additional upward force on the CME.
Due to the cavitation farther back in the CME wake, however, the hoop
force might be less than the estimate that we derive below, which
might therefore represent an upper bound.

It should be noted here that an ejection's evolution will differ if
reconnection is assumed not to occur.  If the major radius $r$ of an
axisymmetric torus with nonzero poloidal field increases ideally, the
poloidal flux density, $B_p$, must decrease, because the same amount
of poloidal flux is spread out over a larger area.
Accordingly, the hoop force will also decrease.
When reconnection occurs in the wake of an expanding torus, however,
the accretion of additional flux onto the erupting system means that
$B_P$ at its trailing edge will generally not decrease with increasing
$r$ in the same manner as the ideal case.  In particular, the added
flux can affect $B_P$ there in two ways: (1) by increasing magnetic
pressure at the rear of the ejection, which keeps flux compressed; and
(2) increasing the pitch angle at the rear of the ejection, as
less-sheared fields reconnect and accrete onto the CME.  The hoop
force must still weaken as the ejection moves outward, but
reconnection should decrease the rate at which the hoop force weakens
with distance.
We also remark that, due to this reconnection, the magnetic helicity
of the erupting flux system is not conserved.  We discuss some aspects
of helicity evolution in Appendix \ref{app:helicity}.

\subsubsection{Hoop Force on a Torus Segment}
\label{subsubsec:hoop_hoop}

Before we estimate the {\em change} in the hoop force from the
reconnection, we first analyze the hoop force itself.
%
%
We assume that the field at the surface of the torus is purely poloidal,
and derive the inward and outward components of
the hoop force due to external poloidal flux $\Delta \Phi$ in
contact with the segment's annular surface.
The poloidal flux $\Delta \Phi$ at the inner edge of the torus
segment matches a corresponding flux $\Delta \Phi$ at its outer edge,
so the poloidal flux densities are related via
\be \frac{B_{\rm in}}{B_{\rm out}} =
\frac{\Delta\Phi/\Delta A_{\rm in}}{\Delta\Phi/\Delta A_{\rm out}} = 
\frac{\Delta A_{\rm out}}{\Delta A_{\rm in}} =
\frac{\Delta R_{\rm out} \, \Delta \phi (r + R)}
     {\Delta R_{\rm in} \, \Delta \phi (r - R)} = 
\frac{(r + R)}{(r - R)}  \label{eqn:in_out}
~. \ee
Here, $\Delta \Phi$ refers to reconnected flux, but $\Delta \phi$
  refers to an angular interval in the toroidal coordinate, $\phi$.
We remark that we have assumed $\Delta R_{\rm in} = \Delta R_{\rm
    out}$ here; if $\Delta R_{\rm out} > \Delta R_{\rm in}$, then the
  outside magnetic field strength (and pressure force) will be weaker.
Consistent with this expression for $B_{\rm in}/B_{\rm out}$, we model the
variation of the poloidal field with $\theta$ to be
\be B_P(\theta) = B_{\rm mid}\left ( \frac{r}{r + R \cos \theta} \right )
~, \label{eqn:bp_theta} \ee
where $B_{\rm mid}$ is the poloidal field strength at the lateral,
mid-position of the torus ({\it i.e.}, $\theta = \pm \pi/2$).
The outward and inward forces on the inner and outer halves of the
segment, respectively, are given by the magnetic pressure,
$B_P(\theta)^2/8\pi$ on each, projected onto the (outward) radial direction,
integrated over the area of each half.  
Infinitesimal areas on the inner and outer halves of the segment are
\bea dS_{\rm in} &=& (r - R \cos \theta) \, d\phi R \, d\theta 
\label{eqn:ds_inner} \\
dS_{\rm out} &=& (r + R \cos \theta) \, d\phi R \, d\theta 
~. \label{eqn:ds_outer} \eea 

To find the inward force on the outer segment of the torus, we
integrate $\theta$ from $(-\pi/2, \pi/2)$ over the outside surface,
and $\phi$ over the range $\Delta \phi$.  Projection onto the outward
direction introduces a factor of $-\cos(\theta)$.  This yields
(\citealt{Dwight1961}, p.105, integral 446.00)
\bea F_{\rm outer}
&=& \int dS_{\rm out} \frac{B_{\rm mid}^2}{8\pi} \left ( - \frac{r^2 \cos \theta}{(r + R \cos \theta)^2} \right ) \\
&=& - r^2 \, \Delta \phi R \frac{B_{\rm mid}^2}{8\pi} \int_{-\pi/2}^{\pi/2} \, d\theta \frac{\cos \theta}{r + R \cos \theta} \\
&=& r^2 \, \Delta \phi R \frac{B_{\rm mid}^2}{8\pi} \left . \left [ - \frac{\theta}{R} 
  + \frac{2r}{R\sqrt{r^2 - R^2}} \tan^{-1} \left (
  \frac{\sqrt{r - R}}{\sqrt{r + R}} \tan (\theta/2)  \right ) \right ] \right \vert_{-\pi/2}^{\pi/2} \\
&=& r^2 \, \Delta \phi \frac{B_{\rm mid}^2}{2\pi} \left [
\frac{ r }{\sqrt{r^2 - R^2}} \tan^{-1}
\left ( \frac{\sqrt{r - R}}{\sqrt{r + R}} \right ) - \frac{\pi}{4} \right ] \\
&=& r^2 \, \Delta \phi \frac{B_{\rm mid}^2}{4} \left [
  \frac{ r }{\sqrt{r^2 - R^2}} \left ( \frac{2}{\pi} \right ) \tan^{-1}
\left ( \frac{\sqrt{r - R}}{\sqrt{r + R}} \right ) - \frac{1}{2} \right ]
~. \label{eqn:f_outer} \eea
The outward force on the inner segment of the torus involves an
analogous integration, with a projection factor of $+\cos(\theta)$ and
complementary domain, yielding
\be F_{\rm inner}
= r^2 \, \Delta \phi \frac{B_{\rm mid}^2}{4} \left [
  \frac{ r }{\sqrt{r^2 - R^2}} \left ( \frac{2}{\pi} \right ) \tan^{-1}
  \left ( \frac{\sqrt{r + R}}{\sqrt{r - R}} \right )  - \frac{1}{2} \right ]
~. \label{eqn:f_inner} \ee
In the limit that torus's minor radius $R$ were to expand to
approximately its major radius, then the force on the inner surface
would diverge, and the force on the outer surface would remain finite.
The inverse tangents in Equations \ref{eqn:f_outer} and
  \ref{eqn:f_inner} are related via the identity $(\tan^{-1}[x] +
\tan^{-1}[1/x]) = \pi/2,$ so their values are coupled.
%
%
The hoop force on the torus segment is given by the sum of $F_{\rm
  outer}$ and $F_{\rm inner}$, which yields
\bea F_{\rm hoop} &=& F_{\rm inner} + F_{\rm outer} 
= r^2 \, \Delta \phi \frac{B_{\rm mid}^2}{2\pi} \left ( \frac{\pi}{2} \right )
\left [ \frac{ r }{\sqrt{r^2 - R^2}} - 1 \right ] \\
&=& r^2 \, \Delta \phi \frac{B_{\rm mid}^2}{4} 
\left [ \frac{ r }{\sqrt{r^2 - R^2}} - 1 \right ]
~. \label{eqn:f_hoop} \eea
This is manifestly positive (outward).

\begin{figure}[htp] 
  \centerline{\includegraphics[width=0.75\textwidth,clip=true]{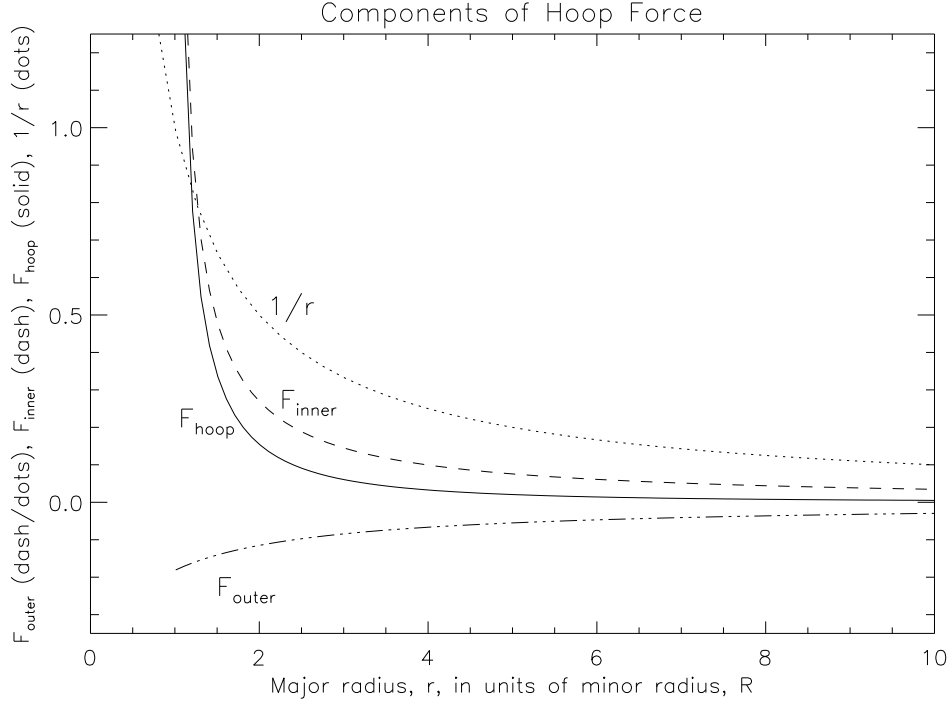}}
  \caption{Plots of the radial dependence of the terms in square
    brackets in Equations \ref{eqn:f_outer}, \ref{eqn:f_inner}, and
    \ref{eqn:f_hoop}, corresponding to the outward $F_{\rm inner}$
        {\it (dashed)}, inward $F_{\rm outer}$ {\it (dash/dots)}, and
        $F_{\rm hoop}$ {\it (solid)}.  For comparison, the curve $1/r$
        is plotted {\it (dotted line)}. }
  \label{fig:hoop_forces}
  \end{figure}

Figure \ref{fig:hoop_forces} plots the radial dependence of the terms
in square brackets in the expressions for $F_{\rm outer}, F_{\rm
  inner},$ and $F_{\rm hoop}$ (from Equations \ref{eqn:f_outer},
\ref{eqn:f_inner}, and \ref{eqn:f_hoop}, respectively) as functions of
major radius $r$.
In the limit that the torus's major radius $r$ expands to
become much larger than its minor radius (so $r \gg R$),
\bea F_{\rm hoop} 
&\simeq& r^2 \, \Delta \phi \frac{B_{\rm mid}^2}{4} \left [
  (1 + (1/2)(R/r)^2) - 1 \right ] \nonumber \\
&\simeq& R^2 \, \Delta \phi \frac{B_{\rm mid}^2}{8} 
~. \label{eqn:f_hoop_bigr} \eea
As shown in Appendix \ref{app:current}, the hoop force per unit length
here scales as $I^2/r$, where $I$ is the total electric current within
the torus, consistent with the result derived by
\citet{Shafranov1966}.  Therefore, {\em if the evolution were ideal,}
the net force outward force per unit length would decrease like $1/r$
if the major radius $r$ grew much greater than its minor radius $R$.

\subsubsection{Change in Hoop Force Due to Reconnection}
\label{subsubsec:dhoop}

How does reconnection affect the hoop force on the ejection?  The
addition of flux should widen the proto-ejection, {\it i.e.}, the minor
radius $R$ of the ejection should increase.
To analyze this effect, we must account for the change in minor radius
by $\Delta R$ as $\Delta\Phi$ is added.  This depends on the flux density
at the proto-ejection's trailing edge. In the frame co-moving with the
ejection, the upward motion of the dipolarizing post-reconnection flux
$\Delta \Phi$ is arrested by the back reaction of magnetic pressure at
that trailing edge.  This implies that magnetic pressures, and
therefore field strengths, are approximately equal at the
proto-ejection's trailing edge and in the arrested flux.  So the
addition of flux $\Delta\Phi$ at the same field strength implies
  \bea \Delta \Phi &=& B_{\rm in} \Delta R \, (r - R) \, \Delta \phi 
  = \left [ B_{\rm mid} \frac{r}{r - R} \right ]
  \Delta R \, (r - R) \, \Delta \phi \\
  &=& B_{\rm mid} \, r \Delta R \, \Delta \phi  \\ 
  \frac{\Delta \Phi}{\Delta R} &=&  B_{\rm mid} \, r \, \Delta \phi  ~, \eea 
where $B_{\rm in}$ is the field strength at the proto-ejection's trailing edge,
which is related to $B_{\rm mid}$ by Equation \ref{eqn:bp_theta}.

The change in hoop force, $\Delta F_{\rm hoop}$, due to the
differential increase in $R$ is then the difference of two versions of
Equation \ref{eqn:f_hoop}, evaluated for minor radii $R+\Delta R$ and
$R$.  Assuming $\Delta R$ is small compared to $(r-R)$, we have
\bea \Delta F_{\rm hoop} &=& \frac{B_{\rm mid}}{4}
  \frac{\Delta \Phi}{\Delta R} r 
\left [ \frac{ r}{\sqrt{r^2 - (R+\Delta R)^2}} -
  \frac{ r }{\sqrt{r^2 - R^2}} \right ] \\
&=& \frac{B_{\rm mid} \, \Delta \Phi}{4 \, \Delta R}r^2
\left [ \frac{ 1}{\sqrt{r^2 - R^2 -2 R \, \Delta R - \Delta R^2}} -
  \frac{ 1 }{\sqrt{r^2 - R^2}  } \right ] \nonumber \\
&=& \frac{B_{\rm mid} \, \Delta \Phi}{4 \, \Delta R}\frac{r^2}{\sqrt{r^2 - R^2}}
\left [ \left (1 -  \frac{2 R \, \Delta R + \Delta R^2}{(r^2 - R^2)}
  \right )^{-1/2} - 1 \right ] \\
&\simeq& \frac{B_{\rm mid} \, \Delta \Phi}{4 \, \Delta R}
\frac{r^2}{\sqrt{r^2 - R^2}} 
\left [ 1 + \frac{R \, \Delta R }{(r^2 - R^2)} - 1 \right ] \\
&\simeq& \frac{B_{\rm mid} \, \Delta \Phi}{4}
\frac{r^2}{\sqrt{r^2 - R^2}}
\frac{R}{(r^2 - R^2)}  \\
&\simeq&
\frac{B_{\rm mid} \, \Delta \Phi}{4}
\frac{r^2 R}{(r^2 - R^2)^{3/2}} \\
&=& \frac{B_{\rm mid} \, \Delta \Phi}{4}
\left [ \frac{(R/r)}{\left ( 1 - (R/r)^2 \right )^{3/2}}  \right ]
~. \label{eqn:df_hoop} \eea
The coefficients in this result are similar in form to those in
Equation \ref{eqn:df_tension}, but the change in force depends on
the ejection's height $r$ and half-width $R$.
\begin{figure}[htp] 
  \centerline{\includegraphics[width=0.75\textwidth,clip=true]{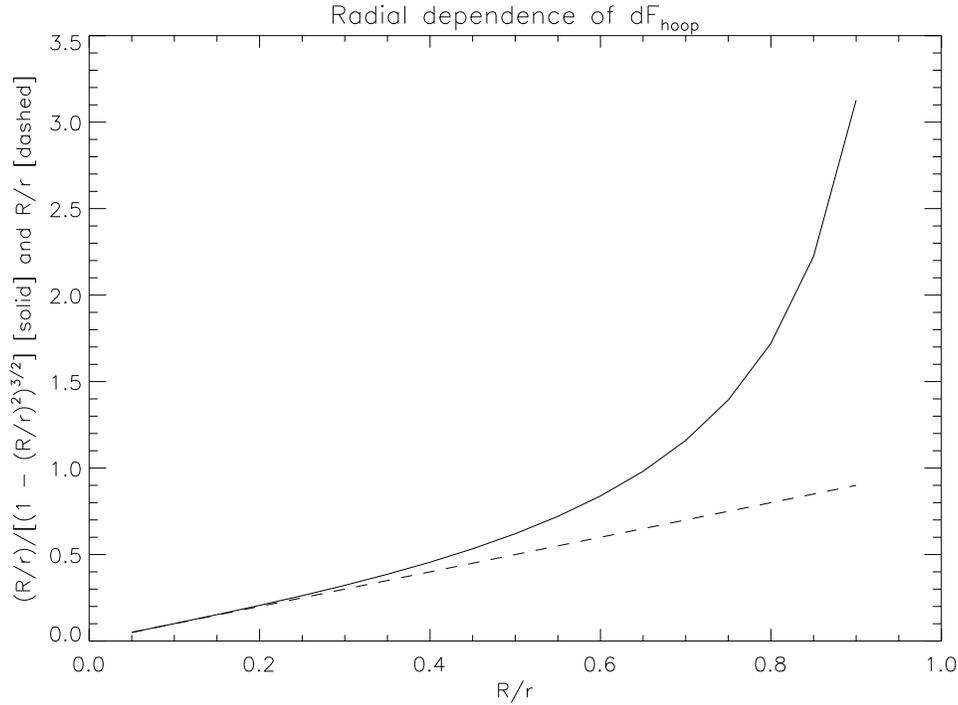}}
  \caption{The {\it solid line} plots the term in square brackets in 
    Equation \ref{eqn:df_hoop} as $(R/r)$ varies.  The {\it dashed line}
    plots $(R/r)$ {\it versus} $(R/r)$. To the extent that CMEs propagate
    with fixed angular width ({\it e.g.}, \citealt{Zhao2002}), the ratio
    $(R/r)$ should be constant as $r$ increases.}
  \label{fig:df_hoop_vs_r}
  \end{figure}
For $r \gg R$, Equation \ref{eqn:df_hoop} implies
\be \Delta F_{{\rm hoop}, r \gg R}
= \frac{B_{\rm mid} \, \Delta \Phi}{4} \left ( \frac{R}{r} \right )
~. \label{eqn:df_hoop_bigr} \ee

The results in Equations \ref{eqn:df_hoop} and
\ref{eqn:df_hoop_bigr} can also be derived by differentiating the
expressions in Equations \ref{eqn:f_hoop} and
\ref{eqn:f_hoop_bigr}, respectively, with respect to $R$, and
expressing the result in terms of $\Delta \Phi$.  

Figure \ref{fig:df_hoop_vs_r} plots the variation with $(R/r)$ of the
term in the square brackets in Equation \ref{eqn:df_hoop}, and
includes the scaling predicted by Equation \ref{eqn:df_hoop_bigr}
for comparison.
To the extent that CMEs evolve with constant angular width ({\it e.g.},
\citealt{Zhao2002}), the ratio $(R/r)$ would remain fixed as major
radius $r$ increases.
The mean and median CME widths in the CDAW catalog dataset referenced
above are 50$^\circ$ and 63$^\circ$, respectively. These probably
overestimate the actual angular widths of typical erupting flux ropes
due to tilts of flux ropes' axes along the line of sight.
So a typical value for $(R/r)$ might be 1/3, corresponding to a torus
cross section that subtends an angle of 37$^\circ$, for which
\be \Delta F_{{\rm hoop}, r = 3R}
= \frac{B_{\rm mid} \, \Delta \Phi}{4} \frac{(1/3)}{(8/9)^{3/2}}
= \frac{B_{\rm mid} \, \Delta \Phi}{4} \frac{9}{64 \sqrt{2}} 
\simeq 0.1 B_{\rm mid} \, \Delta \Phi 
~. \label{eqn:df_hoop_3r} \ee
This is a bit smaller than the change in tension force for a purely
poloidal field, $\Delta F_{\rm tension} \simeq 0.16 B_{\rm tot} \Delta \Phi$
from Equation \ref{eqn:df_tension}, and a bit larger than the
change in tension force for guide-field reconnection with field line
pitch $\omega = 20^\circ$, $\Delta F_{\rm tension} \simeq 0.05 B_{\rm
  tot} \Delta \Phi$ from Equation \ref{eqn:df_guide}.


Equation \ref{eqn:df_hoop} and subsequent results assume that the
reconnected flux $\Delta \Phi$ is completely dipolarized.  The total
change in hoop force should also account for how the hoop force
increases as $\Delta \Phi$ evolves from the pre-dipolarized state to
the post-dipolarized state. In Appendix \ref{app:pre_post_dip}, we
consider this change in force due to dipolarization, and find that the
change in force predicted by Equation \ref{eqn:df_hoop} could be a
lower estimate.

\subsubsection{Toroidal \& Poloidal External Field}
\label{subsubsec:bp_and_bt}

The derivation above has neglected any toroidal component in the
reconnecting field external to the torus. The presence of a toroidal
component, $B_T$, in the reconnecting field alters the relationships
we derived above (Equations \ref{eqn:f_outer}, \ref{eqn:f_inner},
\ref{eqn:f_hoop}, and \ref{eqn:df_hoop}.
The modified inward force on the outer surface, $F_{\rm outer}'$, is given by
\be F_{\rm outer}'
= \int dS_{\rm out} \frac{(B_P^2 + B_T^2)}{8\pi}
~, \label{eqn:fout_phi} \ee
and the modified outward force on the inner surface, $F_{\rm inner}'$,
is given by
\be F_{\rm inner}'
= \int dS_{\rm in} \frac{(B_P^2 + B_T^2)}{8\pi}
~. \label{eqn:finn_phi} \ee
For fixed total field strength $B$, nonzero $B_T$ implies the
poloidal field, $B_P$, is weaker.  We assume, however, that the
dependence of $B_P$ on $\theta$ is unchanged, because Equation
\ref{eqn:in_out} must still apply: every unit of poloidal flux
threading the interior of the torus must map to a larger area exterior
to it.

Without developing a particular model of a field with a toroidal
component, we can, with some basic assumptions, make some
inferences about the effect that a nonzero toroidal field component,
$B_T$, would have on the hoop force.
%
%
\begin{figure}[htp] 
  \centerline{\includegraphics[width=0.9\textwidth,clip=true]{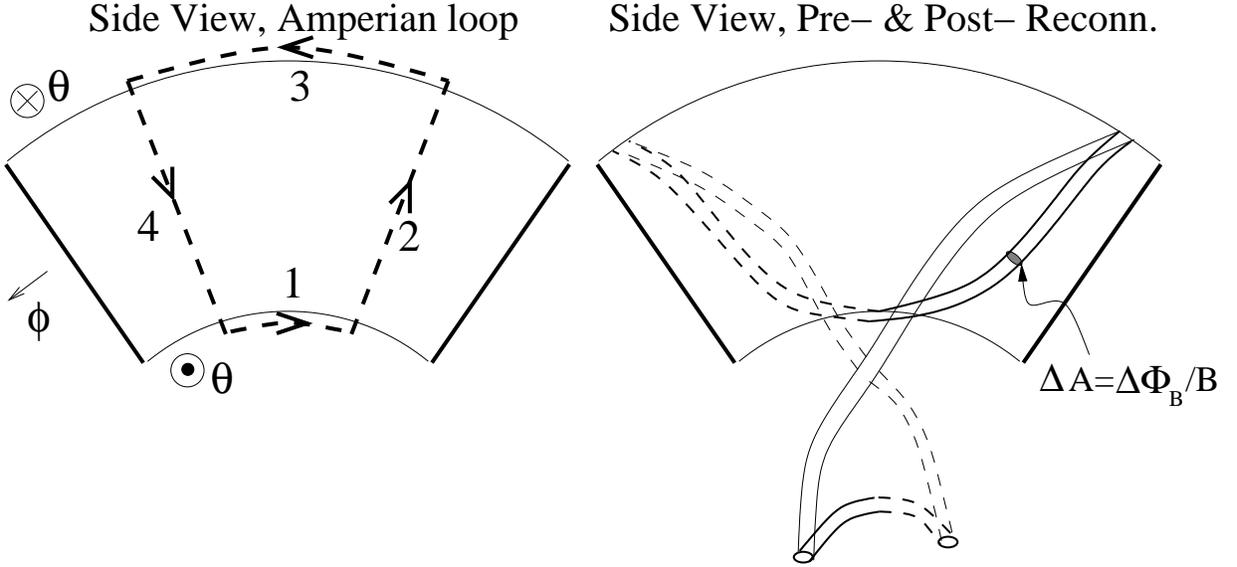}}
  \caption{{\it Left:} The {\it solid lines} depict a cross section of
    a torus segment, viewed from the side, and the {\it dashed lines}
    show an Amp\`erian loop in the plane of the cross section.  {\it
      Right:} The two long, thin flux tubes crossing below the torus
    ({\it solid} being nearer the viewer, {\it dashed} farther) depict
    reconnecting fields with nonzero $B_T$ below a rising ejection.
    Upward-dipolarized, post-reconnection flux (the {\it upper, thick,
      half-dashed, half-solid} flux tube) is in contact with the inner
    surface of the torus over a longer distance than with purely
    poloidal reconnecting field.  The volume integral in Equations
    \ref{eqn:rho_dV} and \ref{eqn:rho_over_B} runs over the length of
    the upward-dipolarized flux tube.}
  \label{fig:b_phi}
  \end{figure}
%
%
We start with the simplifying assumptions that
%
%
(i) fields in the torus and external to it are axisymmetric (invariant
in the toroidal direction),
and (ii) there is no net current across (perpendicular to) the
plane of the torus.  Physically, assumption (ii) is equivalent
to requiring that net charge does not accumulate on either side
of the torus; such an accumulation could not persist.

We now show that these assumptions imply that $B_T$ must also decrease
as the inverse of radial distance from torus center.
Consider the closed path shown in the left panel of Figure
\ref{fig:b_phi}, which lies in the plane of the figure. By
  Amp\'ere's law, the net electric current, $I_n$, normal to the
plane of the figure that is enclosed by the loop is proportional
to $\oint \vec B \cdot d\vec \ell$ around the loop.  Axisymmetry
  implies that the line integrals across the torus, $\int B_r dr$ for
  segment 2 and $-\int B_r dr$ for segment 4, must cancel.  The
  integrations along segments 1 and 3 then imply
  \be I_n \propto \oint \vec B \cdot d\vec \ell =
  \Delta \phi [ r_3 B_T(r_3)  - r_1 B_T(r_1)]
  \label{eqn:ampere} ~. \ee
The radial coefficient of each term implies that, for $I_n$ to vanish,
$B_T$ must decrease as the inverse of distance from the torus center.
The assumed axisymmetry implies that this result is valid even if $I_n
\ne 0$ for $r < r_1$.  In accordance with this result, when comparing
Equations \ref{eqn:fout_phi} and \ref{eqn:finn_phi}, we know that
$dS$ increases linearly with distance from torus center, but $(B_T^2 +
B_P^2)$ decreases quadratically with distance from torus center, so
$|F_{\rm inner}'| > |F_{\rm outer}'|$, and again the hoop force is
outward as in the case of purely poloidal field.

Our argument that $B_T$ should scale a $1/r$ does rest upon
  somewhat restrictive assumptions, which might not apply on the Sun.
In particular, our assumption of patchy reconnection implies that
the assumed axisymmetry is, at best, only approximately correct.

As shown in the right panel of Figure \ref{fig:b_phi}, reconnection of
fields below the torus with nonzero $B_T$ will lead to post
reconnection flux (upper, half-dashed, half-solid tube) in contact
with the inner surface of the torus.  This post-reconnection field
would exert an outward force $F_{\rm inner}'$ on the torus.

\subsection{Comparing Changes in Hoop \& Tension Forces}
\label{subsec:compare}
  
To better understand reconnection-driven changes in forces acting on
CMEs, it is useful to compare the changes in tension and hoop forces.

As a preliminary remark, we note that the presence of axial curvature
of the erupting flux system, which is ultimately responsible for the
hoop force, could, in principle, modify the tension force from the
poloidal field. In Appendix \ref{app:tension}, we show that axial
curvature does not affect the tension of a poloidal flux tube with
$B_P$ given by Equation \ref{eqn:bp_theta}.

Taking the ratio of the change in hoop force for $r \gg R$, from
Equation \ref{eqn:df_hoop_bigr}, to the change in tension force, from
Equation \ref{eqn:df_tension}, yields
  \be \frac{\Delta F_{{\rm hoop}, r \gg R}}{\Delta F_{\rm tension}}
  \simeq \frac{(R/r) \, \Delta \phi B_{\rm mid}/4}
         {\Delta \phi B_{\rm mid}/2\pi}
  \simeq \frac{\pi}{2} \left ( \frac{R}{r} \right )
         ~. \label{eqn:f_ratio} \ee
Hence, the force changes from these effects are similar in magnitude
for $R$ comparable to $r$.

Using instead the change in tension force due to reconnection with a
guide field, from Equation \ref{eqn:df_guide} with a value of
$\omega = 20^\circ$, reduces the tension force by a factor of 1/3.  In
this case, with $(R/r)$ = 1/3 as discussed previously, $\Delta F_{\rm hoop}$
due to the widening of the proto-ejection would be commensurate with
or slightly larger than $\Delta F_{\rm tension}$.

\subsection{Changes in CME Acceleration}
\label{subsec:accel}

We now seek to characterize the change in the CME's acceleration due
to these changes in external forces acting on the eruption.  To do so,
we derive the first-order difference in the time rate of change of
momentum between two instants, $t_1$ and $t_2$, separated in time by
$\Delta t$. Recall that a flux $\Delta \Phi_B$ accreted onto the CME
increases its mass by $\Delta m$ because the flux is frozen to the
plasma outside the diffusion region.  (Here, $\Delta \Phi_B$
  refers to the total flux in the flux tube, not just the flux from
  the reconnecting component of $\vec B$.)  The mass in the segment of
the flux tube that is added to the ejection is
\bea \Delta m &=& \int \rho \, dV = \int \rho \, dA \, dL \label{eqn:rho_dV} \\
&=& \int \frac{\rho}{B} \, \Delta \Phi_B \, dL
= \Delta \Phi_B \int \frac{\rho}{B} \, dL ~. \label{eqn:rho_over_B} \eea
The integral here runs over the upward-dipolarized, thick flux tube in
the right panel of Figure \ref{fig:b_phi}.  (Notice that the presence
of a toroidal field component --- a guide field --- in reconnection
beneath the ejection yields accretion of post-reconnection flux tubes
that are longer than if purely poloidal fields reconnected.)
For field strengths of a few gauss, lengths of a few tens of Mm, and
mass densities near 10$^{-15}$ g cm$^{-3}$ (a few times 10$^{8}$
electrons per cm$^3$), the added mass per unit reconnected flux is
around $\sim 10^{-6}$ g/Mx.  For events with a few times 10$^{21}$ Mx
of reconnected flux \citep{Qiu2005, Kazachenko2017}, the increase in
mass of a proto-ejection by flux accretion would be of order $\sim
10^{15}$ g.  The mean of nonzero CME masses in the CDAW catalog
  is $1.6 \times 10^{15}$ g.

The time rate of change in momentum is given by
\be \frac{d\vec p}{dt} =
\frac{dm_{\rm eff}}{dt} \vec v +
m_{\rm eff} \frac{d\vec v}{dt}
~. \label{eqn:full_dpdt} \ee
Assuming the increase in the CME's effective mass 
is very small compared to the total effective mass, we
neglect $dm_{\rm eff}/dt$ here and in the following steps.
%
%
We then have
\bea
\frac{d\vec p(t_2)}{dt} - \frac{d\vec p(t_1)}{dt}
&=&  \vec F(t_2) - \vec F(t_1) \\
m_{\rm eff} \left ( \frac{d\vec v(t_2)}{dt} - \frac{d\vec v(t_1)}{dt} \right )
  &\simeq&  \Delta \vec F \\
\Delta \vec a &\simeq& \frac{\Delta \vec F} 
          {m_{\rm eff}} \label{eqn:delta_a} ~, \eea
where the changes in tension and hoop forces due to the
reconnection are given by Equations \ref{eqn:df_guide} and
\ref{eqn:df_hoop}, respectively.
For reconnection of a small amount of flux $\Delta \Phi$, both
  Equations \ref{eqn:df_guide} and \ref{eqn:df_hoop} scale as
the product of the reconnecting field strength multiplied by the
reconnected flux,
\be \Delta a \propto \frac{B \Delta \Phi}{m_{\rm eff}}
~. \label{eqn:da_vs_bdphi} \ee
Hence, these assumptions predict that the rate of {\em acceleration}
of a CME increases linearly with reconnected flux.

How significant is the acceleration from reconnection-driven changes
in the forces (equations \ref{eqn:df_guide} and \ref{eqn:df_hoop}
acting on a CME?  
Our formalism analyzes the change in Lorentz forces from
reconnection of a discrete amount of flux, $\Delta \Phi$.
In Section \ref{sec:outflow}, we considered reconnection for a total field
strength of 10G with a 3G poloidal component.
The SADs identified by \citep{Savage2011} and \citet{McKenzie2011}
suggest that patchy reconnection occurs over areas on the order of 10
Mm$^2$.
Accordingly, we consider reconnection of a 10 Mm$^2$ patch of 3G
field, corresponding to $\Delta \Phi = 3 \times 10^{17}$ Mx, and an
effective CME mass $m_{\rm eff} = 2 \times 10^{15}$ g.
For the tension force, the contribution to Equation
\ref{eqn:delta_a} from \ref{eqn:df_guide} alone, with $\omega =
20^\circ$, would imply an increase in acceleration of $\sim 24$ cm
s$^{-2}$.
For the hoop force, with $(R/r) = 3$, the increase in acceleration
would be $\sim$ 38 cm s$^{-2}$.
%
%
Roughly, the two effects both act to accelerate the ejection with
similar magnitude, and the combined $\Delta a$ from both is on the
order of $0.6$ m s$^{-2}$.

\citet{Qiu2005}, \citet{Kazachenko2017}, and \citet{Gopalswamy2017b}
report typical reconnection fluxes 10$^{21}$ Mx for the CMEs in their
samples. \citet{Qiu2005} also suggest that this could be an
underestimate, because not all reconnected flux exhibits ribbon
emission that is detectable by their methods.
It should be noted, however, that their approach might produce an
overestimate: if the reconnection is patchy, and both footpoints in
each polarity produce ribbon emission captured by their method, then
some reconnected flux is double counted.  These two effects ---
missing some flux from weakly emitting footpoints, and double-counting
some flux --- might offset each other, so we assume the flux estimates
from \citet{Qiu2005} approximate the true reconnected flux.
Simply summing $3 \times 10^3$ similar reconnection events, each of $3
\times 10^{17}$ Mx, to reconnect $\sim 10^{21}$ Mx of flux would yield
a total acceleration near 2000 m s$^{-2}$.  Also, this analysis
assumes that each reconnection event accelerates the {\em final} mass
of the CME --- but the CME mass should initially be smaller, and
  grow as reconnection accretes mass onto it.  Moreover, several
events studied by \citet{Qiu2005}, \citet{Kazachenko2017}, and
\citet{Gopalswamy2017b} have substantially higher ribbon fluxes.

For comparison, using data from the LASCO C1 coronagraph
\citep{Brueckner1995} before it failed, \citet{Zhang2006} report
accelerations for 50 CMEs, with a mean value of $300$ m s$^{-2}$, and
a standard deviation of $600$ m s$^{-2}$.  \citet{Vrsnak2006}
  reports a few CMEs with accelerations near 1000 m s$^{-2}$.
%
%
%
%
%
The accelerations implied by our crude estimates of Lorentz
force changes due to reconnection are much larger than those observed
in most CMEs, an issue we revisit shortly.
It should be clear from Equations \ref{eqn:df_guide} and
\ref{eqn:df_hoop} that this too-high acceleration does not
depend upon our choice that $\Delta \Phi = 3 \times 10^{17}$ Mx --- the key
factors are the typical field strength of the post-reconnection field
as it accretes onto the ejection, and the total amount of reconnected
flux over the course of the eruption.

We now consider our model of reconnection-driven changes in forces on
CMEs in the context of observations that show CMEs' {\em speeds} scale
with the reconnected flux.
Equation \ref{eqn:da_vs_bdphi} suggests that CME {\em
  accelerations}, not speeds, should scale with
the amount of reconnected flux.
This inconsistency with observations might be reconciled by the
field-strength dependence in Equation \ref{eqn:da_vs_bdphi}, which
implies that the added force on the ejection from reconnected flux
will diminish with increasing height, because field strengths decrease
with height.

While we have focused on changes in forces on an ejection due to
reconnection, Lorentz forces from external fields not directly
involved in reconnection also act on the ejection, both to drive and
to resist the ejection's rise.  In addition to weakening forces from
reconnected fields, the decrease of magnetic field strengths as the
ejection moves outward will also reduce all other Lorentz forces
acting on it.  Equations \ref{eqn:f_outer} and \ref{eqn:f_inner}
were derived for a small section of a torus, but are also valid for
external hoop forces acting on a longer torus section in the absence
of reconnection.
\begin{figure}[ht] 
  \centerline{\includegraphics[width=0.70\textwidth]{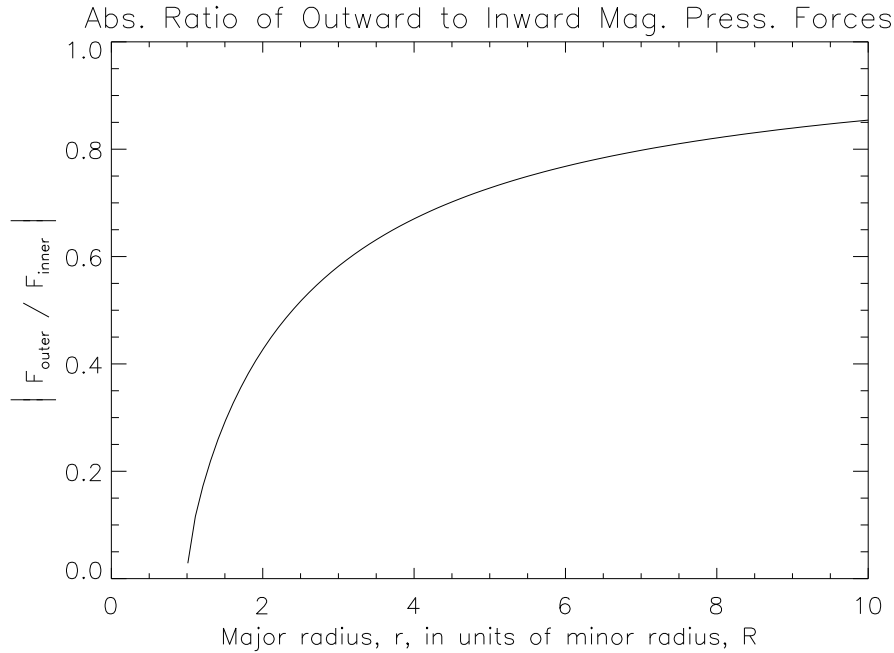}}
    \caption{This plot shows the ratio of the inward magnetic pressure
      force, $|F_{\rm outer}|$, which acts on the ejection's outer
      face, to the outward magnetic pressure force, $|F_{\rm inner}|$,
      which acts on its inner face.  This ratio increases from near
      zero at $(r/R) \simeq 1$ to about 0.85 for large $(r/R)$. }
\label{fig:fouter_to_finner}
\end{figure} 
The ratio of inward to outward hoop forces, plotted as $|F_{\rm
  outer}/F_{\rm inner}|$ in Figure \ref{fig:fouter_to_finner}, aids
consideration of how the total hoop force on the ejection changes as
the ejection rises.  At initiation of the eruption, $r$ is nearest
$R$, so $|F_{\rm outer}/F_{\rm inner}|$ is smallest, and the
acceleration is largest.  As $r$ increases relative to $R$, however,
$|F_{\rm inner}|$ decreases relative to $|F_{\rm outer}|$, so the
external forces slowing the ejection increase substantially relative
to the external forces driving it.  Consistent with this,
\citet{Vrsnak2006} estimated the accelerations of flare sprays,
eruptive prominences, and the leading edges of white-light CMEs, and
found a trend of decreasing acceleration with distance; see especially
his Figure 2.  The effect of decreasing field strength with height
could be investigated using MHD models of erupting CMEs.

Another factor that we have ignored is drag force on the CME.
\citet{Vrsnak2008} analyzed CME accelerations, and found drag to be
dynamically significant over the height range that they analyzed, (2
-- 30) $R_\odot$.  They expressed the drag force as
\be F_{\rm drag} \sim \rho_{\rm  amb} (v - v_{\rm amb})^2 C_D A ~, 
\ee
where $C_D$ is the dimensionless drag coefficient, $\rho_{\rm amb}$ is
the density of the ambient plasma through which the CME propagates,
$v_{\rm amb}$ is the speed of the ambient plasma, and $A$ is the
effective area the ejection perpendicular to its relative motion with
respect to the background plasma.
In the outer corona and heliosphere, $v_{\rm amb}$ is the solar wind
speed. In assessing drag forces on H-$\alpha$ flare surges within 0.25
$R_\odot$, \citet{Vrsnak2006} noted that drag forces can be relatively
strong there, because the ambient plasma speed is zero and ambient
densities are highest.
%
%
Increases in CME speed and area due to flux accretion would, all else
equal, increase drag forces on the CME.
(We have also ignored changes in gravity as the ejection rises,
because that the gravitational energies of many CMEs are dwarfed by
their kinetic energies --- {\it e.g.}, \citealt{Emslie2012}.)


\section{Summary \& Conclusions}
\label{sec:conc}


Several observations indicate that magnetic reconnection in
flare-associated CMEs plays a central role in the development such
ejections.  (1) Emission patterns consistent with reconnection are
common. (2) Estimates of reconnected flux inferred from these emission
patterns are correlated with CME speeds, so a causal link between
reconnection and CME accelerations is plausible.  (3) The masses of
CMEs are correlated with their speeds.
(4) Estimates of reconnected flux are correlated with fitted poloidal
fluxes in CME flux ropes.  Taken together, these observations
implicate reconnection in the formation and acceleration of some CMEs.

We noted that reconnection tends to add mass to CMEs.  This occurs
because reconnection joins external flux with the ejection as it
rises, and plasma is frozen to this flux outside of the small
diffusion region where the reconnection occurs.  If reconnection had
no other effect besides adding mass to the erupting flux system, the
ejection would slow. We expect, however, that reconnection should also
affect the Lorentz forces acting on the ejection.

Accordingly, we then sought to quantify, using order-of-magnitude
estimates, how reconnection might affect the dynamics of a rising
ejection. First, we estimated the momentum flux into a rising ejection
due to outflow from the flare reconnection underneath it.  We found
that the reconnection outflow might be responsible for $\sim 10$\% of
the final momentum of a moderate-speed CME, although the uncertainties
in this estimate are large --- it is plausible that momentum transfer
from the reconnection outflow supplies significantly more or less than
this.

Second, we considered simple models of the magnetic structure of an
ejection before and after reconnection of a small amount of flux
$\Delta \Phi$.  We started by estimating the resulting changes in the
net magnetic tension on the ejection, and found that downward magnetic
tension was reduced by an amount proportional to $B \Delta \Phi$,
where $B$ is the strength of the reconnecting field component and
$\Delta \Phi$ is the amount of reconnected flux.  We then estimated
the change in upward magnetic pressure, due to the increased hoop
force from accretion of post-reconnection flux onto the ejection,
which was also proportional to $B \Delta \Phi$, where $B$ is the total
field strength.

A key result of our analysis is that changes in tension and hoop
forces due to reconnection are commensurate.  This implies that a
tether-cutting model of dynamics misses half of the story: ``hoop
adding'' is just as important.  This also highlights a significant
difference between breakout reconnection above a proto-ejection, which
reduces downward tension from overlying fields, and flare reconnection
below a proto-ejection, which both reduces downward tension forces and
adds hoop forces. Because the added hoop force is on the order of the
reduced tension force, if the reconnecting fields were the same at
each location, then flare reconnection would produce about twice the
change in Lorentz force as breakout reconnection.  But the decrease in
field strength with height together with the field-strength dependence
of the change in force imply that flare reconnection should much more
strongly affect a proto-ejection's acceleration than breakout
reconnection.  This accords with the results reported by
\citet{Karpen2012}.

For reasonable choices of reconnecting field strengths and CME mass,
we estimated the change in CME acceleration due to changes in Lorentz
forces alone.
This yielded rough estimates of CME accelerations near 2000 m
s$^{-2}$. Such accelerations are, however, larger than the observed
accelerations of most CMEs.  We hypothesize that the decrease in
Lorentz forces on the CME with increasing height above the solar
surface ({\it e.g.}, in Figure \ref{fig:fouter_to_finner}), combined
with drag on the ejection, both neglected in our model, can reconcile
the discrepancy between observed CME accelerations and those predicted
by our model.  Another possibility, discussed in Section \ref{subsec:accel}
above, is that the reconnected flux estimated from flare ribbons and
magnetograms, by double counting flux in each polarity, overestimates
the actual reconnected flux.

We hope the rough estimates made here will be tested with data from
MHD simulations of ejections.  In particular, it would be illuminating
to compare the contributions of momentum flux from the reconnection
outflow, reduction in magnetic tension, and increase in hoop force to
the overall momentum budgets of model CMEs.

Acknowledgments:
The author is especially grateful for the extraordinary effort made by
the dedicated and insightful referee, who carefully reviewed the
manuscript through multiple revisions.  Several substantive changes
suggested by the referee were adopted, and these greatly improved the
paper.
We acknowledge funding from the National Science Foundation's Solar
Terrestrial program under award NSF AGS 1548732, NSF's SHINE program
via award NSF AGS 1622495, and NASA's Heliophysics - Guest
Investigator's program via H-GI ODDE NNX15AN68G.
The CDAW CME catalog is generated and maintained at the CDAW Data
Center by NASA and The Catholic University of America in cooperation
with the Naval Research Laboratory. SOHO is a project of international
cooperation between ESA and NASA.
The author is grateful to the U.S. taxpayers for providing the funds
necessary to perform this work. \\

Disclosure of Potential Conflicts of Interest: 
The author declares that he has no conflicts of interest.

\appendix

\section{Evolution of Magnetic Helicity}
\label{app:helicity}


Magnetic reconnection not only alters the Lorentz forces acting
  on an erupting flux rope, it also changes the rope's magnetic
  helicity ({\it e.g.}, \citealt{Berger1998}).  To consider the effect of
  reconnection on the ejection's helicity, we idealize the erupting
  flux system as a toroidal rope of purely azimuthal flux, $\Phi_T$,
formed by the ejection's core, surrounded by purely poloidal
flux, $\Phi_P$.  The mutual helicity of the two flux systems, $|H_{\rm
  mut}| = |2\Phi_P \Phi_T|$, quantifies linkages between the two flux
systems.  If the poloidal or azimuthal flux systems contained
internal, ``self'' linkages, then the total helicity would be the sum
of these self helicities and the mutual helicity.  Because
  magnetic helicity is an ideal MHD invariant, no ideal process can
lead to the increase of poloidal flux wrapping around the azimuthal
core field.  (Ideal processes could, however, cause poloidal flux to
bunch up, leading to regions with stronger $B_P$.)

\begin{figure}[htp] 
  \centerline{\includegraphics[width=\textwidth,clip=true]{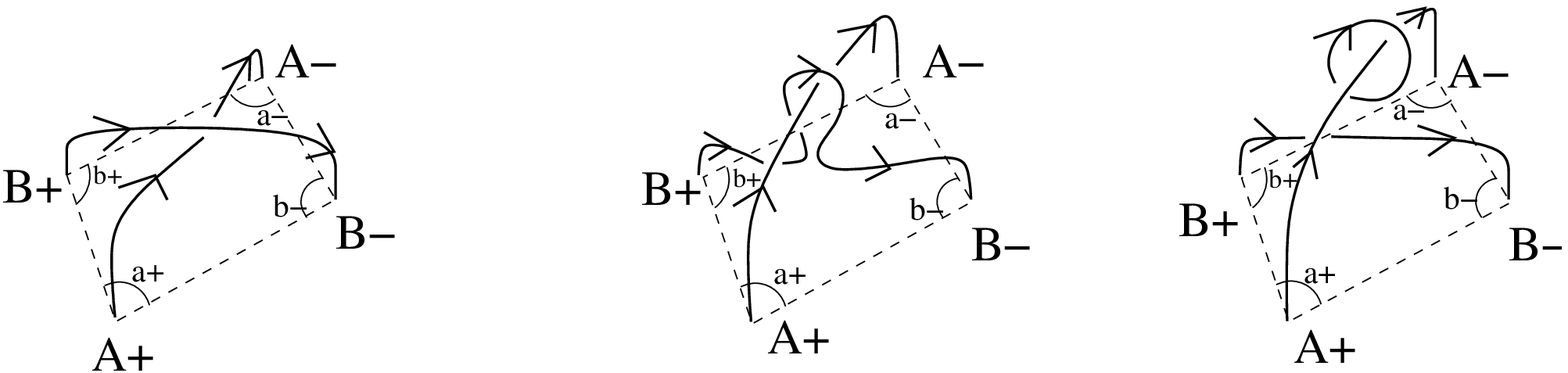}}
  \caption{{\it Left:} Flux systems A and B, each assumed untwisted, have a
    mutual helicity that is greater than zero, due to their positive
    crossing number.  {\it Middle:} System A is erupting upward through B.
    {\it Right:} Flux from B has reconnected with itself, and thereby formed
    two flux systems, one that is closed and linked to system A and
    one that is anchored at the photosphere.  The mutual helicity
    between A and the flux from system B that is anchored at the
    photosphere has changed sign, since the sign of their crossing has
    changed.  Globally, helicity is conserved here, though it has been
    transferred between mutual and self helicities.}
  \label{fig:helicity}
  \end{figure}

When reconnection occurs in the wake of the outward-moving toroidal
rope, however, it adds flux that wraps around the rope.  Helicity is
approximately globally conserved in fast reconnection ({\it e.g.},
\citealt{Berger1999}), and this increase in the helicity of the
erupting flux system is offset by an opposing change in mutual
helicity between the erupting system and surrounding fields.
Figure \ref{fig:helicity} illustrates key aspects of this evolution.
For simplicity, both flux tubes shown are assumed to have no internal
twist; relaxing this assumption would not substantively alter results
of the analysis.
Essentially, some of the mutual helicity between the rising, toroidal
system (A in the figure) and the initially overlying, external flux
system (B in the figure) is transformed into self helicity of the
erupting system (toroid plus reconnected flux that encircles it)
\citep{Berger1998}.

This transfer of helicity between the erupting and background flux
systems may be seen by considering the sign of the crossing between
the external flux and toroidal flux, which changes when the rising
flux passes through the external field by driving the latter to
reconnect with itself.  Initially (left panel), their mutual
  helicity is $H_i = +\Phi_A \Phi_B (b_+ + b_-)/\pi$, and in the final
  state (right panel), their mutual helicity is $H_f = -\Phi_A
  \Phi_B(a_+ + a_-)/\pi$.  (\citet{Demoulin2006} review the sign
  convention for angles: the mutual helicity is defined from footpoint
  angles of the overlying loop; angles swing from the footpoint with
  the same sign as the apex loop's footpoint toward the opposite-sign
  footpoint; and counterclockwise is positive.)  The difference in
  mutual helicities is
\be \Delta H_{\rm mut} = H_f - H_i =
-\Phi_A \Phi_B (a_+ + a_- + b_+ + b_-)/\pi = -2 \Phi_A \Phi_B
~. \label{eqn:delta_h} \ee
The difference in mutual helicities from the changed crossing sign has
been added to the self helicity of the erupting system: the flux that
reconnected to encircle the rising flux contributes to the erupting
system's self helicity.  The linking number of fluxes in the erupting
system is +1, so the helicity within that system is $+2 \Phi_A
\Phi_B$.

More detailed studies of helicity evolution due to reconnection in an
eruption with reconnection have been undertaken ({\it e.g.},
\citealt{Priest2017}).

%

\section{Large-$r$ Scaling of Hoop Force}
\label{app:current}

From Equation \ref{eqn:f_hoop_bigr}, the hoop force $F_{\rm hoop}$
on a length of torus $r \, \Delta \phi$ scales as $R^2 \, \Delta \phi (B_{\rm
  mid}^2/8)$ for $r \gg R$.  The force per unit length therefore
scales as
\be \frac{F_{\rm hoop}}{r \, \Delta \phi} \sim \frac{R^2 B_{\rm mid}^2}{r}
~. \label{eqn:hoop_scaling} \ee

\citet{Shafranov1966} derived an expression for the hoop force per
unit length in terms of the total azimuthal electric current, $I$,
finding that the force per unit length scaled as $I^2/r$.
In our model, with the poloidal field in Equation
\ref{eqn:bp_theta}, electric current along the torus's axis is
\bea I &=& \frac{c}{4\pi} \oint \vec B \cdot d\vec L
= \frac{c}{4\pi}
\int_{-\pi}^{\pi} \frac{B_{\rm mid} \, r}{r + R \cos \theta} R \, d\theta \\
&=& \frac{c r R B_{\rm mid}}{4\pi} \left [ \left .
\frac{2}{\sqrt{r^2 - R^2}} \tan^{-1} \left ( 
\frac{\sqrt{r - R}}{\sqrt{r + R}} \tan (\theta /2)
\right ) \right |_{-\pi}^{\pi} \right ] \\
&=& \frac{c r R B_{\rm mid}}{4\pi} \frac{2 \pi}{\sqrt{r^2 - R^2}} 
= \frac{c r R B_{\rm mid}}{2 \sqrt{r^2 - R^2}} 
~. \label{eqn:tot_current} \eea
Therefore, in our model, $I^2/r$ varies with $r$ as 
\be I^2/r
\sim \frac{r R^2 B_{\rm mid}^2}{(r^2 - R^2)} 
= \frac{R^2 B_{\rm mid}^2}{r (1 - (R/r)^2)} \ee
In the $r \gg R$ limit, then, 
\be I^2/r
\sim \frac{R^2 B_{\rm mid}^2}{r} ~. \ee
This matches the scaling of our force per unit length, from Equation
\ref{eqn:hoop_scaling}.  Figure \ref{fig:hoop_forces} shows a
comparison of $1/r$ (dotted line) with the radial dependence of the
term in square brackets in the expression for $F_{\rm hoop}$ in
Equation \ref{eqn:f_hoop} (solid line).

\section{Pre- to Post-Dipolarization Change in Hoop Force}
\label{app:pre_post_dip}

Another aspect of the change in hoop force is more subtle: the
  total change in force on the proto-ejection by the addition of
  reconnected flux $\Delta \Phi$ is likely larger than the value given
  in Equation \ref{eqn:df_hoop}.  To see why, we must consider how
  the net force exerted by the reconnected field on the ejection
  changes as a result of the dipolarization.

\begin{figure}[htp] 
  \centerline{\includegraphics[width=0.70\textwidth]{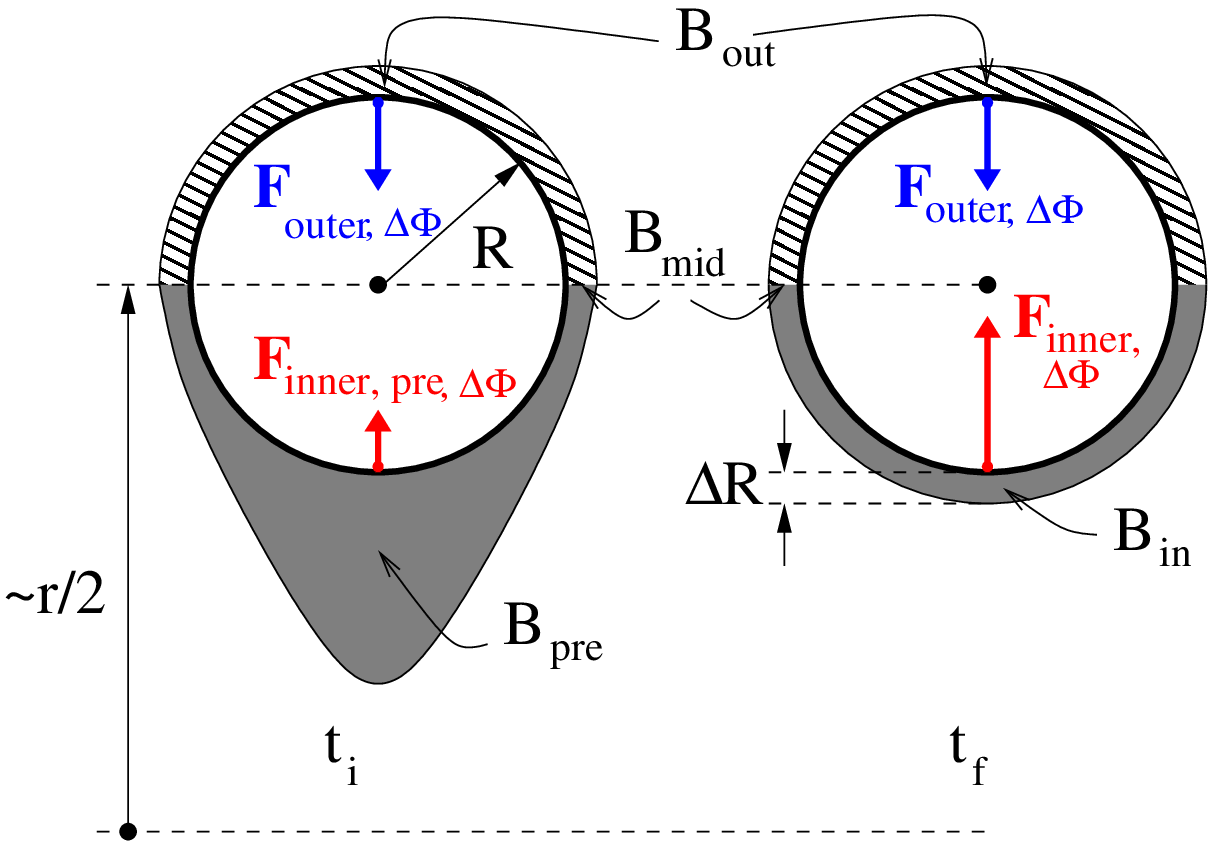}}
     \caption{{\it Left:} Prior to complete dipolarization of
       post-reconnection flux, the flux density behind the ejection,
       $B_{\rm pre}$, is less than the inner-edge flux density $B_{\rm
         in}$ assumed in deriving Equation \ref{eqn:f_inner}.  The
       distance from the reconnection site to the trailing edge of the
       proto-ejection is assumed to be $\sim r/2$. The inward pressure
       force on the proto-ejection's outer surface due to the
       reconnected flux, $F_{{\rm outer},\Delta \Phi}$, assumed to be
       unchanging, is depicted as a {\it blue vector}.  The
       pre-dipolarization outward pressure force on its inner surface
       due to the reconnected flux, $F_{{\rm inner,pre},\Delta \Phi}$,
       is depicted as a {\it red vector}.  For the configuration shown, at
       time $t_i$, the net pressure force from the annulus of
       reconnected flux is inward (negative). {\it Right:} After complete
       dipolarization at $t_f$, the outward pressure force from the
       reconnected flux (denoted $F_{{\rm inner},\Delta \Phi}$) is
       larger, and the net (outward) pressure force from the annulus
       of reconnected flux is strongest, and equal to $\Delta F_{\rm
         hoop}$ from Equation \ref{eqn:df_hoop}.  The total {\em
         change} in the magnetic pressure force on the proto-ejection
       is the net pressure force at $t_f$ minus the net pressure force
       at $t_i$, $\Delta F_{fi}$ and this difference can exceed the
       value for $\Delta F_{\rm hoop}$.}
  \label{fig:b_trailing}
  \end{figure}
%

We define $t_i$ to be an instant after this flux has reconnected
but before it has fully dipolarized, and $t_f$ to be after full
dipolarization.
At $t_i$, the field strength on the inner side of the proto-ejection
is weaker than assumed in deriving Equation \ref{eqn:f_inner}, as
shown in the left magnetic field configuration in Figure
\ref{fig:b_trailing}.  Consequently, the pre-dipolarization, inner-surface
component of the hoop force, $F_{\rm inner,pre}$, is weaker than
$F_{\rm inner}$ derived above.
We assume that flux arching over the proto-ejection's leading edge is
unchanged by the reconnection, so always acts on the proto-ejection
with the same inward pressure force $F_{\rm outer}$.
This implies that the {\em instantaneous} hoop force acting on the
proto-ejection due to just the annulus of reconnected flux is
strongest after complete dipolarization, as depicted in the right
panel of Figure \ref{fig:b_trailing}.  This force is equal to the
value given in Equation \ref{eqn:df_hoop},
which can also be understood as the net pressure force due to only the
annulus of reconnected flux $\Delta \Phi$ that wraps around the
proto-ejection,
  \be \Delta F_{\rm hoop} =
  F_{{\rm inner}, \Delta \Phi} + F_{{\rm outer}, \Delta \Phi}
  \label{eqn:dphi_only} ~, \ee
where the $\Delta \Phi$ subscripts indicate that these forces are due
only to the reconnected flux, and $F_{{\rm outer}, \Delta \Phi}$ is
negative.

This instantaneous force, however, is not the same as the {\em change}
in pressure force acting on the proto-ejection due to the accretion of
reconnected flux.
The change in pressure force on the ejection from $t_i$ to $t_f$ is
the difference between the net forces at $t_f$ and $t_i$, 
\bea \Delta F_{fi}
&=& (F_{{\rm inner}, \Delta \Phi} + F_{{\rm outer}, \Delta \Phi})
- (F_{{\rm inner, pre}, \Delta \Phi} + F_{{\rm outer}, \Delta \Phi})
\label{eqn:fi_1st} \\
&=& (F_{{\rm inner}, \Delta \Phi} - F_{{\rm inner, pre}, \Delta \Phi})
\label{eqn:fi_2nd} ~, \eea
where the subscripts denote the outward and inward forces (subscripted
inner and outer, respectively) on the proto-ejection from the
reconnected flux $\Delta \Phi$.
It can be seen that if $|F_{{\rm inner, pre}, \Delta \Phi}| < |F_{{\rm
  outer}, \Delta \Phi}|$, then $\Delta F_{fi} > \Delta F_{\rm hoop}$
from Equation \ref{eqn:df_hoop}.
We remark that, much like the magnetic tension force from a flux tube
$\Delta \Phi$ that is inward prior to tether cutting reconnection, the
net force due to inward and outward magnetic pressures from a flux
tube $\Delta \Phi$ might also be inward in the pre-dipolarization
state ({\it i.e.}, $(F_{{\rm inner, pre}, \Delta \Phi} + F_{{\rm outer},
  \Delta \Phi}) < 0$ in Equation \ref{eqn:fi_1st}.

How much weaker would the pre-dipolarization magnetic field, $B_{\rm
  pre}$, have to be for the pre-dipolarization force to be inward? As
a first step to addressing this question, we use Equation
\ref{eqn:f_inner} to crudely approximate the outward pressure force
prior to complete dipolarization, $F_{{\rm inner,pre},\Delta \Phi}$,
to be
\be  F_{{\rm inner,pre}, \Delta \Phi}
\simeq (B_{\rm pre}/B_{\rm in})^2 F_{{\rm inner}, \Delta \Phi}
~. \label{eqn:inner_pre} \ee
We expect that the behavior of $F_{{\rm inner}, \Delta \Phi}$ follows
$F_{{\rm inner}}$, and similarly that $F_{{\rm outer}, \Delta \Phi}$
follows $F_{{\rm outer}}$.
It is helpful to refer again to Figure \ref{fig:fouter_to_finner},
which plots the ratio of $|F_{\rm outer}/F_{\rm inner}|$ as a function
of major radius $r$, in units of minor radius $R$.
For $(r/R)$ near unity, $|F_{\rm outer}| << |F_{\rm inner}|$, but
$|F_{\rm outer}|$ is more comparable to $|F_{\rm inner}|$ at larger
$(r/R)$.  For the ratio $(r/R) = 3$ assumed above, $|F_{\rm
  outer}/F_{\rm inner}|$ is near 0.6, implying that a ratio of
$(B_{\rm pre}/B_{\rm inner})$ of 0.7, for instance, would
yield an initially inward force:
\be F_{{\rm inner,pre}, \Delta \Phi} + F_{{\rm outer}, \Delta \Phi}
\simeq 0.49 F_{{\rm inner}, \Delta \Phi} - 0.6 F_{{\rm inner}, \Delta \Phi} < 0
~. \label{eqn:negative_dF} \ee
A flux density of 0.7 times the dipolarized value would occur if the
average thickness of the shaded region in Figure \ref{fig:b_trailing}
were more than about $(0.7)^{-1} \simeq 1.4$ times the final
thickness, $\Delta R$.

It might be suggested that having an $F_{{\rm inner,pre}, \Delta
  \Phi}$ weaker than $F_{{\rm outer}, \Delta \Phi}$ is somehow
unphysical, and that $(F_{{\rm inner}, \Delta \Phi} + F_{{\rm outer},
  \Delta \Phi})$ must for some reason always be outward for an
accelerating CME.  But our analysis does not consider all forces on an
ejection, just forces from a single flux tube $\Delta \Phi$, and
forces from individual tubes may oppose the ejection. As noted above,
a similar analysis is implicit in the tether-cutting model, which
supposes that many individual tethers oppose an eruption's upward
motion, but nonetheless that eruption can have a net outward force.

How much larger is the change in hoop force $\Delta F_{fi}$ than the
estimate in Equation \ref{eqn:df_hoop}?  It is problematic to
estimate the diminished pressure force on the inner surface of the
ejection prior to complete dipolarization without assuming a model of
the field structure in reconnection outflow, which is beyond the scope
of this analysis.

\section{Tension Force for Torus}
\label{app:tension}

We revisit the change in tension estimated in Equation
  \ref{eqn:df_tension} above to investigate the effect, if any, of
  the axial curvature that gives rise to the hoop force, since this
  could also affect the change in tension from $B_P$. To get the
total (sum of upward and downward) tension, we now use Equation
\ref{eqn:bp_theta} in Equation \ref{eqn:df_tension0} and integrate
over $[-\pi,\pi]$,
\bea \Delta F_{\rm tension}
&=& \frac{1}{4\pi} \int dV \, \hat z \cdot (\vec B \cdot \grad) \vec B \\
&=& \frac{1}{4\pi}  \int R \, d\theta  \, dA \,
B_P(\theta) \frac{\hat z }{R} \cdot
\frac{\partial (B_P(\theta) \hat b)}{\partial \theta} \\
&=& \frac{\Delta \Phi}{4\pi}  \int \, d\theta \, \hat z  \cdot
\frac{\partial ( B_P(\theta) \hat b)}{\partial \theta} \\
&=& \frac{\Delta \Phi}{4\pi} 
\int_{-\pi}^{\pi} \, d\theta \left (
- \frac{\partial B_P(\theta)}{\partial \theta} \sin  \theta
- B_P(\theta) \cos \theta 
\right ) \\
&=& \frac{- \Delta \Phi B_{\rm mid}}{4\pi} 
\int_{-\pi}^{\pi} d\theta \left (
\frac{r R \sin^2  \theta}{(r + R \cos  \theta)^2} +
\frac{r \cos  \theta}{r + R \cos  \theta}  
\right ) \\
&=& \frac{- \Delta \Phi B_{\rm mid}}{4\pi} 
\int_{-\pi}^{\pi} d\theta \left (
\frac{r R + r^2 \cos  \theta}{(r + R \cos  \theta)^2} 
\right ) \\
&=& \frac{- \Delta \Phi B_{\rm mid}}{4\pi} 
\int_{-\pi}^{\pi} d\theta \left (
\frac{(r/R) + (r/R)^2 \cos  \theta}{(r/R + \cos  \theta)^2} 
\right ) \\
&=& \frac{- \Delta \Phi B_{\rm mid} \, c}{4\pi} 
\int_{-\pi}^{\pi} d\theta \left (
\frac{1 + c \cos  \theta}{(c + \cos  \theta)^2} 
\right )  ~, \label{eqn:f_inward_tension2} \eea 
where, as before, the unit vector of the poloidal field is given by
$\hat b = \hat \theta = (\cos \theta \hat x - \sin \theta \hat z)$,
$dV = dA \, R \,d\theta$, $\Delta \Phi = B_P \, \Delta A$ is the flux that
reconnects, and in the last line we have defined $c = r/R$.

From \citet{Dwight1961} (p. 106), we first evaluate
\bea
\int_{-\pi}^{\pi} d\theta 
\frac{1}{(c + \cos  \theta)^2}
&=& \left . \frac{\sin \theta}{(1 - c^2)(c + \cos \theta)}
\right \vert_{-\pi}^\pi
- \frac{c}{1 - c^2} \int_{-\pi}^{\pi} d\theta \frac{1}{(c + \cos  \theta)} \\
&=& \frac{c}{c^2 - 1} \int_{-\pi}^{\pi} d\theta \frac{1}{(c + \cos  \theta)}
\label{eqn:app_integral_1} \eea 
Second, and also from \citet{Dwight1961} (p. 106), we evaluate
\bea
\int_{-\pi}^{\pi} d\theta 
\frac{c \cos \theta }{(c + \cos  \theta)^2}
&=& \left . \frac{c^2 \sin \theta}{(c^2 - 1)(c + \cos \theta)}
\right \vert_{-\pi}^\pi
- \frac{c}{c^2 - 1} \int_{-\pi}^{\pi} d\theta \frac{1}{(c + \cos  \theta)} \\
&=& - \frac{c}{c^2 - 1} \int_{-\pi}^{\pi} d\theta \frac{1}{(c + \cos  \theta)}
\label{eqn:app_integral_2} \eea 
Equations \ref{eqn:app_integral_1} and \ref{eqn:app_integral_2}
are equal and opposite.

This cancellation is reasonable, given that our model poloidal field
is essentially circular, so the curvature is uniform around the flux
tube over which the force density is integrated.  It is true that, due
to the higher flux density on the inner section of the flux tube, the
force density is higher there. But the force density is not integrated
over equal volumes for the inner and outer segments; it is over
different volumes threaded by equal amounts of flux.





\begin{thebibliography}{88}
\ifx\bisbn     \undefined \def\bisbn  #1{ISBN #1}\fi
\ifx\binits    \undefined \def\binits#1{#1}\fi
\ifx\bauthor   \undefined \def\bauthor#1{#1}\fi
\ifx\batitle   \undefined \def\batitle#1{#1}\fi
\ifx\bjtitle   \undefined \def\bjtitle#1{\textit{#1}}\fi
\ifx\bvolume   \undefined \def\bvolume#1{\textbf{#1}}\fi
\ifx\byear     \undefined \def\byear#1{#1}\fi
\ifx\bissue    \undefined \def\bissue#1{#1}\fi
\ifx\bfpage    \undefined \def\bfpage#1{#1}\fi
\ifx\blpage    \undefined \def\blpage #1{#1}\fi
\ifx\burl      \undefined \def\burl#1{\textsf{#1}}\fi
\ifx\href      \undefined \def\href#1#2{\textsf{#2}}\fi
\ifx\betal     \undefined \def\betal{\textit{et al.}}\fi
\ifx\bctitle   \undefined \def\bctitle#1{#1}\fi
\ifx\beditor   \undefined \def\beditor#1{#1}\fi
\ifx\bbtitle   \undefined \def\bbtitle#1{\textit{#1}}\fi
\ifx\bedition  \undefined \def\bedition#1{#1}\fi
\ifx\bseriesno \undefined \def\bseriesno#1{\textbf{#1}}\fi
\ifx\blocation \undefined \def\blocation#1{#1}\fi
\ifx\bsertitle \undefined \def\bsertitle#1{\textit{#1}}\fi
\ifx\bsnm      \undefined \def\bsnm#1{#1}\fi
\ifx\bsuffix   \undefined \def\bsuffix#1{#1}\fi
\ifx\bparticle \undefined \def\bparticle#1{#1}\fi
\ifx\barticle  \undefined \def\barticle#1{}\fi
\ifx\binstitute  \undefined \def\binstitute#1{#1}\fi
\ifx\bpublisher  \undefined \def\bpublisher#1{#1}\fi
\ifx\doiurl    \undefined
  \def\doiurl#1{\href{http://dx.doi.org/#1}{\textsf{DOI}}}\fi
\ifx\arxivurl  \undefined
  \def\arxivurl#1{\href{http://arxiv.org/abs/#1}{\textsf{arXiv}}}\fi
\ifx\adsurl    \undefined
  \def\adsurl#1{\href{http://adsabs.harvard.edu/abs/#1}{\textsf{ADS}}}\fi
\ifx\botherref \undefined \def\botherref#1{}\fi
\ifx\url       \undefined \def\url#1{\textsf{#1}}\fi
\ifx\bchapter  \undefined \def\bchapter#1{}\fi
\ifx\bbook     \undefined \def\bbook#1{}\fi
\ifx\bcomment  \undefined \def\bcomment#1{#1}\fi
\ifx\oauthor   \undefined \def\oauthor#1{#1}\fi
\ifx\citeauthoryear \undefined\def \citeauthoryear#1{#1}\fi
\ifx\endbibitem\undefined \def\endbibitem{}\fi
\ifx\bconflocation  \undefined \def\bconflocation#1{#1} \fi

\bibitem[\protect\citeauthoryear{{Amari} \textit{et~al.}}{2010}]{Amari2010}
\begin{barticle}
\bauthor{\bsnm{{Amari}}, \binits{T.}},
\bauthor{\bsnm{{Aly}}, \binits{J.}},
\bauthor{\bsnm{{Mikic}}, \binits{Z.}},
\bauthor{\bsnm{{Linker}}, \binits{J.}}:
\byear{2010},
\batitle{{Coronal Mass Ejection Initiation: On the Nature of the Flux
  Cancellation Model}}.
\bjtitle{\apjl}
\bvolume{717},
\bfpage{L26}.
\doiurl{10.1088/2041-8205/717/1/L26}.
\end{barticle}
\endbibitem

\bibitem[\protect\citeauthoryear{{Antiochos}, {DeVore}, and
  {Klimchuk}}{1999}]{Antiochos1999a}
\begin{barticle}
\bauthor{\bsnm{{Antiochos}}, \binits{S.K.}},
\bauthor{\bsnm{{DeVore}}, \binits{C.R.}},
\bauthor{\bsnm{{Klimchuk}}, \binits{J.A.}}:
\byear{1999},
\batitle{{A Model for Solar Coronal Mass Ejections}}.
\bjtitle{ApJ}
\bvolume{510},
\bfpage{485}.
\end{barticle}
\endbibitem

\bibitem[\protect\citeauthoryear{{Anzer}}{1978}]{Anzer1978}
\begin{barticle}
\bauthor{\bsnm{{Anzer}}, \binits{U.}}:
\byear{1978},
\batitle{{Can coronal loop transients be driven magnetically}}.
\bjtitle{\solphys}
\bvolume{57},
\bfpage{111}.
\doiurl{10.1007/BF00152048}.
\adsurl{1978SoPh...57..111A}.
\end{barticle}
\endbibitem

\bibitem[\protect\citeauthoryear{{Anzer} and {Pneuman}}{1982}]{Anzer1982}
\begin{barticle}
\bauthor{\bsnm{{Anzer}}, \binits{U.}},
\bauthor{\bsnm{{Pneuman}}, \binits{G.W.}}:
\byear{1982},
\batitle{{Magnetic reconnection and coronal transients}}.
\bjtitle{\solphys}
\bvolume{79},
\bfpage{129}.
\doiurl{10.1007/BF00146978}.
\adsurl{1982SoPh...79..129A}.
\end{barticle}
\endbibitem

\bibitem[\protect\citeauthoryear{{Aschwanden} and
  {Alexander}}{2001}]{Aschwanden2001}
\begin{barticle}
\bauthor{\bsnm{{Aschwanden}}, \binits{M.J.}},
\bauthor{\bsnm{{Alexander}}, \binits{D.}}:
\byear{2001},
\batitle{{Flare Plasma Cooling from 30 MK down to 1 MK modeled from Yohkoh,
  GOES, and TRACE observations during the Bastille Day Event (14 July 2000)}}.
\bjtitle{\solphys}
\bvolume{204},
\bfpage{91}.
\doiurl{10.1023/A:1014257826116}.
\adsurl{2001SoPh..204...91A}.
\end{barticle}
\endbibitem

\bibitem[\protect\citeauthoryear{{Bein} \textit{et~al.}}{2012}]{Bein2012}
\begin{barticle}
\bauthor{\bsnm{{Bein}}, \binits{B.M.}},
\bauthor{\bsnm{{Berkebile-Stoiser}}, \binits{S.}},
\bauthor{\bsnm{{Veronig}}, \binits{A.M.}},
\bauthor{\bsnm{{Temmer}}, \binits{M.}},
\bauthor{\bsnm{{Vr{\v s}nak}}, \binits{B.}}:
\byear{2012},
\batitle{{Impulsive Acceleration of Coronal Mass Ejections. II. Relation to
  Soft X-Ray Flares and Filament Eruptions}}.
\bjtitle{\apj}
\bvolume{755},
\bfpage{44}.
\doiurl{10.1088/0004-637X/755/1/44}.
\adsurl{2012ApJ...755...44B}.
\end{barticle}
\endbibitem

\bibitem[\protect\citeauthoryear{{Benz}}{2008}]{Benz2008}
\begin{barticle}
\bauthor{\bsnm{{Benz}}, \binits{A.O.}}:
\byear{2008},
\batitle{{Flare Observations}}.
\bjtitle{Living Reviews in Solar Physics}
\bvolume{5},
\bfpage{1}.
\end{barticle}
\endbibitem

\bibitem[\protect\citeauthoryear{{Berger}}{1999}]{Berger1999}
\begin{barticle}
\bauthor{\bsnm{{Berger}}, \binits{M.A.}}:
\byear{1999},
\batitle{{Magnetic helicity in space physics}}.
\bjtitle{Washington DC American Geophysical Union Geophysical Monograph Series}
\bvolume{111},
\bfpage{1}.
\doiurl{10.1029/GM111p0001}.
\adsurl{1999GMS...111....1B}.
\end{barticle}
\endbibitem

\bibitem[\protect\citeauthoryear{{Berger} \textit{et~al.}}{1998}]{Berger1998}
\begin{barticle}
\bauthor{\bsnm{{Berger}}, \binits{T.E.}},
\bauthor{\bsnm{{Loefdahl}}, \binits{M.G.}},
\bauthor{\bsnm{{Shine}}, \binits{R.S.}},
\bauthor{\bsnm{{Title}}, \binits{A.M.}}:
\byear{1998},
\batitle{{Measurements of Solar Magnetic Element Motion from High-Resolution
  Filtergrams}}.
\bjtitle{ApJ}
\bvolume{495},
\bfpage{973}.
\doiurl{10.1086/305309}.
\adsurl{1998ApJ...495..973B}.
\end{barticle}
\endbibitem

\bibitem[\protect\citeauthoryear{{Brueckner}
  \textit{et~al.}}{1995}]{Brueckner1995}
\begin{barticle}
\bauthor{\bsnm{{Brueckner}}, \binits{G.E.}},
\bauthor{\bsnm{{Howard}}, \binits{R.A.}},
\bauthor{\bsnm{{Koomen}}, \binits{M.J.}},
\bauthor{\bsnm{{Korendyke}}, \binits{C.M.}},
\bauthor{\bsnm{{Michels}}, \binits{D.J.}},
\bauthor{\bsnm{{Moses}}, \binits{J.D.}},
\bauthor{\bsnm{{Socker}}, \binits{D.G.}},
\bauthor{\bsnm{{Dere}}, \binits{K.P.}},
\bauthor{\bsnm{{Lamy}}, \binits{P.L.}},
\bauthor{\bsnm{{Llebaria}}, \binits{A.}},
\bauthor{\bsnm{{Bout}}, \binits{M.V.}},
\bauthor{\bsnm{{Schwenn}}, \binits{R.}},
\bauthor{\bsnm{{Simnett}}, \binits{G.M.}},
\bauthor{\bsnm{{Bedford}}, \binits{D.K.}},
\bauthor{\bsnm{{Eyles}}, \binits{C.J.}}:
\byear{1995},
\batitle{{The Large Angle Spectroscopic Coronagraph (LASCO)}}.
\bjtitle{\solphys}
\bvolume{162},
\bfpage{357}.
\doiurl{10.1007/BF00733434}.
\end{barticle}
\endbibitem

\bibitem[\protect\citeauthoryear{{Cargill} \textit{et~al.}}{1996}]{Cargill1996}
\begin{barticle}
\bauthor{\bsnm{{Cargill}}, \binits{P.J.}},
\bauthor{\bsnm{{Chen}}, \binits{J.}},
\bauthor{\bsnm{{Spicer}}, \binits{D.S.}},
\bauthor{\bsnm{{Zalesak}}, \binits{S.T.}}:
\byear{1996},
\batitle{{Magnetohydrodynamic simulations of the motion of magnetic flux tubes
  through a magnetized plasma}}.
\bjtitle{\jgr}
\bvolume{101},
\bfpage{4855}.
\doiurl{10.1029/95JA03769}.
\adsurl{1996JGR...101.4855C}.
\end{barticle}
\endbibitem

\bibitem[\protect\citeauthoryear{{Casini} \textit{et~al.}}{2003}]{Casini2003}
\begin{barticle}
\bauthor{\bsnm{{Casini}}, \binits{R.}},
\bauthor{\bsnm{{L{\'o}pez Ariste}}, \binits{A.}},
\bauthor{\bsnm{{Tomczyk}}, \binits{S.}},
\bauthor{\bsnm{{Lites}}, \binits{B.W.}}:
\byear{2003},
\batitle{{Magnetic Maps of Prominences from Full Stokes Analysis of the He I D3
  Line}}.
\bjtitle{\apjl}
\bvolume{598},
\bfpage{L67}.
\doiurl{10.1086/380496}.
\adsurl{2003ApJ...598L..67C}.
\end{barticle}
\endbibitem

\bibitem[\protect\citeauthoryear{{Chandra} \textit{et~al.}}{2009}]{Chandra2009}
\begin{barticle}
\bauthor{\bsnm{{Chandra}}, \binits{R.}},
\bauthor{\bsnm{{Schmieder}}, \binits{B.}},
\bauthor{\bsnm{{Aulanier}}, \binits{G.}},
\bauthor{\bsnm{{Malherbe}}, \binits{J.M.}}:
\byear{2009},
\batitle{{Evidence of Magnetic Helicity in Emerging Flux and Associated
  Flare}}.
\bjtitle{\solphys}
\bvolume{258},
\bfpage{53}.
\doiurl{10.1007/s11207-009-9392-z}.
\adsurl{2009SoPh..258...53C}.
\end{barticle}
\endbibitem

\bibitem[\protect\citeauthoryear{{Chen}}{1996}]{Chen1996}
\begin{barticle}
\bauthor{\bsnm{{Chen}}, \binits{J.}}:
\byear{1996},
\batitle{{Theory of prominence eruption and propagation: Interplanetary
  consequences}}.
\bjtitle{\jgr}
\bvolume{101}(\bissue{10}),
\bfpage{27499}.
\end{barticle}
\endbibitem

\bibitem[\protect\citeauthoryear{{Chen} \textit{et~al.}}{2006}]{Chen2006}
\begin{barticle}
\bauthor{\bsnm{{Chen}}, \binits{J.}},
\bauthor{\bsnm{{Marqu{\'e}}}, \binits{C.}},
\bauthor{\bsnm{{Vourlidas}}, \binits{A.}},
\bauthor{\bsnm{{Krall}}, \binits{J.}},
\bauthor{\bsnm{{Schuck}}, \binits{P.W.}}:
\byear{2006},
\batitle{{The Flux-Rope Scaling of the Acceleration of Coronal Mass Ejections
  and Eruptive Prominences}}.
\bjtitle{\apj}
\bvolume{649},
\bfpage{452}.
\doiurl{10.1086/506466}.
\adsurl{2006ApJ...649..452C}.
\end{barticle}
\endbibitem

\bibitem[\protect\citeauthoryear{{Compagnino}, {Romano}, and
  {Zuccarello}}{2017}]{Compagnino2017}
\begin{barticle}
\bauthor{\bsnm{{Compagnino}}, \binits{A.}},
\bauthor{\bsnm{{Romano}}, \binits{P.}},
\bauthor{\bsnm{{Zuccarello}}, \binits{F.}}:
\byear{2017},
\batitle{{A Statistical Study of CME Properties and of the Correlation Between
  Flares and CMEs over Solar Cycles 23 and 24}}.
\bjtitle{\solphys}
\bvolume{292},
\bfpage{5}.
\doiurl{10.1007/s11207-016-1029-4}.
\adsurl{2017SoPh..292....5C}.
\end{barticle}
\endbibitem

\bibitem[\protect\citeauthoryear{{Demoulin}, {Pariat}, and
  {Berger}}{2006}]{Demoulin2006}
\begin{barticle}
\bauthor{\bsnm{{Demoulin}}, \binits{P.}},
\bauthor{\bsnm{{Pariat}}, \binits{E.}},
\bauthor{\bsnm{{Berger}}, \binits{M.A.}}:
\byear{2006},
\batitle{{Basic Properties of Mutual Magnetic Helicity}}.
\bjtitle{\solphys}
\bvolume{233},
\bfpage{3}.
\doiurl{10.1007/s11207-006-0010-z}.
\adsurl{2006SoPh..233....3D}.
\end{barticle}
\endbibitem

\bibitem[\protect\citeauthoryear{{Deng} and {Welsch}}{2017}]{Deng2017}
\begin{barticle}
\bauthor{\bsnm{{Deng}}, \binits{M.}},
\bauthor{\bsnm{{Welsch}}, \binits{B.T.}}:
\byear{2017},
\batitle{{The Roles of Reconnected Flux and Overlying Fields in CME Speeds}}.
\bjtitle{\solphys}
\bvolume{292},
\bfpage{17}.
\doiurl{10.1007/s11207-016-1036-5}.
\adsurl{2017SoPh..292...17D}.
\end{barticle}
\endbibitem

\bibitem[\protect\citeauthoryear{{Dwight}}{1961}]{Dwight1961}
\begin{bbook}
\bauthor{\bsnm{{Dwight}}, \binits{H.B.}}:
\byear{1961},
\bbtitle{{Tables of integrals and other mathematical data (Fourth Ed.)}}.
\end{bbook}
\endbibitem

\bibitem[\protect\citeauthoryear{{Emslie} \textit{et~al.}}{2012}]{Emslie2012}
\begin{barticle}
\bauthor{\bsnm{{Emslie}}, \binits{A.G.}},
\bauthor{\bsnm{{Dennis}}, \binits{B.R.}},
\bauthor{\bsnm{{Shih}}, \binits{A.Y.}},
\bauthor{\bsnm{{Chamberlin}}, \binits{P.C.}},
\bauthor{\bsnm{{Mewaldt}}, \binits{R.A.}},
\bauthor{\bsnm{{Moore}}, \binits{C.S.}},
\bauthor{\bsnm{{Share}}, \binits{G.H.}},
\bauthor{\bsnm{{Vourlidas}}, \binits{A.}},
\bauthor{\bsnm{{Welsch}}, \binits{B.T.}}:
\byear{2012},
\batitle{{Global Energetics of Thirty-eight Large Solar Eruptive Events}}.
\bjtitle{\apj}
\bvolume{759},
\bfpage{71}.
\doiurl{10.1088/0004-637X/759/1/71}.
\end{barticle}
\endbibitem

\bibitem[\protect\citeauthoryear{{Feng} \textit{et~al.}}{2015}]{Feng2015}
\begin{barticle}
\bauthor{\bsnm{{Feng}}, \binits{L.}},
\bauthor{\bsnm{{Wang}}, \binits{Y.}},
\bauthor{\bsnm{{Shen}}, \binits{F.}},
\bauthor{\bsnm{{Shen}}, \binits{C.}},
\bauthor{\bsnm{{Inhester}}, \binits{B.}},
\bauthor{\bsnm{{Lu}}, \binits{L.}},
\bauthor{\bsnm{{Gan}}, \binits{W.}}:
\byear{2015},
\batitle{{Why Does the Apparent Mass of a Coronal Mass Ejection Increase?}}
\bjtitle{\apj}
\bvolume{812},
\bfpage{70}.
\doiurl{10.1088/0004-637X/812/1/70}.
\adsurl{2015ApJ...812...70F}.
\end{barticle}
\endbibitem

\bibitem[\protect\citeauthoryear{{Fletcher} and {Hudson}}{2008}]{Fletcher2008}
\begin{barticle}
\bauthor{\bsnm{{Fletcher}}, \binits{L.}},
\bauthor{\bsnm{{Hudson}}, \binits{H.S.}}:
\byear{2008},
\batitle{{Impulsive Phase Flare Energy Transport by Large-Scale Alfv{\'e}n
  Waves and the Electron Acceleration Problem}}.
\bjtitle{\apj}
\bvolume{675},
\bfpage{1645}.
\doiurl{10.1086/527044}.
\adsurl{2008ApJ...675.1645F}.
\end{barticle}
\endbibitem

\bibitem[\protect\citeauthoryear{{Forbes}}{2000}]{Forbes2000}
\begin{barticle}
\bauthor{\bsnm{{Forbes}}, \binits{T.G.}}:
\byear{2000},
\batitle{{A review on the genesis of coronal mass ejections}}.
\bjtitle{\jgr}
\bvolume{105},
\bfpage{23153}.
\doiurl{10.1029/2000JA000005}.
\adsurl{2000JGR...10523153F}.
\end{barticle}
\endbibitem

\bibitem[\protect\citeauthoryear{{Forbes} and {Priest}}{1984}]{Forbes1984}
\begin{barticle}
\bauthor{\bsnm{{Forbes}}, \binits{T.G.}},
\bauthor{\bsnm{{Priest}}, \binits{E.R.}}:
\byear{1984},.
\bjtitle{NASA Reference Publ.}
\bvolume{1120},
\bfpage{1}.
\end{barticle}
\endbibitem

\bibitem[\protect\citeauthoryear{{Forbes} \textit{et~al.}}{2006}]{Forbes2006}
\begin{barticle}
\bauthor{\bsnm{{Forbes}}, \binits{T.G.}},
\bauthor{\bsnm{{Linker}}, \binits{J.A.}},
\bauthor{\bsnm{{Chen}}, \binits{J.}},
\bauthor{\bsnm{{Cid}}, \binits{C.}},
\bauthor{\bsnm{{K{\'o}ta}}, \binits{J.}},
\bauthor{\bsnm{{Lee}}, \binits{M.A.}},
\bauthor{\bsnm{{Mann}}, \binits{G.}},
\bauthor{\bsnm{{Miki{\'c}}}, \binits{Z.}},
\bauthor{\bsnm{{Potgieter}}, \binits{M.S.}},
\bauthor{\bsnm{{Schmidt}}, \binits{J.M.}},
\bauthor{\bsnm{{Siscoe}}, \binits{G.L.}},
\bauthor{\bsnm{{Vainio}}, \binits{R.}},
\bauthor{\bsnm{{Antiochos}}, \binits{S.K.}},
\bauthor{\bsnm{{Riley}}, \binits{P.}}:
\byear{2006},
\batitle{{CME Theory and Models}}.
\bjtitle{\ssr}
\bvolume{123},
\bfpage{251}.
\doiurl{10.1007/s11214-006-9019-8}.
\adsurl{2006SSRv..123..251F}.
\end{barticle}
\endbibitem

\bibitem[\protect\citeauthoryear{Freidberg}{1987}]{Freidberg1987}
\begin{bbook}
\bauthor{\bsnm{Freidberg}, \binits{J.P.}}:
\byear{1987},
\bbtitle{Ideal magnetohydrodynamics},
\bpublisher{Plenum},
\blocation{New York}.
\end{bbook}
\endbibitem

\bibitem[\protect\citeauthoryear{Furth, Killeen, and
  Rosenbluth}{1963}]{Furth1963}
\begin{barticle}
\bauthor{\bsnm{Furth}, \binits{H.P.}},
\bauthor{\bsnm{Killeen}, \binits{J.}},
\bauthor{\bsnm{Rosenbluth}, \binits{M.N.}}:
\byear{1963},
\batitle{Finite-resistivity instabilities of a sheet pinch}.
\bjtitle{Phys.~Fluids}
\bvolume{6}(\bissue{4}),
\bfpage{459}.
\end{barticle}
\endbibitem

\bibitem[\protect\citeauthoryear{{Goff} \textit{et~al.}}{2007}]{Goff2007}
\begin{barticle}
\bauthor{\bsnm{{Goff}}, \binits{C.P.}},
\bauthor{\bsnm{{van Driel-Gesztelyi}}, \binits{L.}},
\bauthor{\bsnm{{D{\'e}moulin}}, \binits{P.}},
\bauthor{\bsnm{{Culhane}}, \binits{J.L.}},
\bauthor{\bsnm{{Matthews}}, \binits{S.A.}},
\bauthor{\bsnm{{Harra}}, \binits{L.K.}},
\bauthor{\bsnm{{Mandrini}}, \binits{C.H.}},
\bauthor{\bsnm{{Klein}}, \binits{K.L.}},
\bauthor{\bsnm{{Kurokawa}}, \binits{H.}}:
\byear{2007},
\batitle{{A Multiple Flare Scenario where the Classic Long-Duration Flare Was
  Not the Source of a CME}}.
\bjtitle{\solphys}
\bvolume{240},
\bfpage{283}.
\doiurl{10.1007/s11207-007-0260-4}.
\adsurl{2007SoPh..240..283G}.
\end{barticle}
\endbibitem

\bibitem[\protect\citeauthoryear{{Gopalswamy}
  \textit{et~al.}}{2017a}]{Gopalswamy2017b}
\begin{botherref}
\oauthor{\bsnm{{Gopalswamy}}, \binits{N.}},
\oauthor{\bsnm{{Akiyama}}, \binits{S.}},
\oauthor{\bsnm{{Yashiro}}, \binits{S.}},
\oauthor{\bsnm{{Xie}}, \binits{H.}}:
2017a,
{Coronal Flux Ropes and their Interplanetary Counterparts}.
\textit{ArXiv e-prints}.
\adsurl{2017arXiv170508912G}.
\end{botherref}
\endbibitem

\bibitem[\protect\citeauthoryear{{Gopalswamy}
  \textit{et~al.}}{2017b}]{Gopalswamy2017a}
\begin{barticle}
\bauthor{\bsnm{{Gopalswamy}}, \binits{N.}},
\bauthor{\bsnm{{Yashiro}}, \binits{S.}},
\bauthor{\bsnm{{Akiyama}}, \binits{S.}},
\bauthor{\bsnm{{Xie}}, \binits{H.}}:
\byear{2017}b,
\batitle{{Estimation of Reconnection Flux Using Post-eruption Arcades and Its
  Relevance to Magnetic Clouds at 1 AU}}.
\bjtitle{\solphys}
\bvolume{292},
\bfpage{65}.
\doiurl{10.1007/s11207-017-1080-9}.
\adsurl{2017SoPh..292...65G}.
\end{barticle}
\endbibitem

\bibitem[\protect\citeauthoryear{{Green} \textit{et~al.}}{2007}]{Green2007}
\begin{barticle}
\bauthor{\bsnm{{Green}}, \binits{L.M.}},
\bauthor{\bsnm{{Kliem}}, \binits{B.}},
\bauthor{\bsnm{{T{\"o}r{\"o}k}}, \binits{T.}},
\bauthor{\bsnm{{van Driel-Gesztelyi}}, \binits{L.}},
\bauthor{\bsnm{{Attrill}}, \binits{G.D.R.}}:
\byear{2007},
\batitle{{Transient Coronal Sigmoids and Rotating Erupting Flux Ropes}}.
\bjtitle{\solphys}
\bvolume{246},
\bfpage{365}.
\doiurl{10.1007/s11207-007-9061-z}.
\adsurl{2007SoPh..246..365G}.
\end{barticle}
\endbibitem

\bibitem[\protect\citeauthoryear{{Handy} \textit{et~al.}}{1999}]{Handy1999}
\begin{barticle}
\bauthor{\bsnm{{Handy}}, \binits{B.N.}},
\bauthor{\bsnm{{Acton}}, \binits{L.W.}},
\bauthor{\bsnm{{Kankelborg}}, \binits{C.C.}},
\bauthor{\bsnm{{Wolfson}}, \binits{C.J.}},
\bauthor{\bsnm{{Akin}}, \binits{D.J.}},
\bauthor{\bsnm{{Bruner}}, \binits{M.E.}},
\bauthor{\bsnm{{Caravalho}}, \binits{R.}},
\bauthor{\bsnm{{Catura}}, \binits{R.C.}},
\bauthor{\bsnm{{Chevalier}}, \binits{R.}},
\bauthor{\bsnm{{Duncan}}, \binits{D.W.}},
\bauthor{\bsnm{{Edwards}}, \binits{C.G.}},
\bauthor{\bsnm{{Feinstein}}, \binits{C.N.}},
\bauthor{\bsnm{{Freeland}}, \binits{S.L.}},
\bauthor{\bsnm{{Friedlaender}}, \binits{F.M.}},
\bauthor{\bsnm{{Hoffmann}}, \binits{C.H.}},
\bauthor{\bsnm{{Hurlburt}}, \binits{N.E.}},
\bauthor{\bsnm{{Jurcevich}}, \binits{B.K.}},
\bauthor{\bsnm{{Katz}}, \binits{N.L.}},
\bauthor{\bsnm{{Kelly}}, \binits{G.A.}},
\bauthor{\bsnm{{Lemen}}, \binits{J.R.}},
\bauthor{\bsnm{{Levay}}, \binits{M.}},
\bauthor{\bsnm{{Lindgren}}, \binits{R.W.}},
\bauthor{\bsnm{{Mathur}}, \binits{D.P.}},
\bauthor{\bsnm{{Meyer}}, \binits{S.B.}},
\bauthor{\bsnm{{Morrison}}, \binits{S.J.}},
\bauthor{\bsnm{{Morrison}}, \binits{M.D.}},
\bauthor{\bsnm{{Nightingale}}, \binits{R.W.}},
\bauthor{\bsnm{{Pope}}, \binits{T.P.}},
\bauthor{\bsnm{{Rehse}}, \binits{R.A.}},
\bauthor{\bsnm{{Schrijver}}, \binits{C.J.}},
\bauthor{\bsnm{{Shine}}, \binits{R.A.}},
\bauthor{\bsnm{{Shing}}, \binits{L.}},
\bauthor{\bsnm{{Strong}}, \binits{K.T.}},
\bauthor{\bsnm{{Tarbell}}, \binits{T.D.}},
\bauthor{\bsnm{{Title}}, \binits{A.M.}},
\bauthor{\bsnm{{Torgerson}}, \binits{D.D.}},
\bauthor{\bsnm{{Golub}}, \binits{L.}},
\bauthor{\bsnm{{Bookbinder}}, \binits{J.A.}},
\bauthor{\bsnm{{Caldwell}}, \binits{D.}},
\bauthor{\bsnm{{Cheimets}}, \binits{P.N.}},
\bauthor{\bsnm{{Davis}}, \binits{W.N.}},
\bauthor{\bsnm{{Deluca}}, \binits{E.E.}},
\bauthor{\bsnm{{McMullen}}, \binits{R.A.}},
\bauthor{\bsnm{{Warren}}, \binits{H.P.}},
\bauthor{\bsnm{{Amato}}, \binits{D.}},
\bauthor{\bsnm{{Fisher}}, \binits{R.}},
\bauthor{\bsnm{{Maldonado}}, \binits{H.}},
\bauthor{\bsnm{{Parkinson}}, \binits{C.}}:
\byear{1999},
\batitle{{The transition region and coronal explorer}}.
\bjtitle{\solphys}
\bvolume{187},
\bfpage{229}.
\doiurl{10.1023/A:1005166902804}.
\adsurl{1999SoPh..187..229H}.
\end{barticle}
\endbibitem

\bibitem[\protect\citeauthoryear{{Hinterreiter}
  \textit{et~al.}}{2018}]{Hinterreiter2018}
\begin{barticle}
\bauthor{\bsnm{{Hinterreiter}}, \binits{J.}},
\bauthor{\bsnm{{Veronig}}, \binits{A.M.}},
\bauthor{\bsnm{{Thalmann}}, \binits{J.K.}},
\bauthor{\bsnm{{Tschernitz}}, \binits{J.}},
\bauthor{\bsnm{{P{\"o}tzi}}, \binits{W.}}:
\byear{2018},
\batitle{{Statistical Properties of Ribbon Evolution and Reconnection Electric
  Fields in Eruptive and Confined Flares}}.
\bjtitle{\solphys}
\bvolume{293},
\bfpage{38}.
\doiurl{10.1007/s11207-018-1253-1}.
\adsurl{2018SoPh..293...38H}.
\end{barticle}
\endbibitem

\bibitem[\protect\citeauthoryear{{Hudson}}{2011}]{Hudson2011}
\begin{barticle}
\bauthor{\bsnm{{Hudson}}, \binits{H.S.}}:
\byear{2011},
\batitle{{Global Properties of Solar Flares}}.
\bjtitle{\ssr}
\bvolume{158},
\bfpage{5}.
\doiurl{10.1007/s11214-010-9721-4}.
\end{barticle}
\endbibitem

\bibitem[\protect\citeauthoryear{{Inoue} \textit{et~al.}}{2018}]{Inoue2018}
\begin{barticle}
\bauthor{\bsnm{{Inoue}}, \binits{S.}},
\bauthor{\bsnm{{Kusano}}, \binits{K.}},
\bauthor{\bsnm{{B{\"u}chner}}, \binits{J.}},
\bauthor{\bsnm{{Sk{\'a}la}}, \binits{J.}}:
\byear{2018},
\batitle{{Formation and dynamics of a solar eruptive flux tube}}.
\bjtitle{Nature Communications}
\bvolume{9},
\bfpage{174}.
\doiurl{10.1038/s41467-017-02616-8}.
\adsurl{2018NatCo...9..174I}.
\end{barticle}
\endbibitem

\bibitem[\protect\citeauthoryear{{Kahler}}{1992}]{Kahler1992}
\begin{barticle}
\bauthor{\bsnm{{Kahler}}, \binits{S.W.}}:
\byear{1992},
\batitle{{Solar flares and coronal mass ejections}}.
\bjtitle{Ann.~Rev.~Astron.~Astrophys.}
\bvolume{30},
\bfpage{113}.
\doiurl{10.1146/annurev.aa.30.090192.000553}.
\adsurl{1992ARA\%26A..30..113K}.
\end{barticle}
\endbibitem

\bibitem[\protect\citeauthoryear{{Karpen}, {Antiochos}, and
  {DeVore}}{2012}]{Karpen2012}
\begin{barticle}
\bauthor{\bsnm{{Karpen}}, \binits{J.T.}},
\bauthor{\bsnm{{Antiochos}}, \binits{S.K.}},
\bauthor{\bsnm{{DeVore}}, \binits{C.R.}}:
\byear{2012},
\batitle{{The Mechanisms for the Onset and Explosive Eruption of Coronal Mass
  Ejections and Eruptive Flares}}.
\bjtitle{\apj}
\bvolume{760},
\bfpage{81}.
\doiurl{10.1088/0004-637X/760/1/81}.
\adsurl{2012ApJ...760...81K}.
\end{barticle}
\endbibitem

\bibitem[\protect\citeauthoryear{{Kazachenko}
  \textit{et~al.}}{2017}]{Kazachenko2017}
\begin{botherref}
\oauthor{\bsnm{{Kazachenko}}, \binits{M.D.}},
\oauthor{\bsnm{{Lynch}}, \binits{B.J.}},
\oauthor{\bsnm{{Welsch}}, \binits{B.T.}},
\oauthor{\bsnm{{Sun}}, \binits{X.}}:
2017,
{A Database of Flare Ribbon Properties From Solar Dynamics Observatory I:
  Reconnection Flux}.
\textit{ArXiv e-prints}.
\adsurl{2017arXiv170405097K}.
\end{botherref}
\endbibitem

\bibitem[\protect\citeauthoryear{{Kliem} and {T{\"o}r{\"o}k}}{2006}]{Kliem2006}
\begin{barticle}
\bauthor{\bsnm{{Kliem}}, \binits{B.}},
\bauthor{\bsnm{{T{\"o}r{\"o}k}}, \binits{T.}}:
\byear{2006},
\batitle{{Torus Instability}}.
\bjtitle{Physical Review Letters}
\bvolume{96}(\bissue{25}),
\bfpage{255002}.
\doiurl{10.1103/PhysRevLett.96.255002}.
\end{barticle}
\endbibitem

\bibitem[\protect\citeauthoryear{{Kusano} \textit{et~al.}}{2012}]{Kusano2012}
\begin{barticle}
\bauthor{\bsnm{{Kusano}}, \binits{K.}},
\bauthor{\bsnm{{Bamba}}, \binits{Y.}},
\bauthor{\bsnm{{Yamamoto}}, \binits{T.T.}},
\bauthor{\bsnm{{Iida}}, \binits{Y.}},
\bauthor{\bsnm{{Toriumi}}, \binits{S.}},
\bauthor{\bsnm{{Asai}}, \binits{A.}}:
\byear{2012},
\batitle{{Magnetic Field Structures Triggering Solar Flares and Coronal Mass
  Ejections}}.
\bjtitle{\apj}
\bvolume{760},
\bfpage{31}.
\doiurl{10.1088/0004-637X/760/1/31}.
\adsurl{2012ApJ...760...31K}.
\end{barticle}
\endbibitem

\bibitem[\protect\citeauthoryear{{Larson} \textit{et~al.}}{1997}]{Larson1997}
\begin{barticle}
\bauthor{\bsnm{{Larson}}, \binits{D.E.}},
\bauthor{\bsnm{{Lin}}, \binits{R.P.}},
\bauthor{\bsnm{{McTiernan}}, \binits{J.M.}},
\bauthor{\bsnm{{McFadden}}, \binits{J.P.}},
\bauthor{\bsnm{{Ergun}}, \binits{R.E.}},
\bauthor{\bsnm{{McCarthy}}, \binits{M.}},
\bauthor{\bsnm{{R{\` e}me}}, \binits{H.}},
\bauthor{\bsnm{{Sanderson}}, \binits{T.R.}},
\bauthor{\bsnm{{Kaiser}}, \binits{M.}},
\bauthor{\bsnm{{Lepping}}, \binits{R.P.}},
\bauthor{\bsnm{{Mazur}}, \binits{J.}}:
\byear{1997},
\batitle{{Tracing the topology of the October 18-20, 1995, magnetic cloud with
  $\sim 0.1 - 10^{2}$ keV electrons}}.
\bjtitle{\grl}
\bvolume{24},
\bfpage{1911}.
\end{barticle}
\endbibitem

\bibitem[\protect\citeauthoryear{{Leroy}, {Bommier}, and
  {Sahal-Brechot}}{1983}]{Leroy1983}
\begin{barticle}
\bauthor{\bsnm{{Leroy}}, \binits{J.L.}},
\bauthor{\bsnm{{Bommier}}, \binits{V.}},
\bauthor{\bsnm{{Sahal-Brechot}}, \binits{S.}}:
\byear{1983},
\batitle{{The magnetic field in the prominences of the polar crown}}.
\bjtitle{\solphys}
\bvolume{83},
\bfpage{135}.
\doiurl{10.1007/BF00148248}.
\adsurl{1983SoPh...83..135L}.
\end{barticle}
\endbibitem

\bibitem[\protect\citeauthoryear{{Lin}, {Raymond}, and {van
  Ballegooijen}}{2004a}]{Lin2004b}
\begin{barticle}
\bauthor{\bsnm{{Lin}}, \binits{J.}},
\bauthor{\bsnm{{Raymond}}, \binits{J.C.}},
\bauthor{\bsnm{{van Ballegooijen}}, \binits{A.A.}}:
\byear{2004}a,
\batitle{{The Role of Magnetic Reconnection in the Observable Features of Solar
  Eruptions}}.
\bjtitle{\apj}
\bvolume{602},
\bfpage{422}.
\doiurl{10.1086/380900}.
\adsurl{2004ApJ...602..422L}.
\end{barticle}
\endbibitem

\bibitem[\protect\citeauthoryear{{Lin}, {Raymond}, and {van
  Ballegooijen}}{2004b}]{Lin2000}
\begin{barticle}
\bauthor{\bsnm{{Lin}}, \binits{J.}},
\bauthor{\bsnm{{Raymond}}, \binits{J.C.}},
\bauthor{\bsnm{{van Ballegooijen}}, \binits{A.A.}}:
\byear{2004}b,
\batitle{{The Role of Magnetic Reconnection in the Observable Features of Solar
  Eruptions}}.
\bjtitle{\apj}
\bvolume{602},
\bfpage{422}.
\doiurl{10.1086/380900}.
\adsurl{2004ApJ...602..422L}.
\end{barticle}
\endbibitem

\bibitem[\protect\citeauthoryear{{Lin} \textit{et~al.}}{2005}]{Lin2005}
\begin{barticle}
\bauthor{\bsnm{{Lin}}, \binits{J.}},
\bauthor{\bsnm{{Ko}}, \binits{Y.-K.}},
\bauthor{\bsnm{{Sui}}, \binits{L.}},
\bauthor{\bsnm{{Raymond}}, \binits{J.C.}},
\bauthor{\bsnm{{Stenborg}}, \binits{G.A.}},
\bauthor{\bsnm{{Jiang}}, \binits{Y.}},
\bauthor{\bsnm{{Zhao}}, \binits{S.}},
\bauthor{\bsnm{{Mancuso}}, \binits{S.}}:
\byear{2005},
\batitle{{Direct Observations of the Magnetic Reconnection Site of an Eruption
  on 2003 November 18}}.
\bjtitle{\apj}
\bvolume{622},
\bfpage{1251}.
\doiurl{10.1086/428110}.
\adsurl{2005ApJ...622.1251L}.
\end{barticle}
\endbibitem

\bibitem[\protect\citeauthoryear{{Linton} and {Longcope}}{2006}]{Linton2006}
\begin{barticle}
\bauthor{\bsnm{{Linton}}, \binits{M.G.}},
\bauthor{\bsnm{{Longcope}}, \binits{D.W.}}:
\byear{2006},
\batitle{{A Model for Patchy Reconnection in Three Dimensions}}.
\bjtitle{\apj}
\bvolume{642},
\bfpage{1177}.
\doiurl{10.1086/500965}.
\adsurl{2006ApJ...642.1177L}.
\end{barticle}
\endbibitem

\bibitem[\protect\citeauthoryear{{Linton}, {DeVore}, and
  {Longcope}}{2009}]{Linton2009}
\begin{barticle}
\bauthor{\bsnm{{Linton}}, \binits{M.G.}},
\bauthor{\bsnm{{DeVore}}, \binits{C.R.}},
\bauthor{\bsnm{{Longcope}}, \binits{D.W.}}:
\byear{2009},
\batitle{{Patchy reconnection in a Y-type current sheet}}.
\bjtitle{Earth, Planets, and Space}
\bvolume{61},
\bfpage{573}.
\doiurl{10.1186/BF03352925}.
\adsurl{2009EP\%26S...61..573L}.
\end{barticle}
\endbibitem

\bibitem[\protect\citeauthoryear{Longcope}{1996}]{Longcope1996d}
\begin{barticle}
\bauthor{\bsnm{Longcope}, \binits{D.W.}}:
\byear{1996},
\batitle{A model for current ribbon formation and reconnection in a complex
  three-dimensional corona}.
\bjtitle{Solar~Phys.}
\bvolume{169},
\bfpage{91}.
\end{barticle}
\endbibitem

\bibitem[\protect\citeauthoryear{{Longcope} and {Forbes}}{2014}]{Longcope2014}
\begin{barticle}
\bauthor{\bsnm{{Longcope}}, \binits{D.W.}},
\bauthor{\bsnm{{Forbes}}, \binits{T.G.}}:
\byear{2014},
\batitle{{Breakout and Tether-Cutting Eruption Models Are Both Catastrophic
  (Sometimes)}}.
\bjtitle{\solphys}
\bvolume{289},
\bfpage{2091}.
\doiurl{10.1007/s11207-013-0464-8}.
\adsurl{2014SoPh..289.2091L}.
\end{barticle}
\endbibitem

\bibitem[\protect\citeauthoryear{{Lynch} \textit{et~al.}}{2010}]{Lynch2010}
\begin{barticle}
\bauthor{\bsnm{{Lynch}}, \binits{B.J.}},
\bauthor{\bsnm{{Li}}, \binits{Y.}},
\bauthor{\bsnm{{Thernisien}}, \binits{A.F.R.}},
\bauthor{\bsnm{{Robbrecht}}, \binits{E.}},
\bauthor{\bsnm{{Fisher}}, \binits{G.H.}},
\bauthor{\bsnm{{Luhmann}}, \binits{J.G.}},
\bauthor{\bsnm{{Vourlidas}}, \binits{A.}}:
\byear{2010},
\batitle{{Sun to 1 AU propagation and evolution of a slow streamer-blowout
  coronal mass ejection}}.
\bjtitle{Journal of Geophysical Research (Space Physics)}
\bvolume{115}(\bissue{A14}),
\bfpage{7106}.
\doiurl{10.1029/2009JA015099}.
\adsurl{2010JGRA..115.7106L}.
\end{barticle}
\endbibitem

\bibitem[\protect\citeauthoryear{{Lynch} \textit{et~al.}}{2016}]{Lynch2016}
\begin{botherref}
\oauthor{\bsnm{{Lynch}}, \binits{B.J.}},
\oauthor{\bsnm{{Masson}}, \binits{S.}},
\oauthor{\bsnm{{Li}}, \binits{Y.}},
\oauthor{\bsnm{{DeVore}}, \binits{C.R.}},
\oauthor{\bsnm{{Luhmann}}, \binits{J.G.}},
\oauthor{\bsnm{{Antiochos}}, \binits{S.K.}},
\oauthor{\bsnm{{Fisher}}, \binits{G.H.}}:
2016,
{A model for stealth coronal mass ejections}.
\textit{ArXiv e-prints}.
\adsurl{2016arXiv161208323L}.
\end{botherref}
\endbibitem

\bibitem[\protect\citeauthoryear{{Ma} \textit{et~al.}}{2010}]{Ma2010}
\begin{barticle}
\bauthor{\bsnm{{Ma}}, \binits{S.}},
\bauthor{\bsnm{{Attrill}}, \binits{G.D.R.}},
\bauthor{\bsnm{{Golub}}, \binits{L.}},
\bauthor{\bsnm{{Lin}}, \binits{J.}}:
\byear{2010},
\batitle{{Statistical Study of Coronal Mass Ejections With and Without Distinct
  Low Coronal Signatures}}.
\bjtitle{\apj}
\bvolume{722},
\bfpage{289}.
\doiurl{10.1088/0004-637X/722/1/289}.
\adsurl{2010ApJ...722..289M}.
\end{barticle}
\endbibitem

\bibitem[\protect\citeauthoryear{{MacNeice}
  \textit{et~al.}}{2004}]{MacNeice2004}
\begin{barticle}
\bauthor{\bsnm{{MacNeice}}, \binits{P.}},
\bauthor{\bsnm{{Antiochos}}, \binits{S.K.}},
\bauthor{\bsnm{{Phillips}}, \binits{A.}},
\bauthor{\bsnm{{Spicer}}, \binits{D.S.}},
\bauthor{\bsnm{{DeVore}}, \binits{C.R.}},
\bauthor{\bsnm{{Olson}}, \binits{K.}}:
\byear{2004},
\batitle{{A Numerical Study of the Breakout Model for Coronal Mass Ejection
  Initiation}}.
\bjtitle{\apj}
\bvolume{614},
\bfpage{1028}.
\end{barticle}
\endbibitem

\bibitem[\protect\citeauthoryear{{Martin} and {McAllister}}{1996}]{Martin1996}
\begin{bchapter}
\bauthor{\bsnm{{Martin}}, \binits{S.F.}},
\bauthor{\bsnm{{McAllister}}, \binits{A.H.}}:
\byear{1996},
\bctitle{{The Skew of X-ray Coronal Loops Overlying H alpha Filaments}}.
In: \beditor{\bsnm{{Uchida}}, \binits{Y.}},
\beditor{\bsnm{{Kosugi}}, \binits{T.}},
\beditor{\bsnm{{Hudson}}, \binits{H.S.}} (eds.)
\bbtitle{IAU Colloq. 153: Magnetodynamic Phenomena in the Solar Atmosphere -
  Prototypes of Stellar Magnetic Activity},
\bfpage{497}.
\adsurl{1996mpsa.conf..497M}.
\end{bchapter}
\endbibitem

\bibitem[\protect\citeauthoryear{{McKenzie} and {Savage}}{2009}]{McKenzie2009}
\begin{barticle}
\bauthor{\bsnm{{McKenzie}}, \binits{D.E.}},
\bauthor{\bsnm{{Savage}}, \binits{S.L.}}:
\byear{2009},
\batitle{{Quantitative Examination of Supra-arcade Downflows in Eruptive Solar
  Flares}}.
\bjtitle{\apj}
\bvolume{697},
\bfpage{1569}.
\doiurl{10.1088/0004-637X/697/2/1569}.
\adsurl{2009ApJ...697.1569M}.
\end{barticle}
\endbibitem

\bibitem[\protect\citeauthoryear{{McKenzie} and {Savage}}{2011}]{McKenzie2011}
\begin{barticle}
\bauthor{\bsnm{{McKenzie}}, \binits{D.E.}},
\bauthor{\bsnm{{Savage}}, \binits{S.L.}}:
\byear{2011},
\batitle{{Distribution Functions of Sizes and Fluxes Determined from
  Supra-arcade Downflows}}.
\bjtitle{\apjl}
\bvolume{735},
\bfpage{L6}.
\doiurl{10.1088/2041-8205/735/1/L6}.
\adsurl{2011ApJ...735L...6M}.
\end{barticle}
\endbibitem

\bibitem[\protect\citeauthoryear{{Moore} and {Roumeliotis}}{1992}]{Moore1992}
\begin{bchapter}
\bauthor{\bsnm{{Moore}}, \binits{R.L.}},
\bauthor{\bsnm{{Roumeliotis}}, \binits{G.}}:
\byear{1992},
\bctitle{{Triggering of Eruptive Flares - Destabilization of the Preflare
  Magnetic Field Configuration}}.
In: \beditor{\bsnm{{Svestka}}, \binits{Z.}},
\beditor{\bsnm{{Jackson}}, \binits{B.V.}},
\beditor{\bsnm{{Machado}}, \binits{M.E.}} (eds.)
\bbtitle{IAU Colloq. 133: Eruptive Solar Flares},
\bsertitle{Lecture Notes in Physics, Berlin Springer Verlag}
\bseriesno{399},
\bfpage{69}.
\doiurl{10.1007/3-540-55246-4_79}.
\adsurl{1992LNP...399...69M}.
\end{bchapter}
\endbibitem

\bibitem[\protect\citeauthoryear{{Moore} \textit{et~al.}}{2001}]{Moore2001}
\begin{barticle}
\bauthor{\bsnm{{Moore}}, \binits{R.L.}},
\bauthor{\bsnm{{Sterling}}, \binits{A.C.}},
\bauthor{\bsnm{{Hudson}}, \binits{H.S.}},
\bauthor{\bsnm{{Lemen}}, \binits{J.R.}}:
\byear{2001},
\batitle{{Onset of the Magnetic Explosion in Solar Flares and Coronal Mass
  Ejections}}.
\bjtitle{ApJ}
\bvolume{552},
\bfpage{833}.
\end{barticle}
\endbibitem

\bibitem[\protect\citeauthoryear{{Mouschovias} and
  {Poland}}{1978}]{Mouschovias1978}
\begin{barticle}
\bauthor{\bsnm{{Mouschovias}}, \binits{T.C.}},
\bauthor{\bsnm{{Poland}}, \binits{A.I.}}:
\byear{1978},
\batitle{{Expansion and broadening of coronal loop transients - A theoretical
  explanation}}.
\bjtitle{\apj}
\bvolume{220},
\bfpage{675}.
\doiurl{10.1086/155951}.
\adsurl{1978ApJ...220..675M}.
\end{barticle}
\endbibitem

\bibitem[\protect\citeauthoryear{{Poletto} and {Kopp}}{1986}]{Poletto1986}
\begin{bchapter}
\bauthor{\bsnm{{Poletto}}, \binits{G.}},
\bauthor{\bsnm{{Kopp}}, \binits{R.A.}}:
\byear{1986},
\bctitle{{Macroscopic electric fields during two-ribbon flares}}.
In: \beditor{\bsnm{{Neidig}}, \binits{D.F.}} (ed.)
\bbtitle{The lower atmosphere of solar flares; Proceedings of the Solar Maximum
  Mission Symposium, Sunspot, NM, Aug. 20-24, 1985 (A87-26201 10-92). Sunspot,
  NM, National Solar Observatory, 1986, p. 453-465. DOE-sponsored research.},
\bfpage{453}.
\adsurl{1986lasf.conf..453P}.
\end{bchapter}
\endbibitem

\bibitem[\protect\citeauthoryear{{Priest} and {Forbes}}{2000}]{Priest2000}
\begin{bbook}
\bauthor{\bsnm{{Priest}}, \binits{E.R.}},
\bauthor{\bsnm{{Forbes}}, \binits{T.G.}}:
\byear{2000},
\bbtitle{{Magnetic reconnection: MHD theory and applications}},
\bpublisher{Cambridge University Press},
\blocation{New York}.
\adsurl{cgi-bin/nph-bib_query?bibcode=2000mrmt.conf.....P&amp;db_key=AST}.
\end{bbook}
\endbibitem

\bibitem[\protect\citeauthoryear{{Priest} and {Longcope}}{2017}]{Priest2017}
\begin{barticle}
\bauthor{\bsnm{{Priest}}, \binits{E.R.}},
\bauthor{\bsnm{{Longcope}}, \binits{D.W.}}:
\byear{2017},
\batitle{{Flux-Rope Twist in Eruptive Flares and CMEs: Due to Zipper and
  Main-Phase Reconnection}}.
\bjtitle{\solphys}
\bvolume{292},
\bfpage{25}.
\doiurl{10.1007/s11207-016-1049-0}.
\adsurl{2017SoPh..292...25P}.
\end{barticle}
\endbibitem

\bibitem[\protect\citeauthoryear{{Qiu} and {Yurchyshyn}}{2005}]{Qiu2005}
\begin{barticle}
\bauthor{\bsnm{{Qiu}}, \binits{J.}},
\bauthor{\bsnm{{Yurchyshyn}}, \binits{V.B.}}:
\byear{2005},
\batitle{{Magnetic Reconnection Flux and Coronal Mass Ejection Velocity}}.
\bjtitle{\apjl}
\bvolume{634},
\bfpage{L121}.
\doiurl{10.1086/498716}.
\adsurl{2005ApJ...634L.121Q}.
\end{barticle}
\endbibitem

\bibitem[\protect\citeauthoryear{{Qiu} \textit{et~al.}}{2004}]{Qiu2004}
\begin{barticle}
\bauthor{\bsnm{{Qiu}}, \binits{J.}},
\bauthor{\bsnm{{Wang}}, \binits{H.}},
\bauthor{\bsnm{{Cheng}}, \binits{C.Z.}},
\bauthor{\bsnm{{Gary}}, \binits{D.E.}}:
\byear{2004},
\batitle{{Magnetic Reconnection and Mass Acceleration in Flare-Coronal Mass
  Ejection Events}}.
\bjtitle{\apj}
\bvolume{604},
\bfpage{900}.
\doiurl{10.1086/382122}.
\end{barticle}
\endbibitem

\bibitem[\protect\citeauthoryear{{Qiu} \textit{et~al.}}{2007}]{Qiu2007}
\begin{barticle}
\bauthor{\bsnm{{Qiu}}, \binits{J.}},
\bauthor{\bsnm{{Hu}}, \binits{Q.}},
\bauthor{\bsnm{{Howard}}, \binits{T.A.}},
\bauthor{\bsnm{{Yurchyshyn}}, \binits{V.B.}}:
\byear{2007},
\batitle{{On the Magnetic Flux Budget in Low-Corona Magnetic Reconnection and
  Interplanetary Coronal Mass Ejections}}.
\bjtitle{\apj}
\bvolume{659},
\bfpage{758}.
\doiurl{10.1086/512060}.
\adsurl{2007ApJ...659..758Q}.
\end{barticle}
\endbibitem

\bibitem[\protect\citeauthoryear{{Rachmeler}
  \textit{et~al.}}{2010}]{Rachmeler2010}
\begin{barticle}
\bauthor{\bsnm{{Rachmeler}}, \binits{L.A.}},
\bauthor{\bsnm{{Pariat}}, \binits{E.}},
\bauthor{\bsnm{{DeForest}}, \binits{C.E.}},
\bauthor{\bsnm{{Antiochos}}, \binits{S.}},
\bauthor{\bsnm{{T{\"o}r{\"o}k}}, \binits{T.}}:
\byear{2010},
\batitle{{Symmetric Coronal Jets: A Reconnection-controlled Study}}.
\bjtitle{\apj}
\bvolume{715},
\bfpage{1556}.
\doiurl{10.1088/0004-637X/715/2/1556}.
\adsurl{2010ApJ...715.1556R}.
\end{barticle}
\endbibitem

\bibitem[\protect\citeauthoryear{{Robbrecht}, {Patsourakos}, and
  {Vourlidas}}{2009}]{Robbrecht2009}
\begin{barticle}
\bauthor{\bsnm{{Robbrecht}}, \binits{E.}},
\bauthor{\bsnm{{Patsourakos}}, \binits{S.}},
\bauthor{\bsnm{{Vourlidas}}, \binits{A.}}:
\byear{2009},
\batitle{{No Trace Left Behind: STEREO Observation of a Coronal Mass Ejection
  Without Low Coronal Signatures}}.
\bjtitle{\apj}
\bvolume{701},
\bfpage{283}.
\doiurl{10.1088/0004-637X/701/1/283}.
\adsurl{2009ApJ...701..283R}.
\end{barticle}
\endbibitem

\bibitem[\protect\citeauthoryear{{Savage} and {McKenzie}}{2011}]{Savage2011}
\begin{barticle}
\bauthor{\bsnm{{Savage}}, \binits{S.L.}},
\bauthor{\bsnm{{McKenzie}}, \binits{D.E.}}:
\byear{2011},
\batitle{{Quantitative Examination of a Large Sample of Supra-arcade Downflows
  in Eruptive Solar Flares}}.
\bjtitle{\apj}
\bvolume{730},
\bfpage{98}.
\doiurl{10.1088/0004-637X/730/2/98}.
\adsurl{2011ApJ...730...98S}.
\end{barticle}
\endbibitem

\bibitem[\protect\citeauthoryear{{Schmieder}
  \textit{et~al.}}{1996}]{Schmieder1996}
\begin{barticle}
\bauthor{\bsnm{{Schmieder}}, \binits{B.}},
\bauthor{\bsnm{{Demoulin}}, \binits{P.}},
\bauthor{\bsnm{{Aulanier}}, \binits{G.}},
\bauthor{\bsnm{{Golub}}, \binits{L.}}:
\byear{1996},
\batitle{{Differential Magnetic Field Shear in an Active Region}}.
\bjtitle{\apj}
\bvolume{467},
\bfpage{881}.
\doiurl{10.1086/177662}.
\adsurl{1996ApJ...467..881S}.
\end{barticle}
\endbibitem

\bibitem[\protect\citeauthoryear{{Schrijver}
  \textit{et~al.}}{2008}]{Schrijver2008}
\begin{barticle}
\bauthor{\bsnm{{Schrijver}}, \binits{C.J.}},
\bauthor{\bsnm{{De Rosa}}, \binits{M.L.}},
\bauthor{\bsnm{{Metcalf}}, \binits{T.}},
\bauthor{\bsnm{{Barnes}}, \binits{G.}},
\bauthor{\bsnm{{Lites}}, \binits{B.}},
\bauthor{\bsnm{{Tarbell}}, \binits{T.}},
\bauthor{\bsnm{{McTiernan}}, \binits{J.}},
\bauthor{\bsnm{{Valori}}, \binits{G.}},
\bauthor{\bsnm{{Wiegelmann}}, \binits{T.}},
\bauthor{\bsnm{{Wheatland}}, \binits{M.S.}},
\bauthor{\bsnm{{Amari}}, \binits{T.}},
\bauthor{\bsnm{{Aulanier}}, \binits{G.}},
\bauthor{\bsnm{{D{\'e}moulin}}, \binits{P.}},
\bauthor{\bsnm{{Fuhrmann}}, \binits{M.}},
\bauthor{\bsnm{{Kusano}}, \binits{K.}},
\bauthor{\bsnm{{R{\'e}gnier}}, \binits{S.}},
\bauthor{\bsnm{{Thalmann}}, \binits{J.K.}}:
\byear{2008},
\batitle{{Nonlinear Force-free Field Modeling of a Solar Active Region around
  the Time of a Major Flare and Coronal Mass Ejection}}.
\bjtitle{\apj}
\bvolume{675},
\bfpage{1637}.
\doiurl{10.1086/527413}.
\end{barticle}
\endbibitem

\bibitem[\protect\citeauthoryear{{Schuck}}{2010}]{Schuck2010}
\begin{barticle}
\bauthor{\bsnm{{Schuck}}, \binits{P.W.}}:
\byear{2010},
\batitle{{The Photospheric Energy and Helicity Budgets of the Flux-injection
  Hypothesis}}.
\bjtitle{\apj}
\bvolume{714},
\bfpage{68}.
\doiurl{10.1088/0004-637X/714/1/68}.
\end{barticle}
\endbibitem

\bibitem[\protect\citeauthoryear{{Shafranov}}{1966}]{Shafranov1966}
\begin{barticle}
\bauthor{\bsnm{{Shafranov}}, \binits{V.D.}}:
\byear{1966},
\batitle{{Plasma Equilibrium in a Magnetic Field}}.
\bjtitle{Reviews of Plasma Physics}
\bvolume{2},
\bfpage{103}.
\adsurl{1966RvPP....2..103S}.
\end{barticle}
\endbibitem

\bibitem[\protect\citeauthoryear{{Somov} and {Bogachev}}{2003}]{Somov2003}
\begin{barticle}
\bauthor{\bsnm{{Somov}}, \binits{B.V.}},
\bauthor{\bsnm{{Bogachev}}, \binits{S.A.}}:
\byear{2003},
\batitle{{The Betatron Effect in Collapsing Magnetic Traps}}.
\bjtitle{Astronomy Letters}
\bvolume{29},
\bfpage{621}.
\doiurl{10.1134/1.1607500}.
\adsurl{2003AstL...29..621S}.
\end{barticle}
\endbibitem

\bibitem[\protect\citeauthoryear{Spruit}{1981}]{Spruit1981}
\begin{barticle}
\bauthor{\bsnm{Spruit}, \binits{H.C.}}:
\byear{1981},
\batitle{Motion of magnetic flux tubes in the solar convection zone and
  chromosphere}.
\bjtitle{A\&A}
\bvolume{98},
\bfpage{155}.
\end{barticle}
\endbibitem

\bibitem[\protect\citeauthoryear{{Sturrock}}{1989}]{Sturrock1989}
\begin{barticle}
\bauthor{\bsnm{{Sturrock}}, \binits{P.A.}}:
\byear{1989},
\batitle{{The role of eruption in solar flares}}.
\bjtitle{\solphys}
\bvolume{121},
\bfpage{387}.
\doiurl{10.1007/BF00161708}.
\adsurl{1989SoPh..121..387S}.
\end{barticle}
\endbibitem

\bibitem[\protect\citeauthoryear{{Su}, {Golub}, and {Van
  Ballegooijen}}{2007}]{Su2007a}
\begin{barticle}
\bauthor{\bsnm{{Su}}, \binits{Y.}},
\bauthor{\bsnm{{Golub}}, \binits{L.}},
\bauthor{\bsnm{{Van Ballegooijen}}, \binits{A.A.}}:
\byear{2007},
\batitle{{A Statistical Study of Shear Motion of the Footpoints in Two-Ribbon
  Flares}}.
\bjtitle{\apj}
\bvolume{655},
\bfpage{606}.
\doiurl{10.1086/510065}.
\adsurl{2007ApJ...655..606S}.
\end{barticle}
\endbibitem

\bibitem[\protect\citeauthoryear{{Svestka} and {Cliver}}{1992}]{Svestka1992}
\begin{bchapter}
\bauthor{\bsnm{{Svestka}}, \binits{Z.}},
\bauthor{\bsnm{{Cliver}}, \binits{E.W.}}:
\byear{1992},
\bctitle{{History and Basic Characteristics of Eruptive Flares}}.
In: \beditor{\bsnm{{Svestka}}, \binits{Z.}},
\beditor{\bsnm{{Jackson}}, \binits{B.V.}},
\beditor{\bsnm{{Machado}}, \binits{M.E.}} (eds.)
\bbtitle{IAU Colloq. 133: Eruptive Solar Flares},
\bsertitle{Lecture Notes in Physics, Berlin Springer Verlag}
\bseriesno{399},
\bfpage{1}.
\doiurl{10.1007/3-540-55246-4_70}.
\adsurl{1992LNP...399....1S}.
\end{bchapter}
\endbibitem

\bibitem[\protect\citeauthoryear{{Vourlidas}
  \textit{et~al.}}{2000}]{Vourlidas2000}
\begin{barticle}
\bauthor{\bsnm{{Vourlidas}}, \binits{A.}},
\bauthor{\bsnm{{Subramanian}}, \binits{P.}},
\bauthor{\bsnm{{Dere}}, \binits{K.P.}},
\bauthor{\bsnm{{Howard}}, \binits{R.A.}}:
\byear{2000},
\batitle{{Large-Angle Spectrometric Coronagraph Measurements of the Energetics
  of Coronal Mass Ejections}}.
\bjtitle{\apj}
\bvolume{534},
\bfpage{456}.
\doiurl{10.1086/308747}.
\adsurl{2000ApJ...534..456V}.
\end{barticle}
\endbibitem

\bibitem[\protect\citeauthoryear{{Vourlidas}
  \textit{et~al.}}{2002}]{Vourlidas2002}
\begin{bchapter}
\bauthor{\bsnm{{Vourlidas}}, \binits{A.}},
\bauthor{\bsnm{{Buzasi}}, \binits{D.}},
\bauthor{\bsnm{{Howard}}, \binits{R.A.}},
\bauthor{\bsnm{{Esfandiari}}, \binits{E.}}:
\byear{2002},
\bctitle{{Mass and energy properties of LASCO CMEs}}.
In: \beditor{\bsnm{{A.~Wilson}}} (ed.)
\bbtitle{Solar Variability: From Core to Outer Frontiers},
\bsertitle{ESA Special Publication}
\bseriesno{506},
\bfpage{91}.
\adsurl{2002ESASP.506...91V}.
\end{bchapter}
\endbibitem

\bibitem[\protect\citeauthoryear{{Vr{\v s}nak}}{2006}]{Vrsnak2006}
\begin{barticle}
\bauthor{\bsnm{{Vr{\v s}nak}}, \binits{B.}}:
\byear{2006},
\batitle{{Forces governing coronal mass ejections}}.
\bjtitle{Advances in Space Research}
\bvolume{38},
\bfpage{431}.
\doiurl{10.1016/j.asr.2005.03.090}.
\adsurl{2006AdSpR..38..431V}.
\end{barticle}
\endbibitem

\bibitem[\protect\citeauthoryear{{Vr{\v s}nak}, {Vrbanec}, and {{\v
  C}alogovi{\'c}}}{2008}]{Vrsnak2008}
\begin{barticle}
\bauthor{\bsnm{{Vr{\v s}nak}}, \binits{B.}},
\bauthor{\bsnm{{Vrbanec}}, \binits{D.}},
\bauthor{\bsnm{{{\v C}alogovi{\'c}}}, \binits{J.}}:
\byear{2008},
\batitle{{Dynamics of coronal mass ejections. The mass-scaling of the
  aerodynamic drag}}.
\bjtitle{\aap}
\bvolume{490},
\bfpage{811}.
\doiurl{10.1051/0004-6361:200810215}.
\adsurl{2008A\%26A...490..811V}.
\end{barticle}
\endbibitem

\bibitem[\protect\citeauthoryear{{Vr{\v s}nak}
  \textit{et~al.}}{2004}]{Vrsnak2004}
\begin{barticle}
\bauthor{\bsnm{{Vr{\v s}nak}}, \binits{B.}},
\bauthor{\bsnm{{Mari{\v c}i{\'c}}}, \binits{D.}},
\bauthor{\bsnm{{Stanger}}, \binits{A.L.}},
\bauthor{\bsnm{{Veronig}}, \binits{A.}}:
\byear{2004},
\batitle{{Coronal Mass Ejection of 15 May 2001: II. Coupling of the Cme
  Acceleration and the Flare Energy Release}}.
\bjtitle{\solphys}
\bvolume{225},
\bfpage{355}.
\doiurl{10.1007/s11207-004-4995-x}.
\adsurl{2004SoPh..225..355V}.
\end{barticle}
\endbibitem

\bibitem[\protect\citeauthoryear{{Webb} and {Howard}}{2012}]{Webb2012}
\begin{barticle}
\bauthor{\bsnm{{Webb}}, \binits{D.F.}},
\bauthor{\bsnm{{Howard}}, \binits{T.A.}}:
\byear{2012},
\batitle{{Coronal Mass Ejections: Observations}}.
\bjtitle{Living Reviews in Solar Physics}
\bvolume{9},
\bfpage{3}.
\doiurl{10.12942/lrsp-2012-3}.
\adsurl{2012LRSP....9....3W}.
\end{barticle}
\endbibitem

\bibitem[\protect\citeauthoryear{{Williams}
  \textit{et~al.}}{2005}]{Williams2005}
\begin{barticle}
\bauthor{\bsnm{{Williams}}, \binits{D.R.}},
\bauthor{\bsnm{{T{\"o}r{\"o}k}}, \binits{T.}},
\bauthor{\bsnm{{D{\'e}moulin}}, \binits{P.}},
\bauthor{\bsnm{{van Driel-Gesztelyi}}, \binits{L.}},
\bauthor{\bsnm{{Kliem}}, \binits{B.}}:
\byear{2005},
\batitle{{Eruption of a Kink-unstable Filament in NOAA Active Region 10696}}.
\bjtitle{\apjl}
\bvolume{628},
\bfpage{L163}.
\doiurl{10.1086/432910}.
\adsurl{2005ApJ...628L.163W}.
\end{barticle}
\endbibitem

\bibitem[\protect\citeauthoryear{{Yamada} \textit{et~al.}}{1997}]{Yamada1997}
\begin{barticle}
\bauthor{\bsnm{{Yamada}}, \binits{M.}},
\bauthor{\bsnm{{Ji}}, \binits{H.}},
\bauthor{\bsnm{{Hsu}}, \binits{S.}},
\bauthor{\bsnm{{Carter}}, \binits{T.}},
\bauthor{\bsnm{{Kulsrud}}, \binits{R.}},
\bauthor{\bsnm{{Bretz}}, \binits{N.}},
\bauthor{\bsnm{{Jobes}}, \binits{F.}},
\bauthor{\bsnm{{Ono}}, \binits{Y.}},
\bauthor{\bsnm{{Perkins}}, \binits{F.}}:
\byear{1997},
\batitle{{Study of driven magnetic reconnection in a laboratory plasma}}.
\bjtitle{Physics of Plasmas}
\bvolume{4},
\bfpage{1936}.
\doiurl{10.1063/1.872336}.
\adsurl{1997PhPl....4.1936Y}.
\end{barticle}
\endbibitem

\bibitem[\protect\citeauthoryear{{Zhang} and {Dere}}{2006}]{Zhang2006}
\begin{barticle}
\bauthor{\bsnm{{Zhang}}, \binits{J.}},
\bauthor{\bsnm{{Dere}}, \binits{K.P.}}:
\byear{2006},
\batitle{{A Statistical Study of Main and Residual Accelerations of Coronal
  Mass Ejections}}.
\bjtitle{\apj}
\bvolume{649},
\bfpage{1100}.
\doiurl{10.1086/506903}.
\adsurl{2006ApJ...649.1100Z}.
\end{barticle}
\endbibitem

\bibitem[\protect\citeauthoryear{{Zhao} \textit{et~al.}}{2016}]{Zhao2016}
\begin{barticle}
\bauthor{\bsnm{{Zhao}}, \binits{J.}},
\bauthor{\bsnm{{Gilchrist}}, \binits{S.A.}},
\bauthor{\bsnm{{Aulanier}}, \binits{G.}},
\bauthor{\bsnm{{Schmieder}}, \binits{B.}},
\bauthor{\bsnm{{Pariat}}, \binits{E.}},
\bauthor{\bsnm{{Li}}, \binits{H.}}:
\byear{2016},
\batitle{{Hooked Flare Ribbons and Flux-rope-related QSL Footprints}}.
\bjtitle{\apj}
\bvolume{823},
\bfpage{62}.
\doiurl{10.3847/0004-637X/823/1/62}.
\adsurl{2016ApJ...823...62Z}.
\end{barticle}
\endbibitem

\bibitem[\protect\citeauthoryear{{Zhao}, {Plunkett}, and
  {Liu}}{2002}]{Zhao2002}
\begin{barticle}
\bauthor{\bsnm{{Zhao}}, \binits{X.P.}},
\bauthor{\bsnm{{Plunkett}}, \binits{S.P.}},
\bauthor{\bsnm{{Liu}}, \binits{W.}}:
\byear{2002},
\batitle{{Determination of geometrical and kinematical properties of halo
  coronal mass ejections using the cone model}}.
\bjtitle{Journal of Geophysical Research (Space Physics)}
\bvolume{107},
\bfpage{1223}.
\doiurl{10.1029/2001JA009143}.
\adsurl{2002JGRA..107.1223Z}.
\end{barticle}
\endbibitem

\end{thebibliography}

\end{document}